\documentclass[amsmath,superscriptaddress,amssymb,aps,twocolumn,prl]{revtex4}
\usepackage{graphicx}
\usepackage{soul}
\usepackage[colorlinks=true,citecolor=blue,linkcolor=magenta]{hyperref}
\usepackage[usenames]{color}

\newcommand{\ket}[1]{\left\vert #1 \right\rangle}
\newcommand{\bra}[1]{\left\langle #1 \right\vert}
\newcommand{\ketbra}[2]{\ket{ #1}\bra{ #2} }
\newcommand{\bla}[1]{\left( #1 \right)}
\newcommand{\blb}[1]{\left[ #1 \right]}

\def \ket#1{|{#1}\rangle}
\def \bra#1{\langle{#1}|}

\def \be{\begin{equation}}
\def \ee{\end{equation}}
\def \ba{\begin{array}}
\def \ea{\end{array}}
\def \bea{\begin{eqnarray}}
\def \eea{\end{eqnarray}}

\def \l{\left}
\def \r{\right}

\def \p{p}

\renewcommand{\phi}{\varphi}

\newcommand{\ele}[2]{^{#1} \mbox{#2}}

\begin{document}

\title{Towards a large-scale quantum simulator on diamond surface at room temperature}


\author{Jianming Cai}
\affiliation{Institut f\"{u}r Theoretische Physik, Albert-Einstein Allee 11, Universit\"{a}t Ulm, 89069 Ulm, Germany}
\affiliation{Center for Integrated Quantum Science and Technology, Universit\"{a}t Ulm, 89069 Ulm, Germany}
\author{ Alex Retzker}
\affiliation{Institut f\"{u}r Theoretische Physik, Albert-Einstein Allee 11, Universit\"{a}t Ulm, 89069 Ulm, Germany}
\affiliation{Racah Institute of Physics, The Hebrew University of Jerusalem, Jerusalem 91904, Givat Ram, Israel}
\author{Fedor Jelezko}
\affiliation{Center for Integrated Quantum Science and Technology, Universit\"{a}t Ulm, 89069 Ulm, Germany}
\affiliation{Institut f\"{u}r Quantenoptik, Albert-Einstein Allee 11, Universit\"{a}t Ulm, 89069 Ulm, Germany}
\author{Martin B. Plenio}
\affiliation{Institut f\"{u}r Theoretische Physik, Albert-Einstein Allee 11, Universit\"{a}t Ulm, 89069 Ulm, Germany}
\affiliation{Center for Integrated Quantum Science and Technology, Universit\"{a}t Ulm, 89069 Ulm, Germany}

\date{\today}

\begin{abstract}
Strongly-correlated quantum many-body systems exhibits a variety of exotic phases with long-range quantum correlations, such as spin liquids and supersolids. Despite the rapid increase in computational power of modern computers, the numerical simulation of these complex systems becomes intractable even for a few dozens of particles. Feynman's idea of quantum simulators offers an innovative way to bypass this computational barrier. However, the proposed realizations of such devices either require very low temperatures (ultracold gases in optical lattices, trapped ions, superconducting devices) and considerable technological effort, or are extremely hard to scale in practice (NMR, linear optics). In this work, we propose a new architecture for a scalable quantum simulator that can operate at room temperature. It consists of strongly-interacting nuclear spins attached to the diamond surface by its direct chemical treatment, or by means of a functionalized graphene sheet. The initialization, control and read-out of this quantum simulator can be accomplished with nitrogen-vacancy centers implanted in diamond. The system can be engineered to simulate a wide variety of interesting strongly-correlated models with long-range dipole-dipole interactions. Due to the superior coherence time of nuclear spins and nitrogen-vacancy centers in diamond, our proposal offers new opportunities towards large-scale quantum simulation at room temperatures.

\end{abstract}

\maketitle


Many intriguing phenomena in condensed-matter systems originate from the interplay of strong interactions and frustrations. A representative example is frustrated quantum magnetism, where the spins cannot order to minimize all local interactions, and the ground state is highly degenerate \cite{Lacroix2011,Sachdev08}. Together with long-range interactions between non-nearest neighbors, the frustrated quantum models give rise to intriguing quantum phases, e.g. supersolid \cite{Leggett70,Kim04}. Moreover, they can also stabilize the long-sought quantum spin liquid \cite{Balents10,Meng10,Varney11}, which has connections with high-temperature superconductivity \cite{Anderson87}.  However, the properties of these systems have proven to be very hard to understand from numerical calculations, partly due to the combination of long-range quantum correlations and the superposition principle of quantum mechanics \cite{Sachdev01}. This principle implies that the required computational resources grow exponentially with the number of particles, making numerical approaches inefficient. Richard Feynman's idea \cite{Feynman82} of quantum simulation provides a powerful solution to this problem: one could gain deep insight into complex states of matter by experimentally simulating them with another well-controlled quantum system \cite{Lloyd96}. Large scale quantum simulations is expected to be a powerful tool \cite{Buluta09} for the investigation of fundamental problems in condensed-matter physics.

Quantum simulation has attracted extensive research interest in the last decade. Various architectures for quantum simulations have been constructed based on different systems, ranging from ultracold neutral atoms \cite{Bloch12,Simon11,Struck11}, trapped ions \cite{Blatt12,Britton12}, and photonic systems \cite{Wal12} to superconducting circuits in solid-state devices \cite{Houck12}. The still challenging goal is to realize a large-scale quantum simulator that cannot be efficiently solved numerically with classical computers, which will probably require going beyond one-dimensional systems. In this work, we propose a scalable architecture that is of practical interest for large-scale quantum simulation. Our quantum simulator is based on lattices of interacting nuclear spins, which can be fabricated chemically on the diamond surface \cite{Ristein06,Sen09} or by depositing functionalized graphene films \cite{Nair10,Elias09,Yu12}. We propose to implant the nitrogen-vacancy (NV) centers \cite{Bala09} beneath the diamond surface, and exploit them as an efficient control element for cooling, spin-spin interaction engineering, and read-out of the nuclear-spin quantum simulator. 

We explain how this simulator is constructed, and establish schemes for its initialization, control and read-out. We analyze its validity by detailed numerical studies, thus demonstrating the feasibility of our proposal within the current experimental capabilities. 

\begin{figure*}
\begin{center}
\hspace{-1cm}
\includegraphics[width=6.8cm]{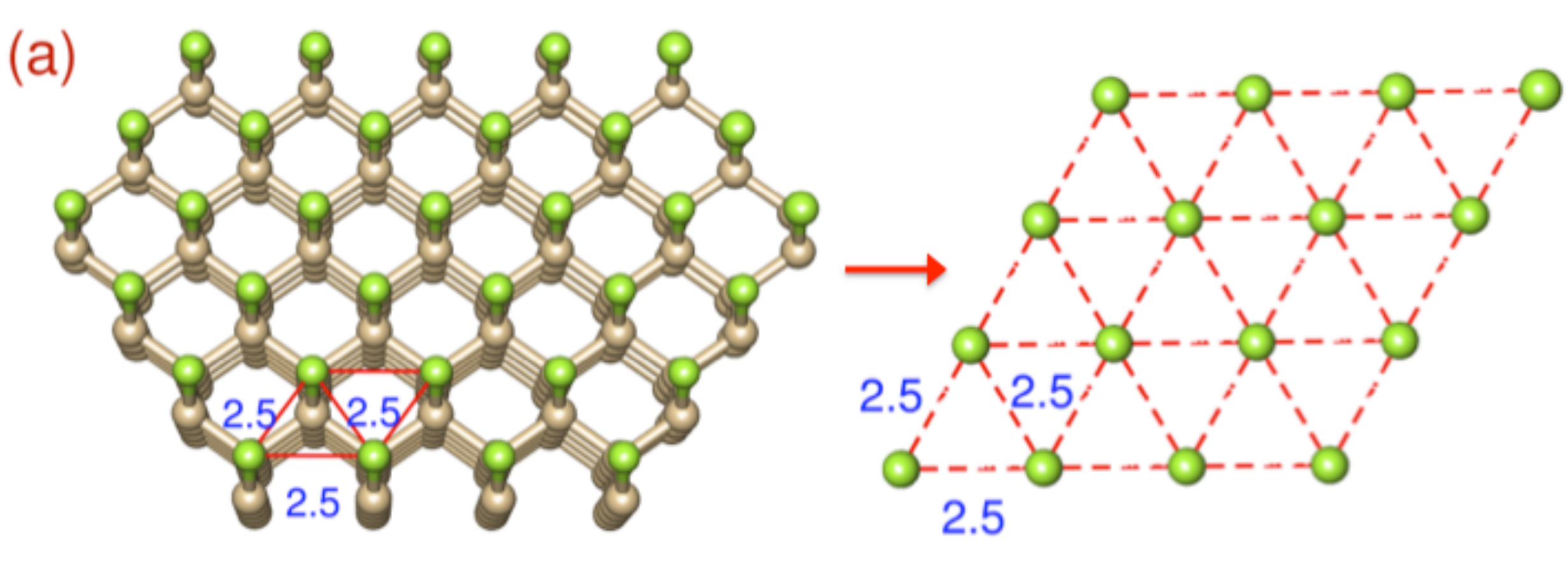}
\hspace{0.4cm}
\includegraphics[width=5.8cm]{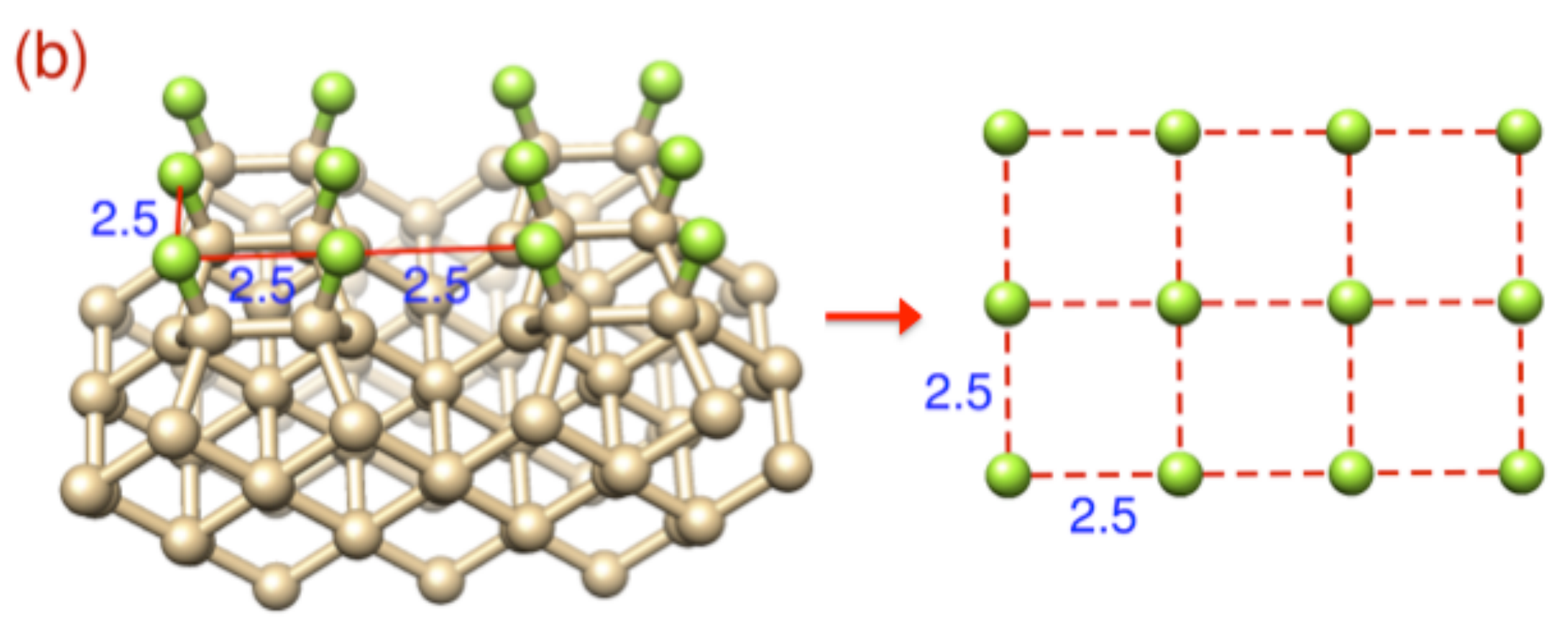}
\hspace{0.4cm}
\includegraphics[width=3cm]{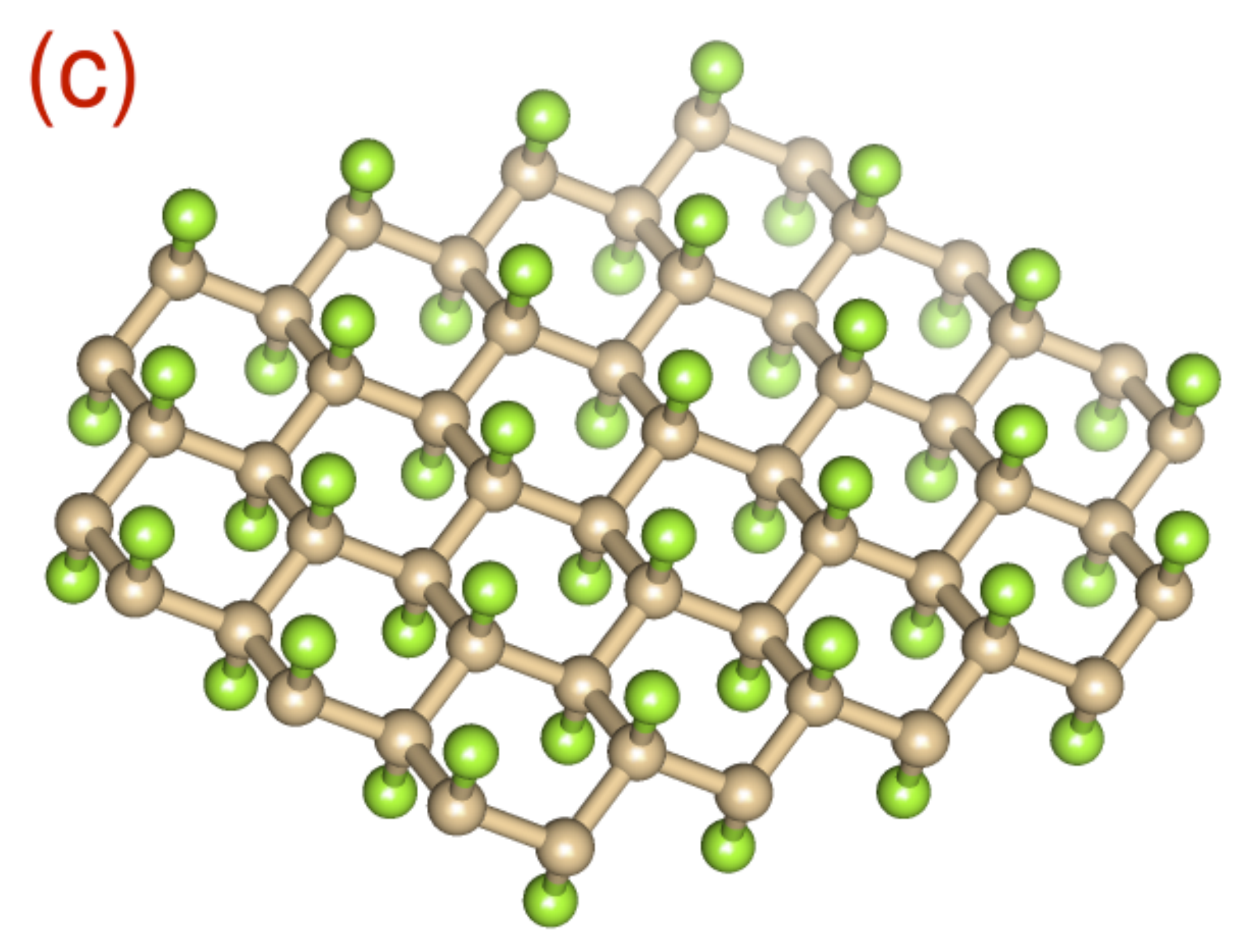}
\end{center}
\caption{{\bf Lattices of fluorine nuclear spin}. (a) A triangular nuclear spin lattice on the fluorine-terminated diamond (111) surface. (b) A rectangular nuclear spin lattice on the fluorine-terminated diamond (100) 2$\times$1 surface. The distance between two nearest neighbour fluorine atoms is about $2.5$ (unit $\mathring{\mbox{A}}$). (c) A double-layer triangular lattice from fluorographene. Yellow atoms represent carbon, and green atoms represents fluorine.}\label{fig:lattice}
\end{figure*}

{\em Construction of Hardware--} We discuss two main paths for the fabrication of the hardware for our quantum simulator. Firstly, large-scale lattices of nuclear spins can be constructed by chemically-controlled termination of diamond surfaces. The fluorine ($^{19}\mbox{F}$ with spin $\frac{1}{2}$), oxygen ($^{17}\mbox{O}$ with spin $\frac{5}{2}$) and hydrogen/hydroxyl group ($^{1}\mbox{H}$ with spin $\frac{1}{2}$ and $^{2}\mbox{H}$ with spin $1$) termination of the diamond surface can be obtained from the process of chemical vapor deposition (CVD) ,or by functionalizing the diamond surface. As a representative example, we will mainly concentrate on fluorine-terminated diamond surface, which has a positive electron affinity. The two most important diamond surfaces are the (111) and (100) surfaces, which constitute the crystal faces of polycrystalline CVD diamond films and can be selectively grown with appropriately controlled process parameters \cite{Ristein06}. The (111) surface of diamond is the natural cleaving plane of diamond, and has one dangling bond per surface carbon atom which is terminated by carbon-fluorine bonds on the fluorine-terminated diamond surface \cite{Smen96}. The fluorine atoms naturally arrange in a triangular lattice with nearest-neighbour distances of about $2.5 \mathring{\mbox{A}}$ \cite{Ristein06,Sen09}, see Fig.\ref{fig:lattice}(a). The (100) surface of diamond shows two dangling bonds per surface carbon atom, which will undergo a reconstruction into a 2$\times$1 geometry with neighboring surface carbon atoms forming a $\pi$-bonded dimer. The remaining dangling bonds are terminated by carbon-fluorine bonds, which leads to a rectangular lattice of fluorine atoms \cite{Ristein06,Sen09}, see Fig.\ref{fig:lattice}(b). Functionalized graphene (fluorographene) provides a double-layer triangular lattice of fluorine atoms, see Fig.\ref{fig:lattice}(c). This can be obtained through the mechanical cleavage of graphite fluoride, or by exposing graphene to atomic fluorine formed by decomposition of xenon difluoride ($\mbox{XeF}_2$) \cite{Nair10}. 

In addition to the nuclear spins, we shall introduce NV centers by shallow implantation a few nanometers below the surface of diamond \cite{Okai12,Ohno1207}. This constitutes a fundamental ingredient of our quantum simulator, since it allows for the initialization and read-out of the nuclear spins. Let us remark that in contrast to graphene, fluorographene exhibits a large band gap of more than $3.5 \mbox{eV}$, which is larger than the optical gap of NV centers in diamond ($\sim 2.9\mbox{eV}$). This avoids the unwanted fluorescence quenching of NV centers and is thus crucial for using the NV centers for the control of the nuclear spin arrays.

\emph{Engineering of Interacting Hamiltonian--} The nuclear spins interact with each other via magnetic dipole-dipole interactions as $V_{ij}=g(r_{ij})[ \mathbf{s}_i \cdot \mathbf{s}_j-3\l( \mathbf{s}_i \cdot\hat{\mathbf{r}}_{ij}\r) \l(\mathbf{s}_j \cdot\hat{\mathbf{r}}_{ij} \r)]$, 
where $\mathbf{s}_i $ are the nuclear spin operators, $g(r_{ij})=(\hbar^2 \mu_0\gamma_i \gamma_j )/(4\pi r_{ij}^3)$, $\gamma_i$, $\gamma_j$ are gyromagnetic ratios of the $i$th and $j$th nuclear spin, which are connected by the vector $\mathbf{r}_{ij}=r_{ij} \hat{\mathbf{r}}_{ij}$. Since nearest-neighbor fluorine nuclear spins are at a distance of $2.5\mathring{\mbox{A}}$, a coupling strength of $g(a)\simeq 6.8 \mbox{kHz}$ is achieved. A strong magnetic field, which leads to energy level shifts exceeding the nuclear spin coupling strength, simplifies the effective Hamiltonian due to the rotating wave approximation(RWA) to an XXZ model
\bea
V_{ij}=g(\mathbf{r}_{ij})\blb{ \mathbf{s}_i^z \mathbf{s}_j^z-\Delta \bla{\mathbf{s}_i^x \mathbf{s}_j^x+\mathbf{s}_i^y \mathbf{s}_j^y}}
\eea
with $\Delta =\frac{1}{2}$. Our calculations verify that such a RWA can already be well satisfied for a magnetic field as low as $750\mbox{G}$ (3$\mbox{MHz}$), see SI for details. We denote the diamond surface, on which the nuclear spin lattice is constructed as the X-Y plane, while the vector perpendicular to it defines the Z axis (spatial axes), and the magnetic field direction as $\hat{m}$ axis which gives the quantization of nuclear spins. The spin-spin interaction strength can be tuned by changing the direction of magnetic field as
\be
g(\mathbf{r}_{ij})=g(r_{ij}) \l(1-3\cos^2\theta_{ij}\r)
\ee
where $\theta_{ij}$ is the relative angle between $\mathbf{r}_{ij}$ and $\hat{m}$. Thus, by changing the direction of the external magnetic fields, we can control the spatial anisotropy and the sign of the interaction strength, which determines whether the interaction is ferromagnetic or anti-ferromagnetic and thereby induces spin frustration. 
\begin{figure}[b]
\begin{center}
\begin{minipage}{6.5cm}
\includegraphics[width=6.5cm]{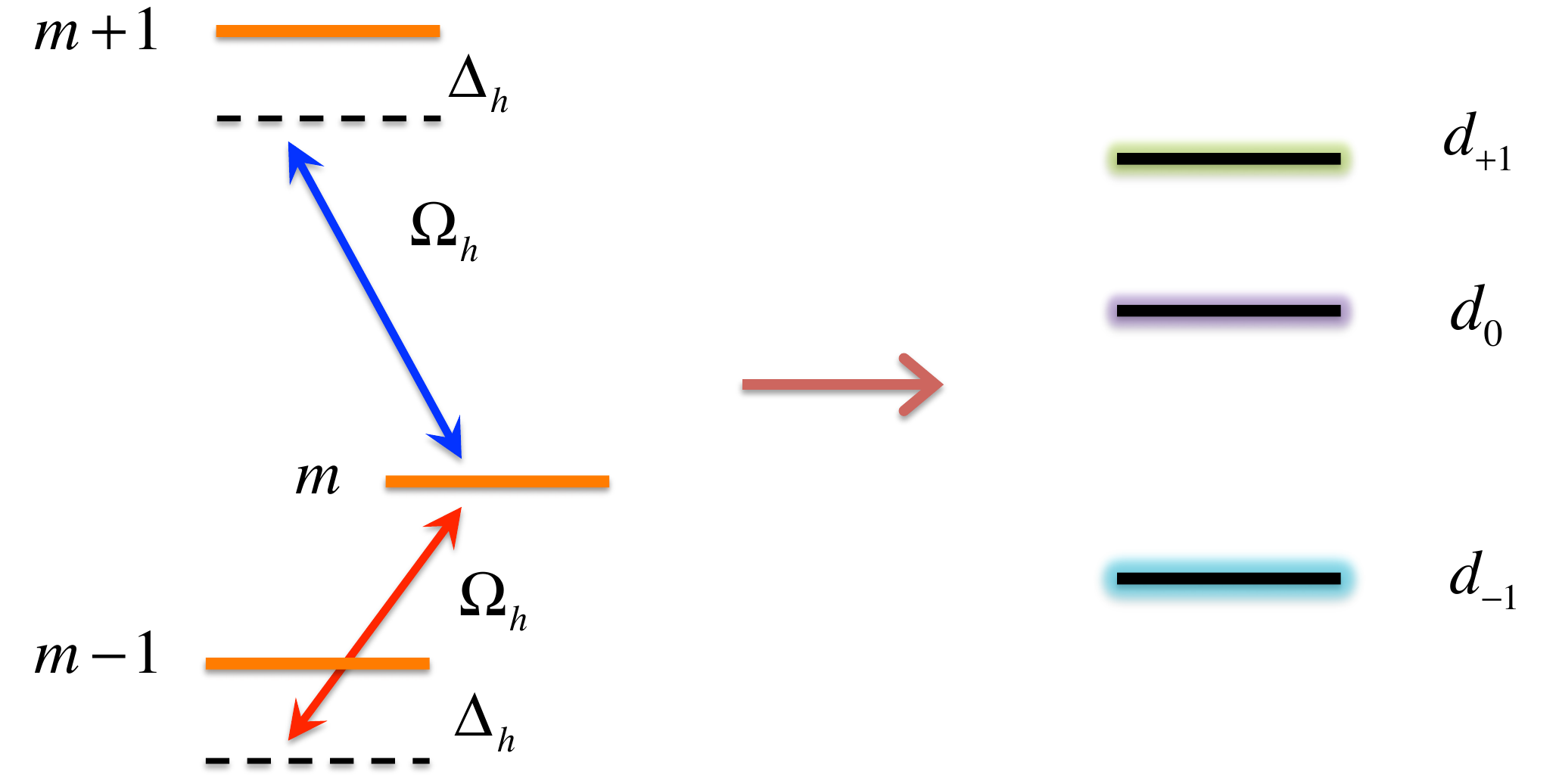}
\end{minipage}
\end{center}
\caption{{\bf Tuning spin anisotropy with dressed states.} Simulation of an effective spin-$\frac{1}{2}$ by the dressed states from a higher spin with radio frequency driving fields (with Rabi frequency $\Omega_h$ and detuning $ \Delta_h$).}\label{fig:JZXY}
\end{figure}
The value of the spin anisotropy $\Delta$ can be tuned with gradient magnetic fields to modulate the hopping coupling ($\mathbf{s}_i^x \mathbf{s}_j^x+\mathbf{s}_i^y \mathbf{s}_j^y$) while keeping the repulsive interaction $\mathbf{s}_i^z \mathbf{s}_j^z$ unchanged \cite{Gmh}. In this way the interplay between geometrical frustration and the effects of quantum fluctuations on the realisation of spin liquid phase could be tested \cite{Gia11}. As an alternative scheme with higher nuclear spin species, for example $^{2}\mbox{H}$ with spin $1$ and $^{17}\mbox{O}$ with spin-$\frac{5}{2}$, one can tune the spin anisotropy $\Delta$ by applying continuous fields corresponding to the nuclear spin transitions $\ket{\mbox{m}-1}\leftrightarrow \ket{\mbox{m}}$ and $\ket{\mbox{m}}\leftrightarrow \ket{\mbox{m}+1}$ with Rabi frequencies $\Omega_h$ and detuning $  \Delta_h$, see Fig.\ref{fig:JZXY}. An effective spin-$\frac{1}{2}$ can be encoded in two of the induced dressed states with the spin anisotropy $\Delta$ flexibly tuned (see SI for details). We remark that spin Hamiltonians for higher spins may directly lead to rich quantum phases, see e.g. \cite{Rossini11,QuantumMagnetism}.

%
\begin{figure}[t]
\begin{center}
\includegraphics[width=6cm]{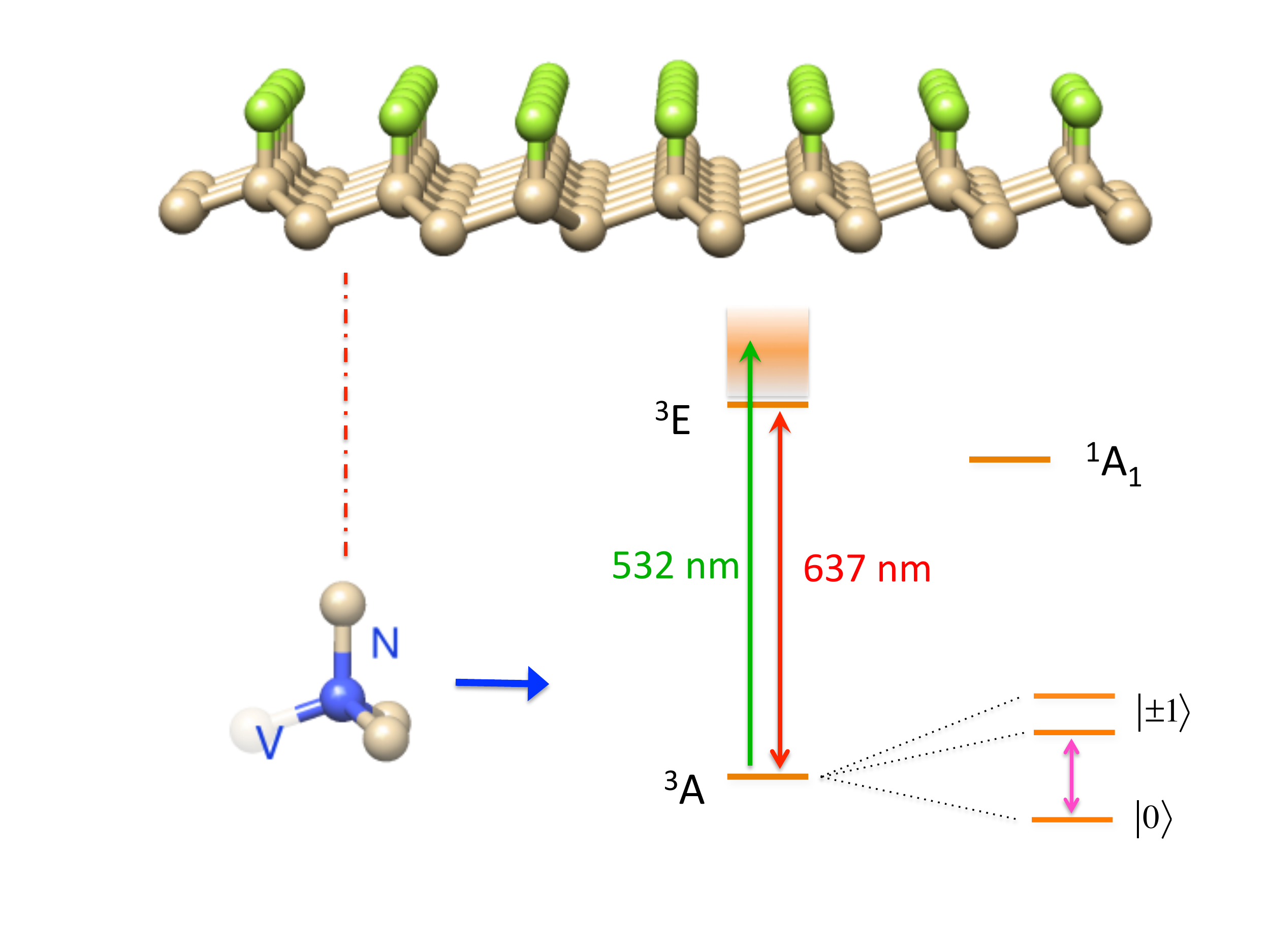}
\end{center}
\caption{{\bf Initialization and read-out of nuclear spins}. NV center lying beneath the nuclear spin layer in bulk diamond is exploited to initialize and measure nuclear spins on diamond surface. The spin-$1$ ground state of NV center can be optically readout via spin-dependent fluorescence, and can be coherently controlled with microwave fields.}\label{fig:NV_Probe}
\end{figure}

\emph{Cooling of Quantum Simulator--} The reliable preparation of the quantum simulator in a low entropy state is a prerequisite for the observation of quantum phases. Here we explain in detail how the nuclear spin lattice can be initialized  to a well defined spin direction using dynamical nuclear polarization via NV  centers in diamond, see Fig.\ref{fig:NV_Probe}. The entire process consists of repetitive cycles. In each cycle, the NV center is first prepared in the $\ket{+}=\sqrt{\frac{1}{2}} (\ket{m_s=0}+\ket{m_s=+1})$ state by optical pumping to the $\ket{m_s=0}$ state using a green laser (532nm) and a subsequent $\frac{\pi}{2}$ microwave pulse. A microwave field (with Rabi frequency $\Omega_{NV}$) on resonance with the electronic transition $\ket{m_s=0}\leftrightarrow\ket{m_s=+1}$ will lock the NV electron spin state. If the Hartmann-Hahn condition with nuclear spins \cite{Hahn62} is satisfied, the NV electron spin polarization will be transferred to the nuclear spins \cite{DS1112} (see SI for details). Due to their proximity, the interactions between nuclear spins exceeds that between the electron spin of the NV center and the nuclear spins. Therefore, it will be advantageous to effectively decouple nuclear spin interactions, which also facilitates estimates of the cooling efficiency. To this end, we apply a radio frequency (RF) field with the amplitude $\Omega_{\mbox{p}}$ whose frequency is detuned from the Larmor frequency of the nuclear spins by $\Delta_{\mbox{p}}$. The eigenstates of the effective local Hamiltonian of nuclear spin $\tilde{H}_i=\bla{\Omega_{\mbox{p}}/2} \mbox{s}_i^{x}+\Delta_{\mbox{p}}  \mbox{s}_i^{z}$ gives a new spin basis $\{\ket{\tilde{\uparrow}},\ket{\tilde{\downarrow}}\}$. Written in such a spin basis, the energy conserving terms of nuclear spin interactions cancel each other when $\Omega_{\mbox{p}}=2\sqrt{2}\Delta_{\mbox{p}} $, and the anti-rotating terms are suppressed as long as the energy mismatch $\omega_{\mbox{f}} =(\Omega_{\mbox{p}}^2+\Delta_{\mbox{p}}^2)^{\frac{1}{2}} \gg g_{ij}$ is fulfilled. Thus, the nuclear spins behave as isolated particles, and the Hartmann-Hahn condition becomes $\Omega_{NV} =(\gamma_N B- \Delta_{\mbox{p}})+ [(\Omega_{\mbox{p}} /2)^2+\Delta_{\mbox{p}} ^2]^{\frac{1}{2}}$, see SI for more details. Such a mechanism for the isolation of the nuclear spins (as we will discuss in the following section) can also reduce the perturbation on the nuclear spin state during the read-out process, which will be beneficial for the accuracy of the measurement.

\begin{figure}[b]
\begin{center}
\hspace{-0.3cm}
\includegraphics[width=4.4cm]{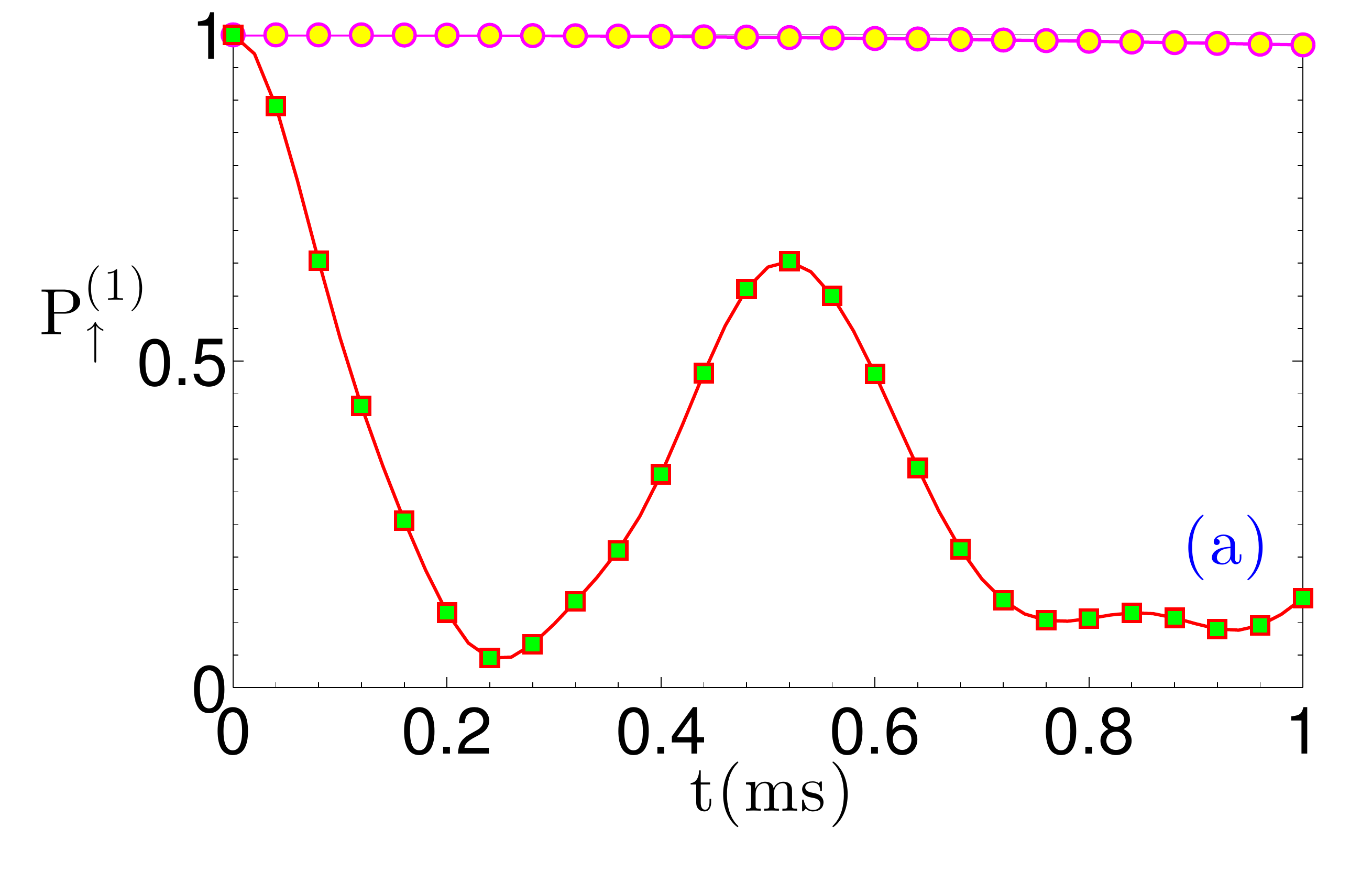}
\hspace{-0.3cm}
\includegraphics[width=4.4cm]{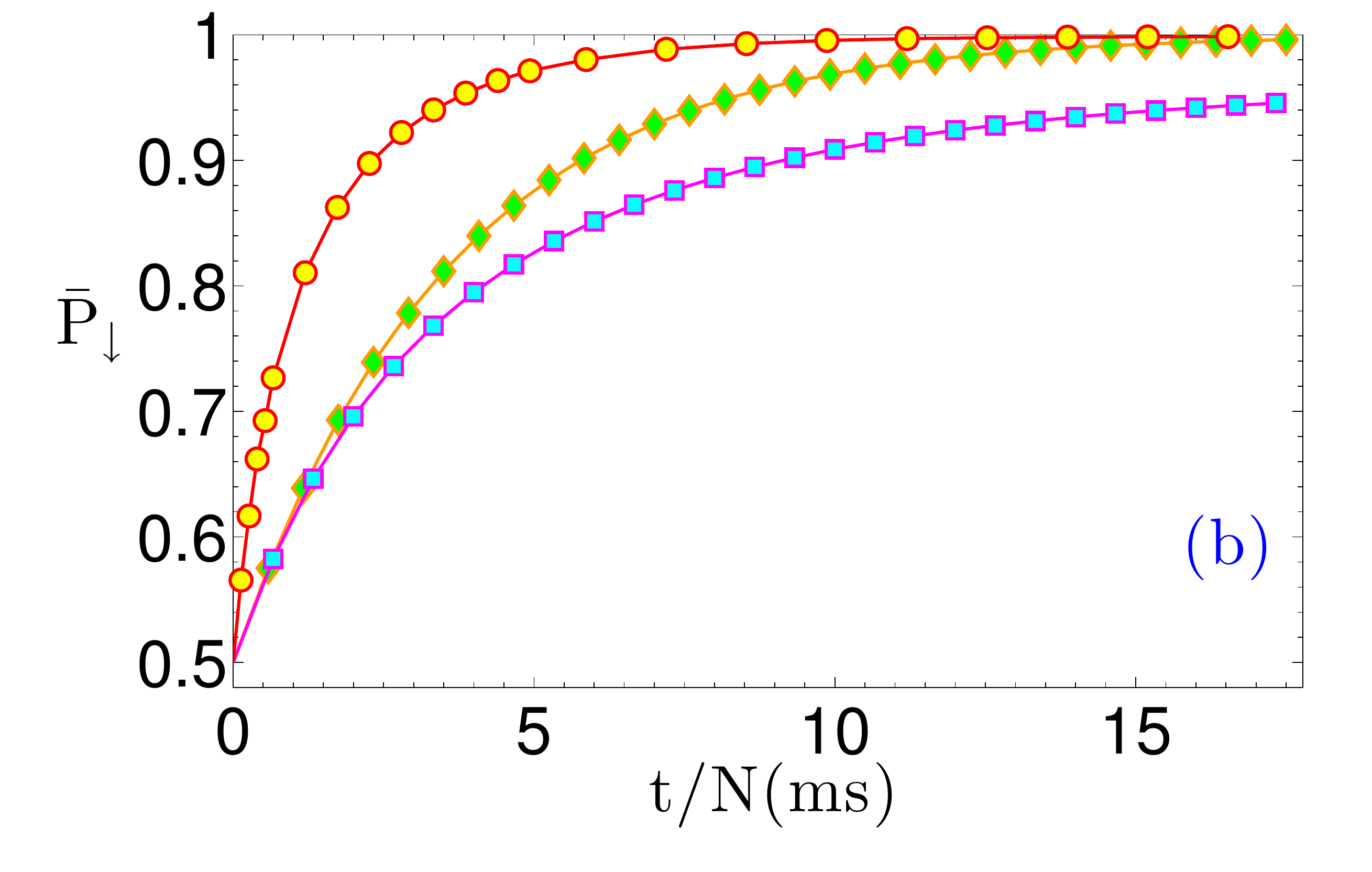}
\end{center}
\caption{{\bf Isolating and initialization of nuclear spins}. (a) The localization (circle, with RF-driving) and de-localization (square, without RF-driving) of spin excitation initially created in one specific site, demonstrating the efficiency of decoupling of nuclear-nuclear interaction in the former case. The driving parameter is $\omega_{\mbox{f}}=160 \mbox{kHz}$. (b) The average nuclear spin down state population $P_{\downarrow}$ as a function of the polarization time $\mbox{t}$: Exact numerical simulation (purple, square) spin temperature approximation (yellow, diamond) with the cycle time $\tau=30 \mu s$; Exact numerical simulation (red, circle) with $\tau=120 \mu s$ and magnetic noise is introduced to remove nuclear spin coherence every 20 cycles. The nuclear spin lattice is $3\times 3$ with a distance of $5\mbox{nm}$ to the NV center. The magnetic field direction is $\hat{m}=(\sqrt{\frac{1}{2}},0,\sqrt{\frac{1}{2}})$.}\label{fig:POL}
\end{figure}

\begin{figure*} 
\begin{center}
\hspace{-0.5cm}
\includegraphics[width=7.5cm]{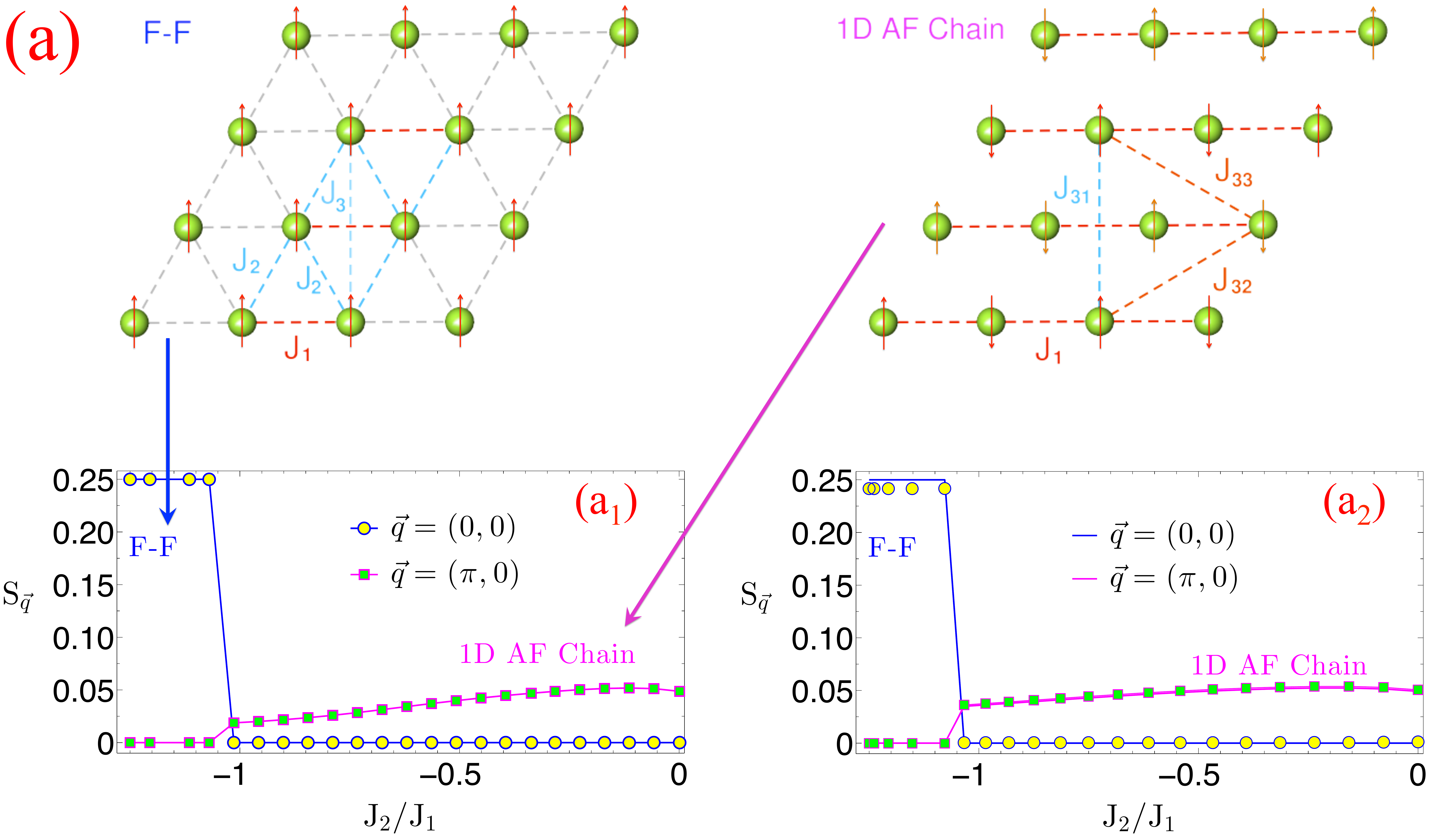}
\hspace{1cm}
\includegraphics[width=7.5cm]{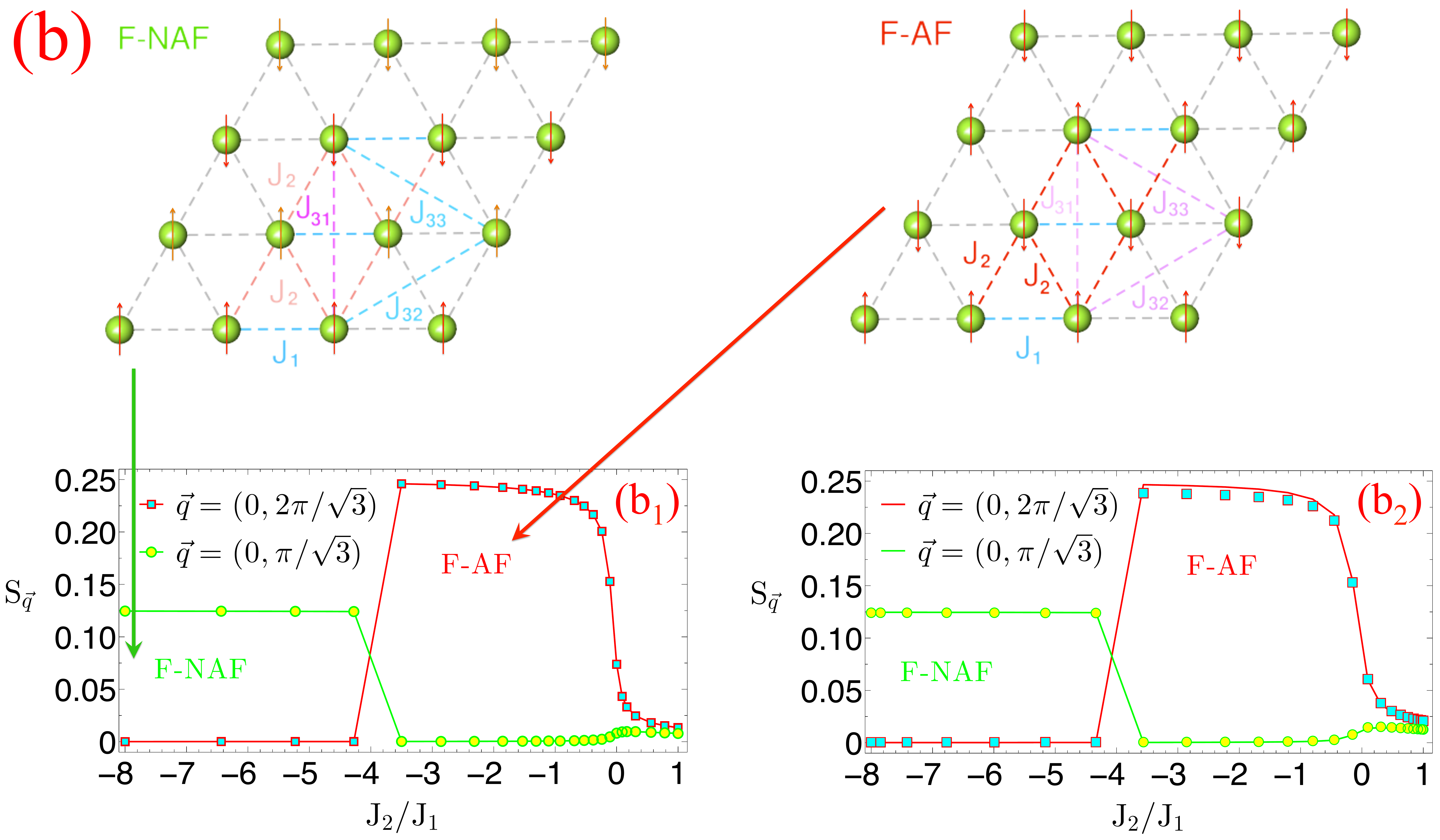}
\end{center}
\caption{{\bf Quantum magnetic phase transitions of the fluorine quantum simulation on a triangular lattice}. (a) Tuning from 1D anti-ferromagnetic chains to ferromagnetic order with the magnetic field in the direction $\hat{m}=(0,\cos\theta,\sin\theta)$. (b) The phase transition from ferromagnetic-antiferromagnetic (F-AF) order to ferromagnetic-(alternative) antiferromagnetic (F-NAF) controlled by the magnetic field in the direction $\hat{m}=(\cos\theta,0,\sin\theta)$. (a1-b1) The spin structure factors are calculated using the Lanczos algorithm  for a 6$\times 4$ nuclear spin lattice under periodic boundary condition. (a2-b2) Comparison of the spin structure factors from exact diagonalization of a $4\times 4$ nuclear spin lattice (lines) and the estimated values (which have been rescaled for visibility) via NV measurements (circles and squares) by numerical simulations. The interaction time (between the electron spin of NV center and the nuclear spins) for read-out is $\tau=60 \mu s$.}\label{fig:QMP}
\end{figure*}

To verify the validity of this idea, we have used a Chebyshev expansion \cite{Raedt03} to calculate polarization dynamics with the exact Hamiltonian of a 3$\times$3 nuclear spin lattice assuming a distance of $5\mbox{nm}$ from the NV center. It can be seen from Fig.\ref{fig:POL}(a) that the coupling between nuclear spins is effectively eliminated. We compare the exact numerical calculation with the results for isolated nuclear spins under the spin temperature approximation in which coherences between nuclear spins are neglected \cite{Christ07}, see Fig.\ref{fig:POL}(b), which show a good agreement. To achieve an ultra-high polarization given by the spin temperature approximation, one can introduce magnetic noise to remove coherence among nuclear spins in between polarization cycles (during which the NV center is polarized to the state $\ket{m_s=0}$ and would not affect nuclear spins). From the spin temperature approximation, one can estimate the polarization rate as $\p_i=(g_i^{\perp})^2 \tau =\epsilon (g_i^{\perp})^2/ \sqrt{\sum_k (g_k^{\perp})^2}\approx \frac{1}{N}(\epsilon g_k^{\perp})$, where $\tau$ is the polarization cycle time, $\epsilon $ is a chosen constant, and $g_k^{\perp}$ is the flip-flop rate between the NV electron spin and the nuclear spin $k$ (see SI for details). The required polarization time scales linearly with the total number of nuclear spins $N$ and the inverse effective temperature. The ultimate polarization efficiency will be limited by the relaxation time $\mbox{T}_1$ of nuclei, which can be as long as a few hours even at room temperature. The polarization cycle time sets the required coherence time of NV centers and nuclear spins to a few hundreds of $\mu s$, which is readily achievable with the current experimental techniques in diamond samples. The polarization efficiency can also be improved by optimizing the polarization cycle time and exploiting several NV centers. Once the nuclear spins have been initialized, by performing an adiabatic passage, one can prepare the system in the ground state of the engineered interacting Hamiltonian (similar to other types of quantum simulator \cite{Islam12}).

\emph{Measurement of Quantum Simulator--} Before discussing the static and dynamical properties of the proposed quantum simulator, let us describe the measurement schemes and demonstrate their viability by means of numerical simulations. The main goal is to measure observables that can provide information about nuclear spin state \cite{Liu12}. It is challenging to measure nuclear spins directly due to their small magnetic moments. However, NV centers implanted beneath the diamond surface will provide the solution as a measurement interface for nuclear spin states \cite{DS1112}. Before the measurement, we apply a RF-pulse to map the nuclear spin basis from $\ket{{\uparrow}}$ and $\ket{{\downarrow}}$ to $\ket{\tilde{\uparrow}}$ and $\ket{\tilde{\downarrow}} $, in which the nuclear spins can be effectively decoupled from each other by a continuous driving field as described in the process of initialization. The NV center is prepared in the state $\ket {\mu}$ ($\mu =\pm$), and is driven to match the Hartmann-Hahn condition between the electron spin of NV center and the nuclear spins. After the electron spin interacts with the nuclear spins for time $\tau$, we measure the population of state $\ket{\nu} $ of the NV center, which is given by $P_\mu^\nu=\tau^2 \sum_i\sum_j  ( g_i^{\perp}   g_j^{\perp} )\langle \mathbf{s}^\mu_i  \mathbf{s}^\nu_j\rangle$ to second-order in $ \tau$, where $\mu, \nu =\pm$. The above observables include both local contributions of individual nuclear spins (for $i=j$) and two-point correlations of nuclear pairs (for $i\neq j$). We can extract information about the average nuclear spin magnetization by $\Delta_M= P_-^+ - P_+^- = 2 \tau^2 \sum_i (g_i^{\perp} )^2 \langle \mathbf{s}_z^i \rangle$, and the transverse correlation as 
$\Delta_{XY}= P_-^ + + P_+^- =2\tau^2 \sum_i\sum_j( g_i^{\perp}   g_j^{\perp} ) \langle \mathbf{s}^x_i  \mathbf{s}^x_j + \mathbf{s}^y_i  \mathbf{s}^y_j\rangle$. The correlation function along the other directions $\Delta_{YZ}$ and $\Delta_{XZ}$ can be obtained by applying the Hadamard operation $O_H=\ketbra{{\uparrow_x}}{{\uparrow}}+\ketbra{{\downarrow_x}}{{\downarrow}}$ with $\ket{{\uparrow_x}}=\sqrt{\frac{1}{2}}(\ket{{\uparrow}}+\ket{{\downarrow}})$, and the phase transformation $O_I=\ketbra{{\uparrow}}{{\uparrow}}+i \ketbra{{\downarrow}}{{\downarrow}}$ on the nuclear spin state before the measurement. To measure observables such as structure factors, which are very important for the study of quantum phase transition and for inferring entanglement properties \cite{Cramer11}, we can use a gradient field, with which the nuclear spin at the position $\vec{\mathbf{r}}_i$ experiences a magnetic field $b_i=\vec{\mathbf{r}}_i \cdot  (b_x,b_y)$ and gains a position-depend phase $\phi_i=\vec{\mathbf{r}}_i \cdot \vec{q}$ where $\vec{q}=\gamma_N t_p (b_x,b_y)$ after a certain time $t_p$. By performing the same measurement as before, we can obtain $\Delta_{XY}^{\vec{q}}=\tau^2  \sum_i\sum_j (g_i^{\perp}   g_j^{\perp} ) \cos{((\vec{\mathbf{r}}_i-\vec{\mathbf{r}}_j)\cdot \vec{q})} \langle \mathbf{s}^x_i  \mathbf{s}^x_j+\mathbf{s}^y_i  \mathbf{s}^y_j\rangle $ and $\Delta_{YZ}^{\vec{q}}$, $\Delta_{XZ}^{\vec{q}}$ in a similar way. We remark that it is possible to generate a gradient field as large as $15 \mbox{G}$ (about 10 times larger than the coupling between fluorine nuclear spins) over the lattice constant ($2.5 \mathring{\mbox{A}}$) by one single magnetic tip with the state-of-art technology \cite{Rugar}. The validity of our measurement scheme is numerically tested in the context of witnessing quantum phase transitions as described in the following section.

\emph{Frustrated Quantum Magnetism --} Our system can simulate quantum spin models with the tunable spin-spin interaction $g(\mathbf{r}_{ij})$: positive $g(\mathbf{r}_{ij})$ correspond to anti-ferromagnetic (AF) interaction, and negative $g(\mathbf{r}_{ij})$ is ferromagnetic (F) interaction. In the triangular lattice of the fluorine simulator, the nearest-neighbor nuclear spin interactions are denoted as $J_{\hat{a}_1}$, $J_{\hat{a}_2}$, $J_{\hat{a}_3}$ with $\hat{a}_1=(1,0,0)$ $\hat{a}_2=(\frac{1}{2},\frac{\sqrt{3}}{2},0)$, $\hat{a}_3=(-\frac{1}{2},\frac{\sqrt{3}}{2},0)$. The long-range interactions and spin frustration make it hard to perform numerical simulations using the quantum Monte Carlo method due to the subtle sign problem \cite{Sign}. Here, we exactly diagonalize the system on a 6$\times 4$ lattice using the Lanczos algorithm under periodic boundary condition to provide evidence for various phases of quantum magnetism.

In the situation where $J_{\hat{a}_1}\equiv J_1=g_a$ is positive (AF) , and $J_{\hat{a}_2}=J_{\hat{a}_3}\equiv J_2=g_a(1-\frac{9}{4} \cos^2 \theta)$ is negative for $\cos\theta \geq \frac{2}{3}$ (the magnetic field direction is $\hat{m}=(0,\cos\theta,\sin\theta)$), the triangle which consists of $J_2 - J_2 - J_1$ is spin frustrated, see Fig.\ref{fig:QMP}(a). For small values of $|J_2/J_1|$, the system consists of 1D (AF) chains with weak intra-chain interaction which induces ferromagnetic order in the sublattice of every two 1D chains and is characterized by the (normalized) spin structure factor $S_{\vec{q}}(\pi,0)=(1/N^2)\langle (\sum_{i} e^{i\vec{q}\cdot \vec{\mathbf{r}}_i} \mathbf{s}_i^z)(\sum_{i} e^{i\vec{q}\cdot \vec{\mathbf{r}}_i} \mathbf{s}_i^z)^\dagger \rangle$ with $\vec{q}=(\pi,0)$. As $|J_2/J_1|$ increases, the competition between anti-ferromagnetic ($J_1$) and ferromagnetic $(J_2)$ interactions lead to the ferromagnetic phase above the point $|J_2/J_1|=1$ corresponding to the spin structure factor $S_{\vec{q}}(0,0)$, see Fig.\ref{fig:QMP}(a$_1$). Note that the largest non-nearest neighbor interaction $J_3$ is ferromagnetic and is essential for the emergence of the ferromagnetic phase which is absent for the short-range model, see SI for details.

For the other situation where $J_{\hat{a}_2}=J_{\hat{a}_3}\equiv J_2=g(a)(1-\frac{3}{4} \cos^2 \theta)$ is always positive (AF), and $J_{\hat{a}_1}\equiv J_1=g(a)(1-3 \cos^2 \theta)$ is negative (F) for $\cos{\theta}>\frac{1}{3}$, where $\theta$ is defined by the magnetic field direction is $\hat{m}=(\cos\theta,0,\sin\theta)$. Considering only the nearest-neighbor interactions, the system is non-frustrated, and is expected to be ferromagnetic in the $\hat{a}_1$, while anti-ferromagnetic in the $\hat{a}_2$ direction (F-AF phase), which is identified by the spin structure factor $S_{\vec{q}}(0,2\pi/\sqrt{3})$. As the value of $\cos\theta$ approaches to $1$, the non-nearest neighbor interactions $J_{31}=1/3\sqrt{3}$, $J_{32}=(1-\frac{9}{4}\cos^2 \theta)/3\sqrt{3}$ become comparable to $J_2$. The competition between them leads to a magnetic phase identified with the spin structure factor $S_{\vec{q}}(0,\pi/\sqrt{3})$. The spins are ferromagnetic in the $\hat{a}_1$ direction, while they are anti-ferromagnetic between next-nearest-neighbor chains in the $\hat{a}_2$ direction, see Fig.\ref{fig:QMP}(b$_1$). 

We also check the feasibility of using NV centers to identify different magnetic phases. We numerically calculate the dynamics during measurement and obtain the estimation of spin structure factors. Due to the limit of computational overhead, we consider an example of a 4$\times$4 nuclear spin lattice with periodic boundary conditions. By applying a gradient field we can generate the relative phases between different nuclear spins corresponding to the spin structure factor, which can then be estimated by the observable $S_{\vec{q}}(q_x,q_y)\propto \Delta_{XZ} (q_x,q_y)+\Delta_{YZ} (q_x,q_y)-\Delta_{XY} (q_x,q_y)$ via NV centers. We find that the estimation is in good agreement with the results from exact diagonalization, and that it can witness quantum phase transitions between different magnetic orders, see Fig.\ref{fig:QMP}(a$_2$-b$_2$) and SI for more details.

\emph{Quantum Superfluid and Supersolid--} The nuclear spin Hamiltonian can be mapped to the hard-core boson model by the Holstein-Primakoff transformation as
\be
H_b=\sum_{\langle i,j \rangle }\blb{ V_{ij} n_i n_j- t_{ij} \bla{ a_i^\dagger a_j+ a _i a_j^\dagger }‚} +\mu\sum_i  n_i
\ee
with $n_i=s_i^z+\frac{1}{2}$ ($n_i=0,1$). Here, the chemical potential is $\mu=(\gamma_N B) - \sum_j  g(\mathbf{r}_{ij}) $, the repulsive interaction is $V_{ij}=  g(\mathbf{r}_{ij})$, and the hopping is $t_{ij}=\frac{\Delta}{2} g(\mathbf{r}_{ij})$. Our proposed system can therefore simulate the hard-core boson model, which demonstrates interesting phases such as (long-range off-diagonal order) superfluid, and moreover a supersolid phase characterized by both long-range off-diagonal and diagonal order. Note that quantum simulation of similar models with polar molecules in optical lattices have been proposed \cite{Micheli06,Gor11} and numerically studied in \cite{Pupillo10,Pollet10} for long range repulsive dipole interaction ($V_{ij}$). Our system possesses long range interactions (both $V_{ij}$ and $t_{ij}$), and thus offers a platform to investigate rich phases of hard-core bosons with potentially novel features. Indeed, these models with frustration and long-range interactions pose considerable challenges to the classical numerical simulations due to the sign problem for 2D systems. 

\begin{figure}[b]
\begin{center}
\hspace{-0.3cm}
\includegraphics[width=4.2cm]{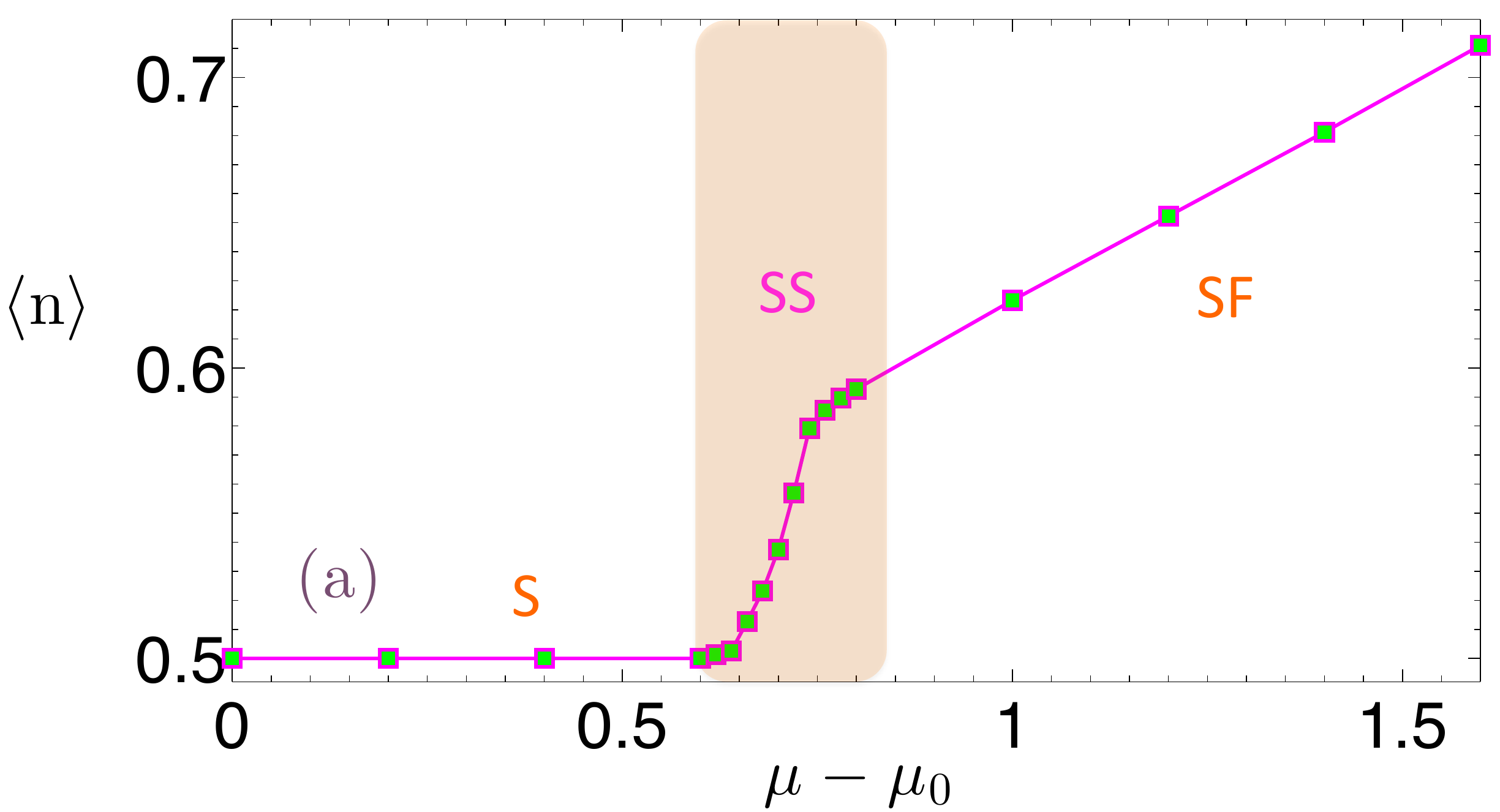}
\includegraphics[width=4.5cm]{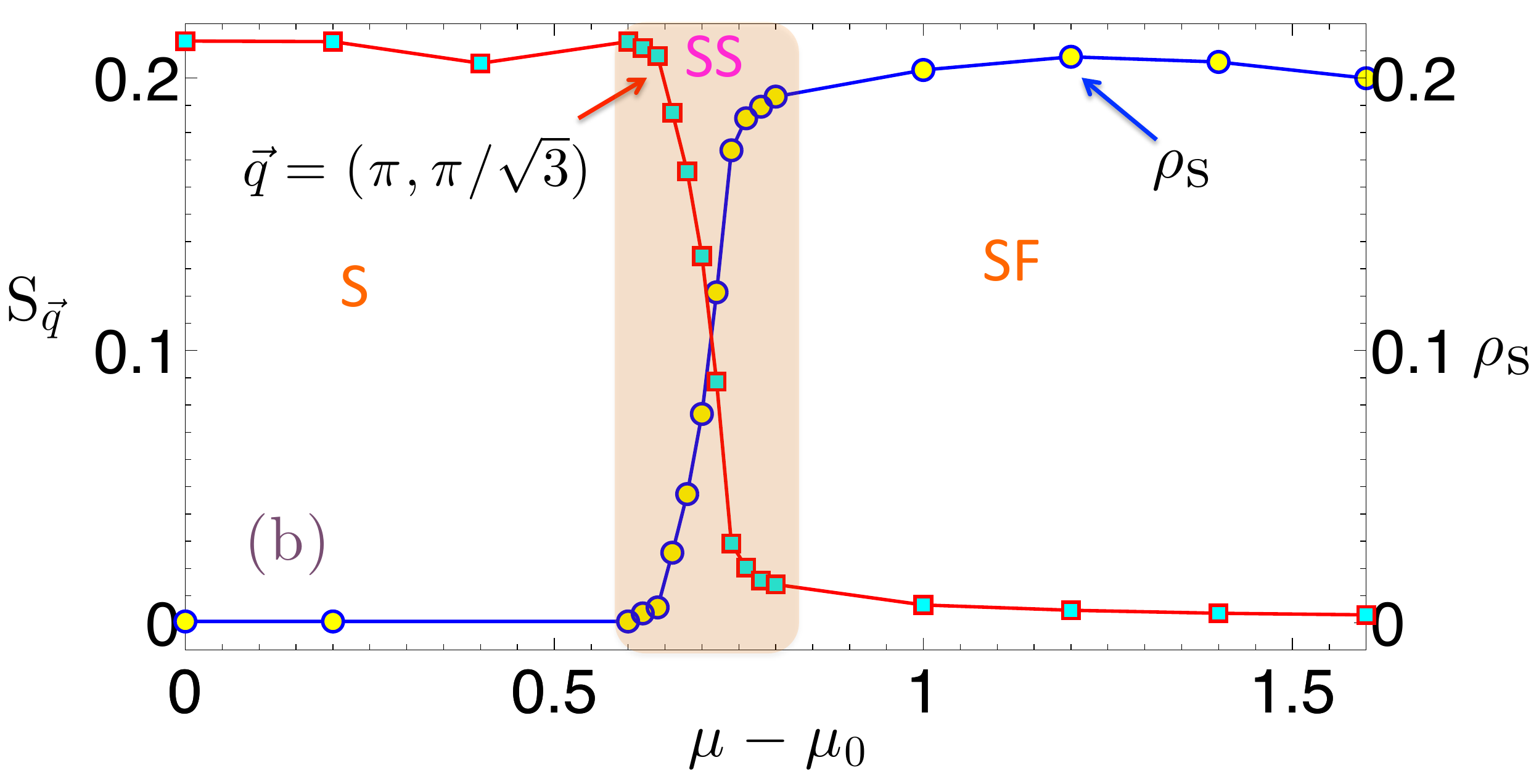}
\end{center}
\caption{{\bf Quantum phases of superfluid (SF) and supersolid (SS)}. (a) The average filling $\langle \mbox{n}\rangle$ as a function of chemical potential $\mu-\mu_0$ (in the unit of $g_a$), where $\mu_0$ is the chemical potential corresponding to half-filling. (b) The superfluid density $\rho_{\mbox{s}}$ and the normalized structure factor $\mbox{S}_{\vec{q}} (\pi,\pi/\sqrt{3})$. The system size is $12\times 12$ and $t/V=0.2$. The magnetic field direction is $\hat{m}=\cos\theta \hat{a}_3 +\sin\theta \hat{Z}$ with $\cos\theta=\sqrt{\frac{1}{3}}$. The temperature used in our simulation is $T/g_a=0.1$.}\label{fig:SFSS}
\end{figure}

With the magnetic field along the direction $\hat{m}=\cos\theta \hat{a}_3 +\sin\theta (0,0,1)=(-\frac{1}{2}\cos\theta,\frac{\sqrt{3}}{2}\cos\theta,\sin\theta)$,  the nearest-neighbor interactions are $V_{\hat{a}_1}=V_{\hat{a}_2}\equiv V_1 =g_a(1-\frac{3}{4}\cos^2\theta)$, and $V_{\hat{a}_3}\equiv V_2=g_a(1-3\cos^2\theta)$. By changing the value of the magnetic direction angle $\theta$, we can gradually tune the geometric frustration as quantified by the ratio $V_2/V_1$. For $\cos\theta\in [0,\sqrt{\frac{1}{3}}]$, 
the values of $V$ for all interactions have the same sign (including long-range interactions), and one can simulate such a model with the directed loop algorithm in the stochastic series expansion representation of the ALPS library \cite{ALPS}. By comparison with the short-range model, it can bee seen that the long-range interaction significantly enhances the superfluidity, see SI for more details. By tuning the anisotropy value, we can observe quantum phase transitions between solid (S), supersolid (SS) and superfluid (SF), see Fig.\ref{fig:SFSS}. One can also use a gradient field to selectively tune hopping interactions. For example, with a gradient field that decreases along the direction $\Delta\hat{b}=\Delta_b (\frac{\sqrt{3}}{2},\frac{1}{2})$, the hopping interaction will be suppressed except in the direction $\hat{a}_3 =(-\frac{1}{2},\frac{\sqrt{3}}{2})$. 

\emph{Outlook---} We have introduced a novel platform  for quantum simulations that incorporates the properties of diamond surfaces with the characteristic of NV centres in diamond. With numerically tractable examples, we have demonstrated that important quantum phases can be observed with the proposed quantum simulator. Due to the flexibility of the method in engineering Hamiltonians, quantum models with novel features will be realizable, which would require more exhaustive classical simulation for understanding the richness of the quantum phases. In addition to the static properties of the quantum phase transitions, the present quantum simulator is also capable of studying quantum many-body dynamics, such as quantum quenches and the generation of multi-particle entanglement, see SI for a simple example. The prerequisite technologies for the implementation of such a quantum simulator, such as the techniques for charge state manipulation of NV centers in (surface functionalized) diamond \cite{Grotz12}, coherent control of NV centers, have been developed already \cite{Neumann08,Jiang09,Fuchs09}. The present proposal will stimulate the interest of material scientists to fabricate other candidate systems, in particular for the simulation of quasi 3D systems (e.g. a number of fluorographene sheets). The relevant tools in this work may also be beneficial for further research in coherent control on surfaces/mono-layer films.

\emph{Acknowledgements--} We are grateful for the valuable communications with Matthias Troyer, Lode Pollet, Barbara Capogrosso-Sansone about the properties of supersolid and QMC simulation with ALPS. We also thank Robert Rosenbach and Javier Almeida for their help in numerical simulations. The work was supported by the Alexander von Humboldt Foundation, the EU Integrating Project Q-ESSENCE, the EU STREP PICC and DIAMANT, the BMBF Verbundprojekt QuOReP, DFG (FOR 1482, FOR 1493, SFB/TR 21) and DARPA. J.-M.C was supported also by a Marie-Curie Intra-European Fellowship (FP7). Computations were performed on the bwGRiD.



\newpage 

\onecolumngrid

\section*{Supplementary Information}

\setcounter{figure}{0}
\setcounter{equation}{0}

\noindent

\begin{figure}[h]
\begin{center}
\includegraphics[width=8cm]{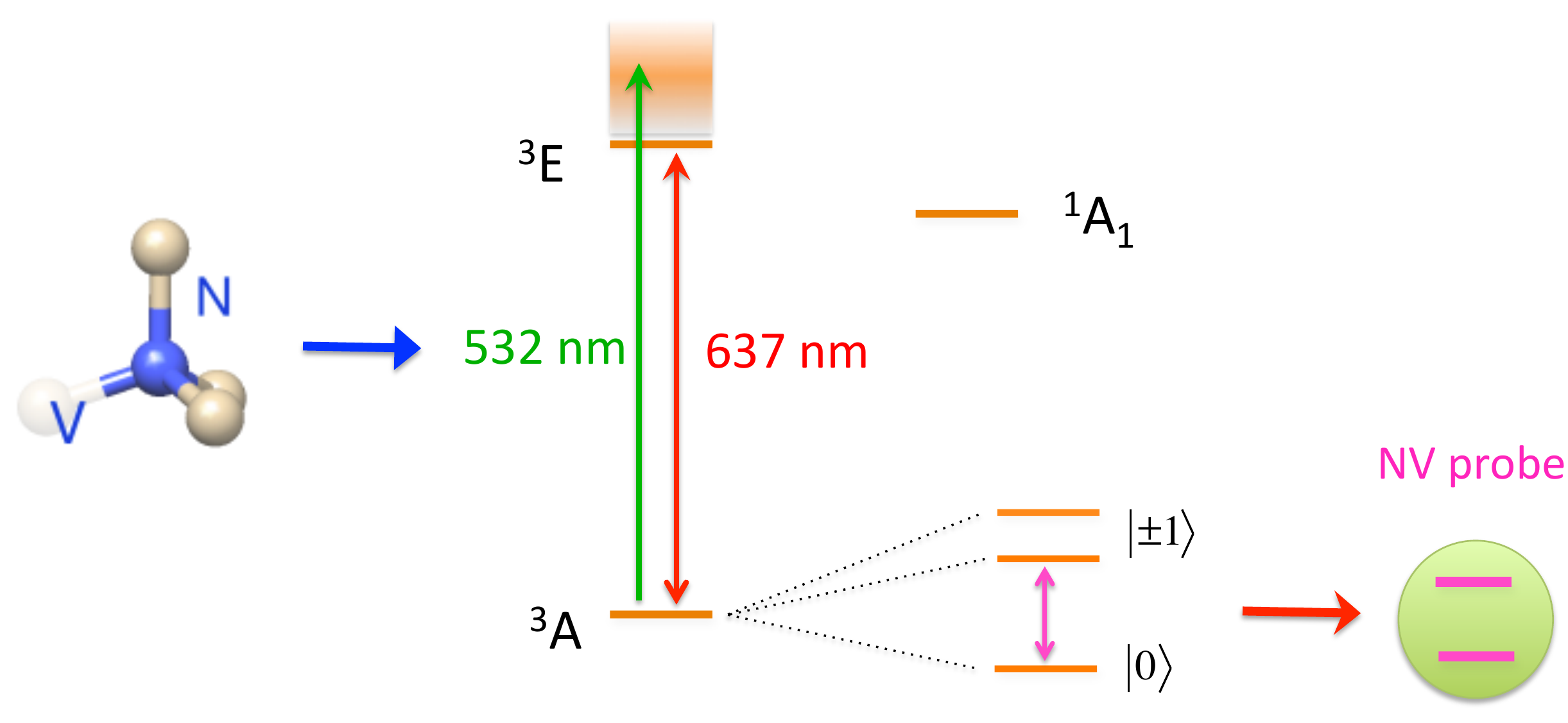}
\end{center}
\caption{The sketch of the energy levels of NV center. The spin-$1$ ground state of NV center ($\ket{m_s=0},\ket{m_s=\pm 1}$) can be optically readout via spin-dependent fluorescence, and can be coherently controlled with microwave fields. We continuously drive the one of its electronic transition $\ket{\mbox{m}_s=-1} \leftrightarrow \ket{\mbox{m}_s=0}$, which provides an effective spin$-\frac{1}{2}$ serving as a quantum probe for target nuclear spins and in the mean time is decoupled from other noise.}\label{fig:NVsetup-s}
\end{figure}

{\it Hamiltonian of NV centers and nuclear spins.---} The system under discussion comprises the electron
spin of the NV center and the nuclear spin array on the diamond surface. The dynamics of these parts and
their mutual interaction is described by the Hamiltonian
\be
    H=H_{\mbox{nv}}^{\mbox{gs}}+H_{\mbox{F}}+H_{\mbox{NV}-\mbox{F}},
\ee
where the individual terms are discussed below. The lower energy manifold of the NV center, made up
of the electron spin states, is described by the Hamiltonian
\be
    H_{\mbox{nv}}^{\mbox{gs}}= D S_z^2-\gamma_e \mathbf{B}\cdot \mathbf{S}
\ee
where $D=2.87\mbox{GHz}$ is the zero field splitting, $\gamma_e$ is the gyromagnetic ratio of the electron 
spin, and $\mathbf{S}$ is the spin-1 operator. Here, $H_{\mbox{nv}}^{\mbox{gs}}$ describes three electron spin 
states, see Fig.\ref{fig:NVsetup-s}, where the degeneracy of the states $\mbox{m}_s=+1$ and $\mbox{m}_s=-1$
is lifted by a magnetic field. Two of the eigenstates $\ket{\mbox{m}_s=-1} \equiv \ket{{\uparrow}}$ 
and $\ket{\mbox{m}_s=0} \equiv \ket{{\downarrow}}$ can thus serve as an effective spin-$\frac{1}{2}$ 
with the following Hamiltonian
\be
    H_{\mbox{nv}}^{\mbox{gs}}=\frac{\omega_{\mbox{nv}}}{2}\sigma_z
\ee
where $\sigma_z=\ketbra{{\uparrow}}{{\uparrow}}-\ketbra{{\downarrow}}{{\downarrow}}$ is the 
Pauli operator. To achieve the goals of initialization, control and measurement of the quantum 
simulator, while at the same time decoupling the electron spin of the NV center from other spin 
species \cite{DS1112-s}, we continuously drive the electron spin of the NV center with a microwave 
field on resonance with the electronic transition $\ket{\mbox{m}_s=-1}\equiv \ket{{\uparrow}} 
\leftrightarrow \ket{\mbox{m}_s=0}\equiv \ket{{\downarrow}}$. The effective Hamiltonian of the NV 
center then becomes
\be
    H_{\mbox{nv}}^{\mbox{gs}}=\frac{\Omega_{\mbox{nv}}}{2}\sigma_x
\label{eq:HNV-x}
\ee
where $\Omega_{\mbox{nv}}$ is the Rabi frequency of microwave driving and $\sigma_x = \ketbra{{\uparrow}}{{\downarrow}}+\ketbra{{\downarrow}}{{\uparrow}}$, see Fig.\ref{fig:NVsetup-s}. 
The Hamiltonian of the nuclear spin quantum simulator on the diamond surface is given by
\be
    H_{\mbox{F}}=\sum_i \gamma_N \mbox{B} \mathbf{s}_i^{z}+\frac{\mu_0}{4\pi} \sum_{i,j} \frac{\gamma_N ^2}{r_{ij}^3} \blb{{\bf s} _i \cdot {\bf s}_j-3\bla{{\bf s}\cdot \hat{\mathbf{r}}_{ij}}\bla{{\bf s}_j\cdot \hat{\mathbf{r}}_{ij}}}
    \label{eq:HF}
\ee
where $\gamma_N$ is the gyromagnetic ratio of nuclear spins, and $\mathbf{r}_{ij}=r_{ij}\hat{\mathbf{r}}_{ij} $ is the vector that connects two nuclear spins. The nuclear spin is quantized along the direction of the magnetic field. The interaction between the NV center electron spin and the nuclear spins is as follows
\be
H_{\mbox{NV}-\mbox{F}}= \frac{\mu_0}{4\pi} \sum_{i} \frac{\gamma_e\gamma_N}{r_{i}^3} \blb{{\bf S} \cdot {\bf s}_i-3\bla{{\bf S}\cdot \hat{r}}\bla{{\bf s}_i\cdot \hat{r}}}.
\ee
The Hamiltonian for other species of nuclear spins, e.g $\ele{13}{C}$ and $\ele{14}{N}$ of NV center, takes a similar form. For an isotopically engineered diamond, the number of $\ele{13}{C}$ can be reduced to $0.01\%$ \cite{Maurer12}, and only very few $\ele{13}{C}$ will affect NV centers. The Hamiltonian for $\ele{14} {N}$ of NV center has a similar form with $\ele{13} {C}$ with the hyperfine coupling $A_0^{\parallel}=2.1\mbox{MHz}$ and $A_0^{\perp}=2.3\mbox{MHz}$ \cite{He93-s}. Note that the use of continuous wave decoupling techniques \cite{CCD11-s} can reduce the impact of noise from impurities such as $\ele{13}{C}$ or $\ele{14}{N}$ further rendering their impact negligible. \\

{\it Hamiltonian engineering for the quantum simulator.---} The Hamiltonian for nuclear spins on the diamond surface lattice is presented in Eq.(\ref{eq:HF}). We apply a radio-frequency field with the amplitude $\Omega_F$ and the frequency $\omega=\gamma_N  \mbox{B} - \omega_F$ where $\omega_F$ denotes the detuning. Therefore, the Hamiltonian of Eq.(\ref{eq:HF}) can be rewritten as
\be
    H_{\mbox{F}}=\sum_i \gamma_N \mbox{B} \mathbf{s}_i^{z}+\frac{\mu_0}{4\pi} \sum_{i,j} \frac{\gamma_N ^2}{r_{ij}^3} \blb{{\bf s} _i \cdot {\bf s}_j-3\bla{{\bf s}_i\cdot \hat{\mathbf{r}}_{ij}}\bla{{\bf s}_j\cdot \hat{\mathbf{r}}_{ij}}} +2 \Omega_F \cos\blb{(\gamma_N  \mbox{B} - \omega_F)t}\sum_i \mathbf{s}_i^{x}
\ee
The effective Hamiltonian in an interaction picture with respect to 
$H_0=\sum_i (\gamma_N  \mbox{B} - \omega_F) \mathbf{s}_i^{z}$ is given by
\bea
    H_{\mbox{F}}&=& 
    \sum_i \bla{\omega_F \mathbf{s}_i^{z} + \Omega_F \mathbf{s}_i^{x} }+\frac{\mu_0}{4\pi} \sum_{i,j} \frac{\gamma_N ^2}{r_{ij}^3}  \bla{1-3( \hat{\mathbf{r}}_{ij}^z)^2} \blb{  {\bf s}_i^z\cdot {\bf s}_j^z - \frac{1}{2} ({\bf s}_i^x\cdot {\bf s}_j^x+{\bf s}_i^y\cdot {\bf s}_j^y) } \\
&\equiv & H_S +\sum_i  \Omega_F \mathbf{s}_i^{x}
\eea
where
\be
H_S=\sum_i \omega_F \mathbf{s}_i^{z} +\frac{\mu_0}{4\pi} \sum_{i,j} \frac{\gamma_N ^2}{r_{ij}^3}  \bla{1-3( \hat{\mathbf{r}}_{ij}^z)^2} \blb{  {\bf s}_i^z\cdot {\bf s}_j^z - \Delta ({\bf s}_i^x\cdot {\bf s}_j^x+{\bf s}_i^y\cdot {\bf s}_j^y) }
\label{HS-11}
\ee
with the spin anisotropy $\Delta=\frac{1}{2}$. Eq.(\ref{HS-11}) is the quantum simulator Hamiltonian, for which we can tune the nuclear spin interaction strength and the spin anisotropy as described in the main text. If the amplitude of the RF field is much larger than the nuclear spin interaction, the ground state of the Hamiltonian will be $\ket{X}\equiv \ket{{\downarrow}_x}\otimes \cdots \ket{{\downarrow}_x} $, which can be prepared with NV centers (more details will be discussed in the following sections). By adiabatically decreasing the amplitude $\Omega_F$, the system will end in the ground state of the new Hamiltonian $H_S$.\\

\begin{figure}[t]
\begin{center}
\includegraphics[width=4.4cm]{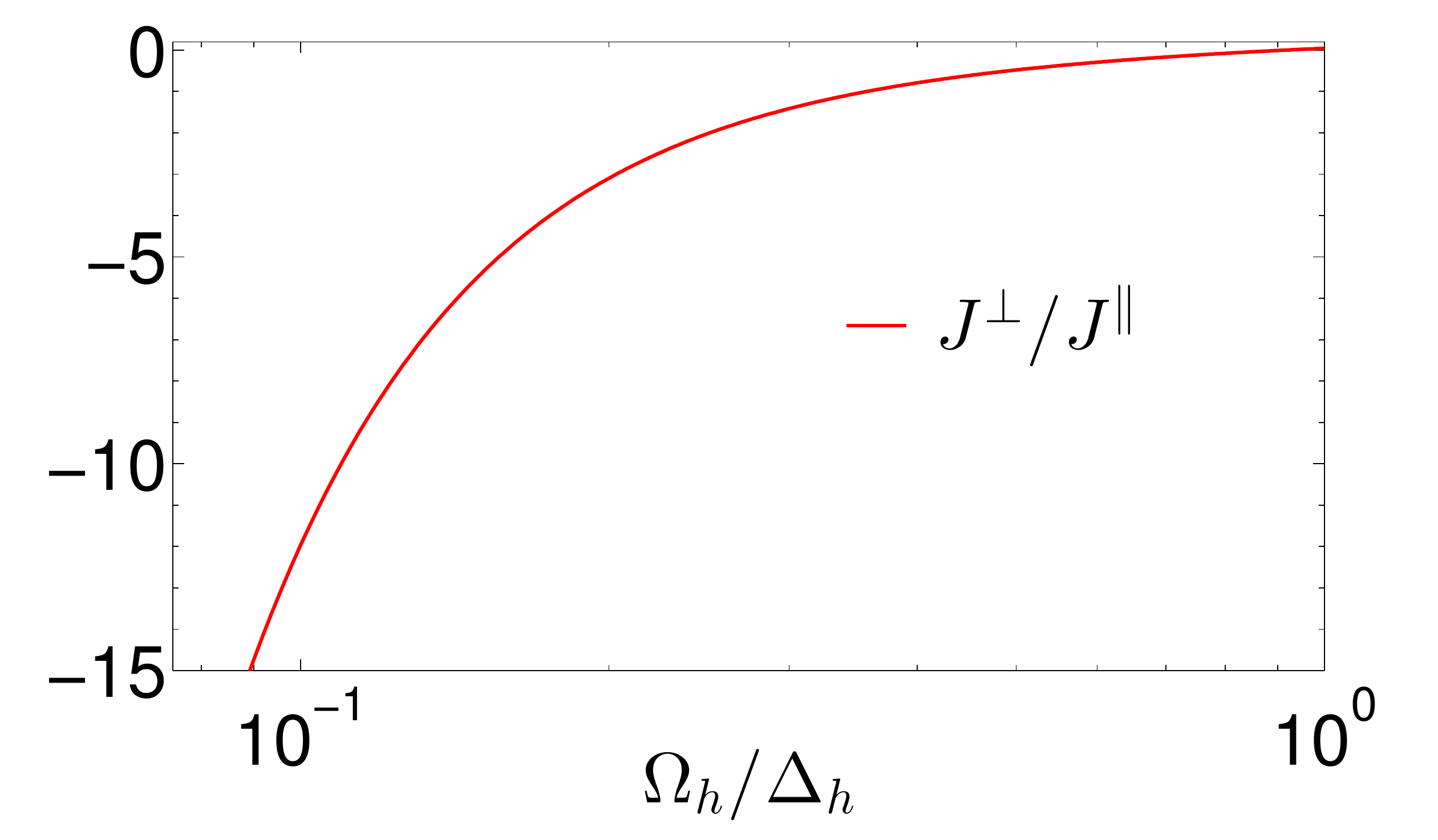}
\includegraphics[width=4.4cm]{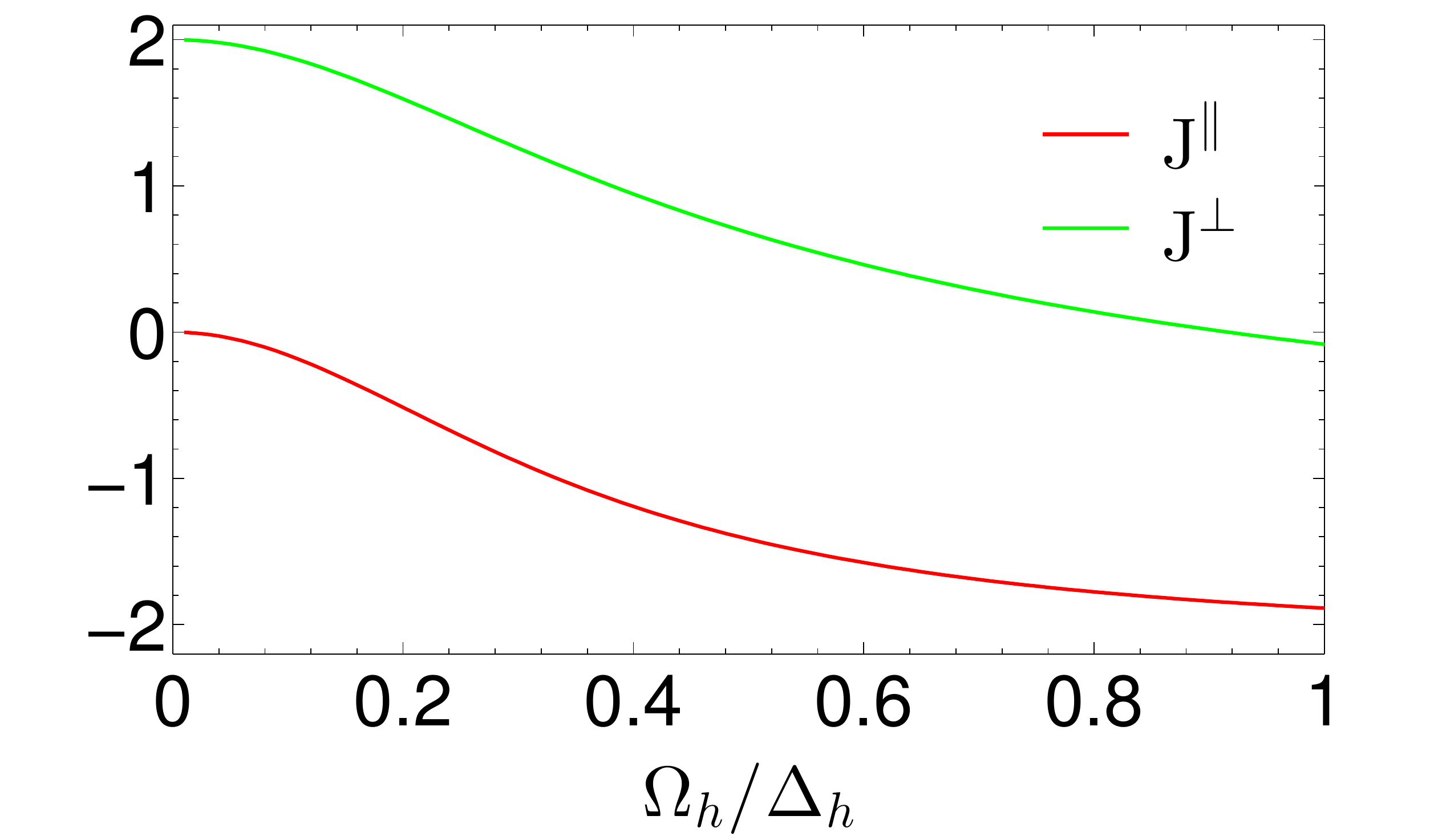}
\includegraphics[width=4.4cm]{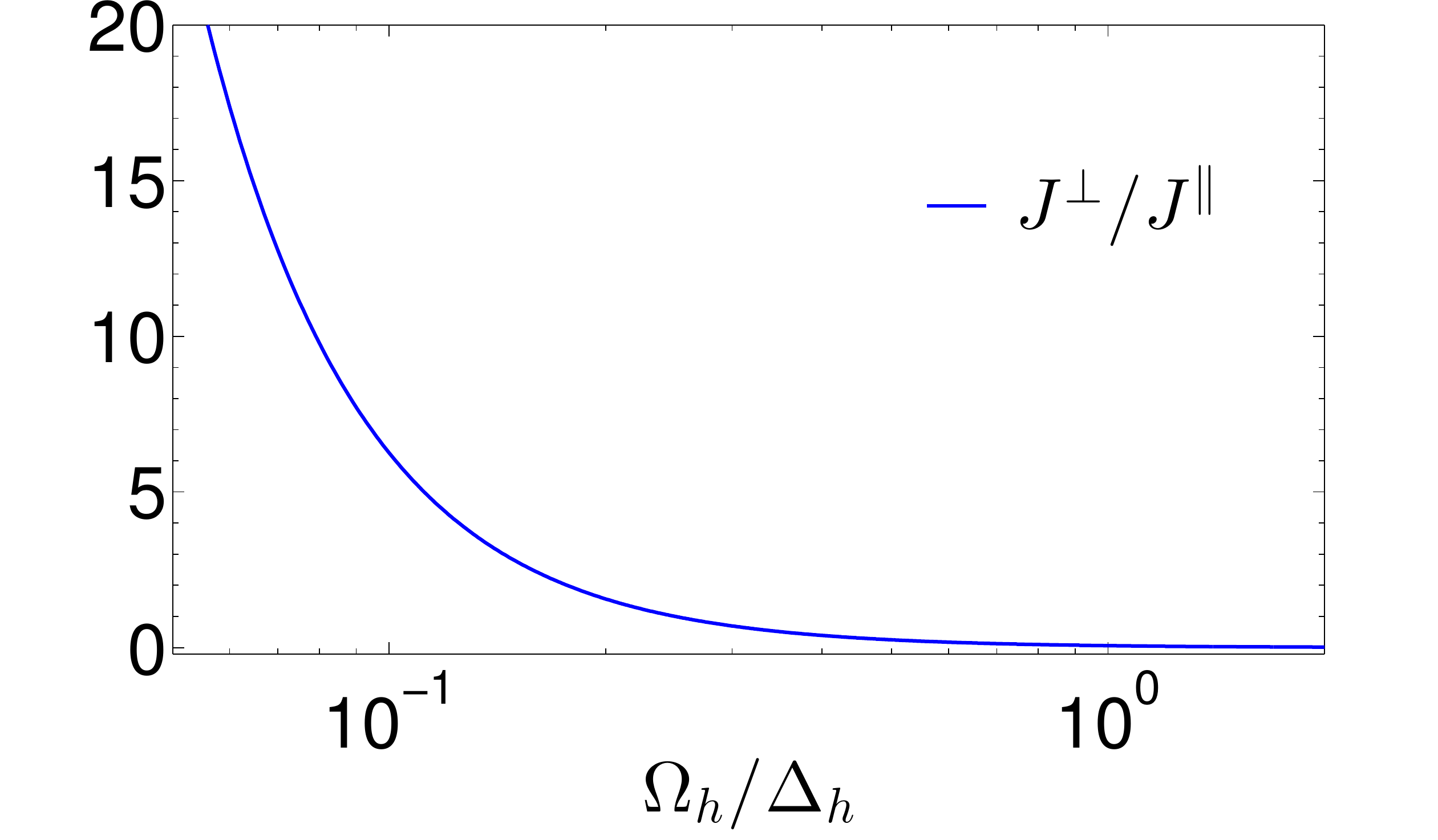}
\includegraphics[width=4.4cm]{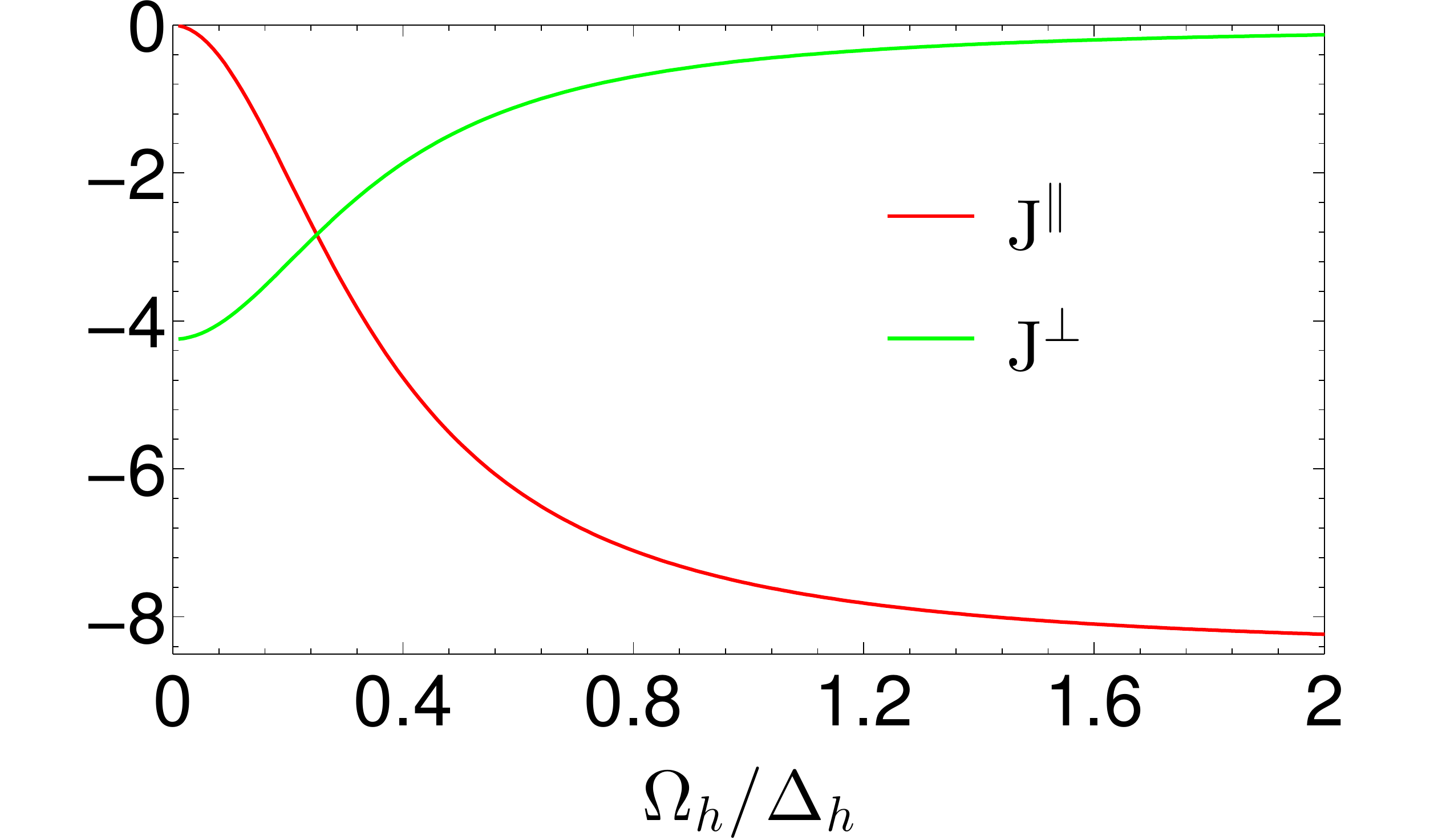}
\end{center}
\caption{The spin anisotropy $J^{\perp}/J^{\parallel}\equiv-\Delta$ as a function of the ratio between the radio frequency field amplitude and the detuning $\Omega_h/\Delta_h$. (a-b) The effective spin-$\frac{1}{2}$ is represented by two of the dressed states $\{ \ket { d_0} ,\ket{d_{+1} }\}$; (c-d) The effective spin-$\frac{1}{2}$ is represented by $\{ \ket { d_{-1}} ,\ket{d_{+1} }\}$ from the original nuclear spin states $\{\ket{-\frac{1}{2}},\ket{+\frac{1}{2}},\ket{+\frac{3}{2}} \}$ of $^{17}\mbox{O}$.}\label{fig:JZXY-s}
\end{figure}

Besides using gradient fields, in combination with higher nuclear spin species (e.g. $^{2}\mbox{H}$ with spin $1$ and $^{17}\mbox{O}$ with spin-$\frac{5}{2}$), it is also possible to tune the value of the spin anisotropy $\Delta$ for an effective spin-$\frac{1}{2}$ Hamiltonian by applying continuous fields with appropriate Rabi frequencies and detunings. The quadrupole together with Zeeman splitting allow us to apply microwave fields to selectively drive the nuclear spin transitions $\ket{\mbox{m}-1}\leftrightarrow \ket{\mbox{m}}$ and $\ket{\mbox{m}}\leftrightarrow \ket{\mbox{m}+1}$ with the detuning $\Delta_h$ and $-\Delta_h$ respectively, see Fig.2 of the main text. The Rabi frequencies are the same and denoted as $\Omega_h$. In the subspace of $\{\ket{\mbox{m}-1},\ket{\mbox{m}},\ket{\mbox{m}+1}\}$, the Hamiltonian in the interaction picture is
\be
    H_O= \left( \begin{array}{ccc}
    \Delta_h & \Omega_h  & 0  \\
    \Omega_h & 0 &  \Omega_h   \\
    0 &  \Omega_h &\Delta_h
    \end{array} \right)
\ee
Its three eigenstates are 
\bea
\ket{d_{-1}}&=&\frac{1}{N_1}\l[ \ket{\mbox{m}+1}- \frac{1+\sqrt{1+8(\Omega_h/\Delta_h)^2}}{2(\Omega_h/\Delta_h)} \ket{\mbox{m}}+\ket{\mbox{m}-1}\r],\\
\ket{d_{0}}&=&\frac{1}{\sqrt{2}} \l( \ket{\mbox{m}+1}-\ket{\mbox{m}-1}\r),\\
\ket{d_{+1}}&=&\frac{1}{N_2}\l[ \ket{\mbox{m}+1}- \frac{1-\sqrt{1+8(\Omega_h/\Delta_h)^2}}{2(\Omega_h/\Delta_h)} \ket{\mbox{m}}+\ket{\mbox{m}-1}\r].
\eea
If we choose the dressed states $\{ \ket {d_0} ,\ket{d_{+1} }\}$ or $\ket { d_{-1}} ,\ket{d_{+1} }\}$ as above to represent an effective spin-$\frac{1}{2}$, and express the nuclear spin interaction in this subspace, the value of the spin anisotropy $J^{\perp}/J^{\parallel}\equiv \Delta $ (with $J^{\perp}$ and $J^{\parallel}$ denotes the strength of coupling ${\bf s}_i^x\cdot {\bf s}_j^x+{\bf s}_i^y\cdot {\bf s}_j^y$ and ${\bf s}_i^z\cdot {\bf s}_j^z$ respectively) will depend on the ratio between the Rabi frequency and the detuning $\Omega_h/\Delta_h$. In Fig.\ref{fig:JZXY-s}, with the example of $^{17}\mbox{O}$, we see that the value of $J^{\perp}/J^{\parallel}$ can be fully tuned in the range of $(-\infty,\infty)$ by changing the value of $\Omega_h/\Delta_h$.  \\

\begin{figure}[b]
\begin{center}
\begin{minipage}{18cm}
\includegraphics[width=5cm]{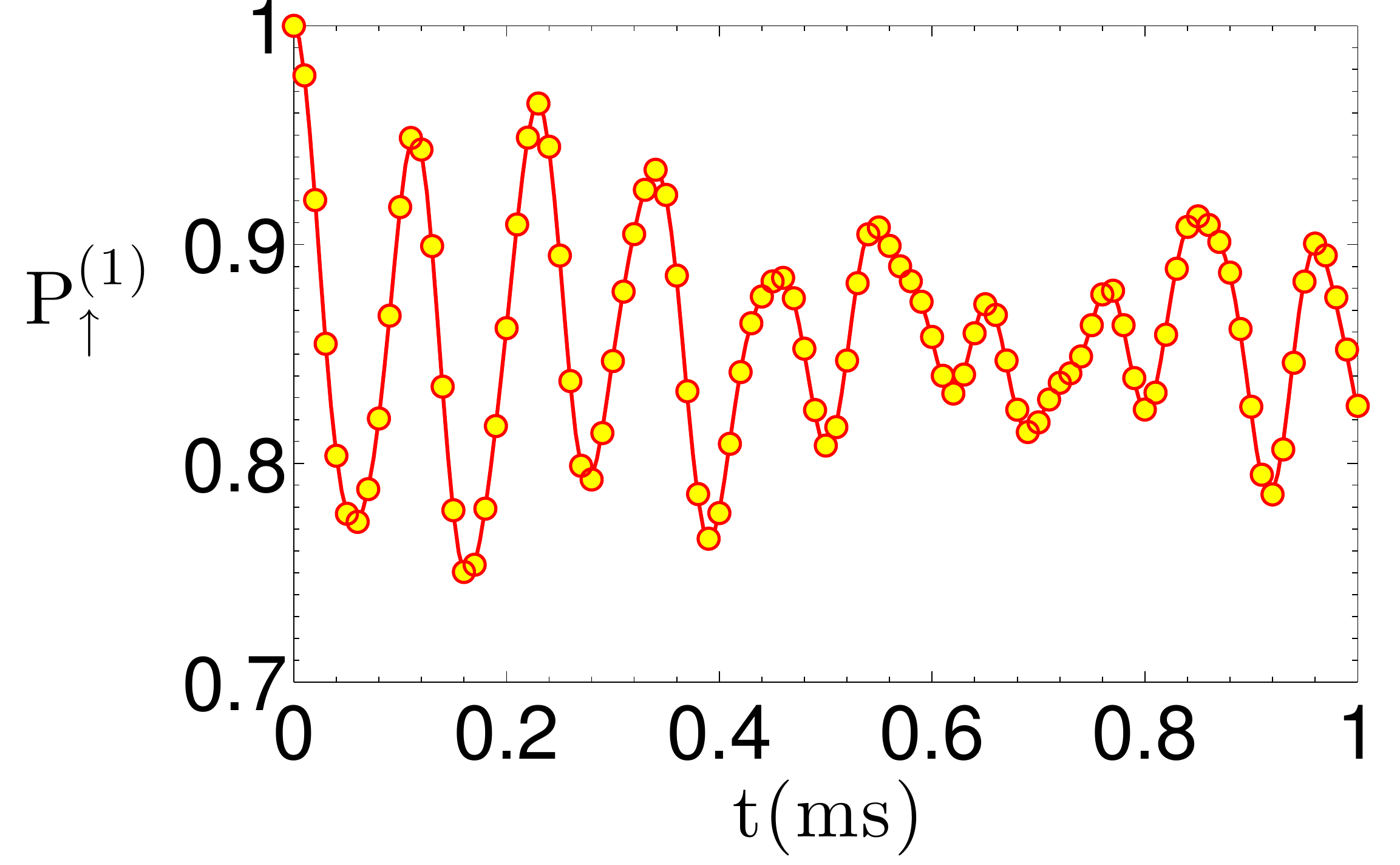}
\hspace{0.2cm}
\includegraphics[width=5cm]{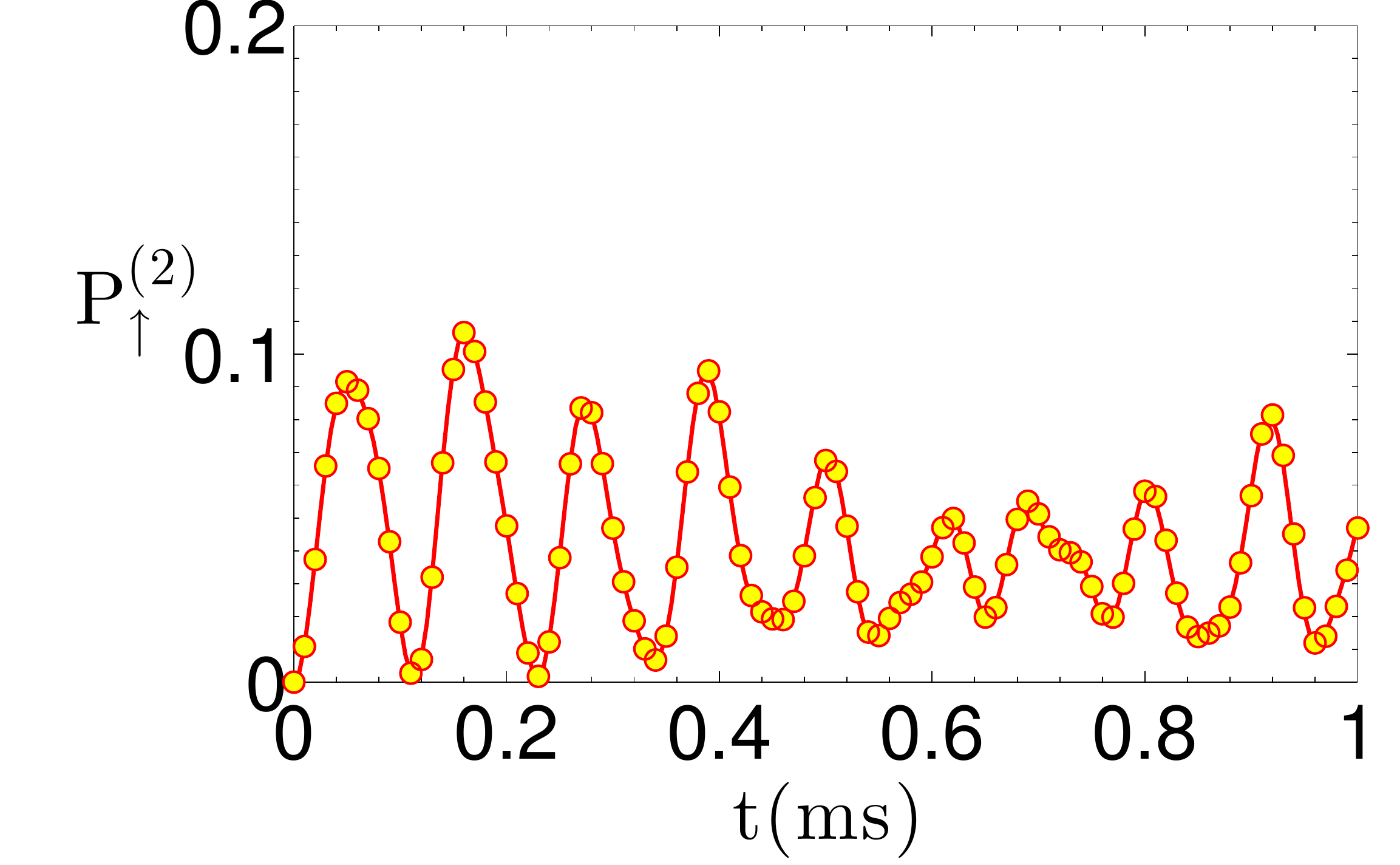}
\hspace{0.2cm}
\includegraphics[width=5cm]{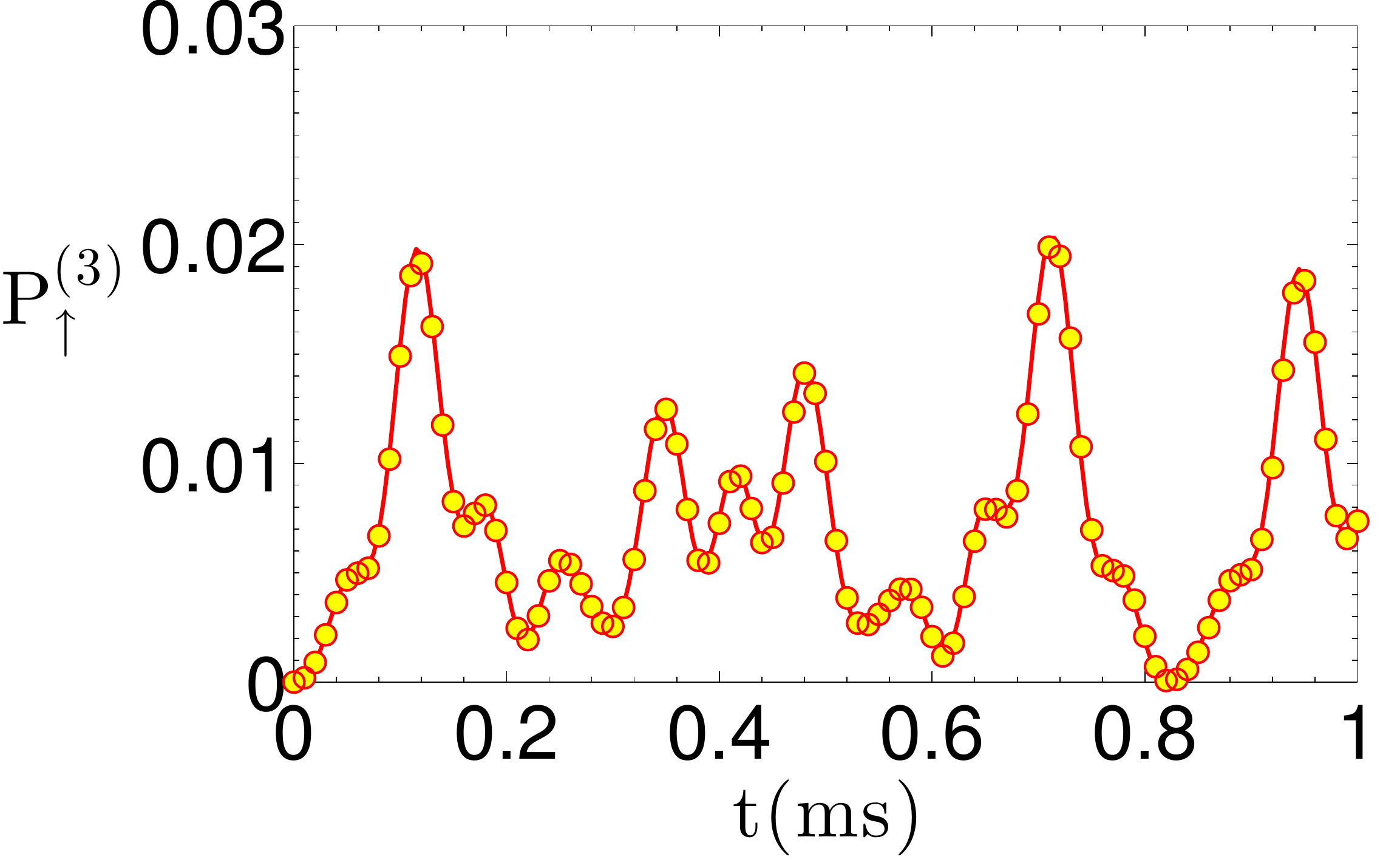}
\end{minipage}
\begin{minipage}{18cm}
\includegraphics[width=5cm]{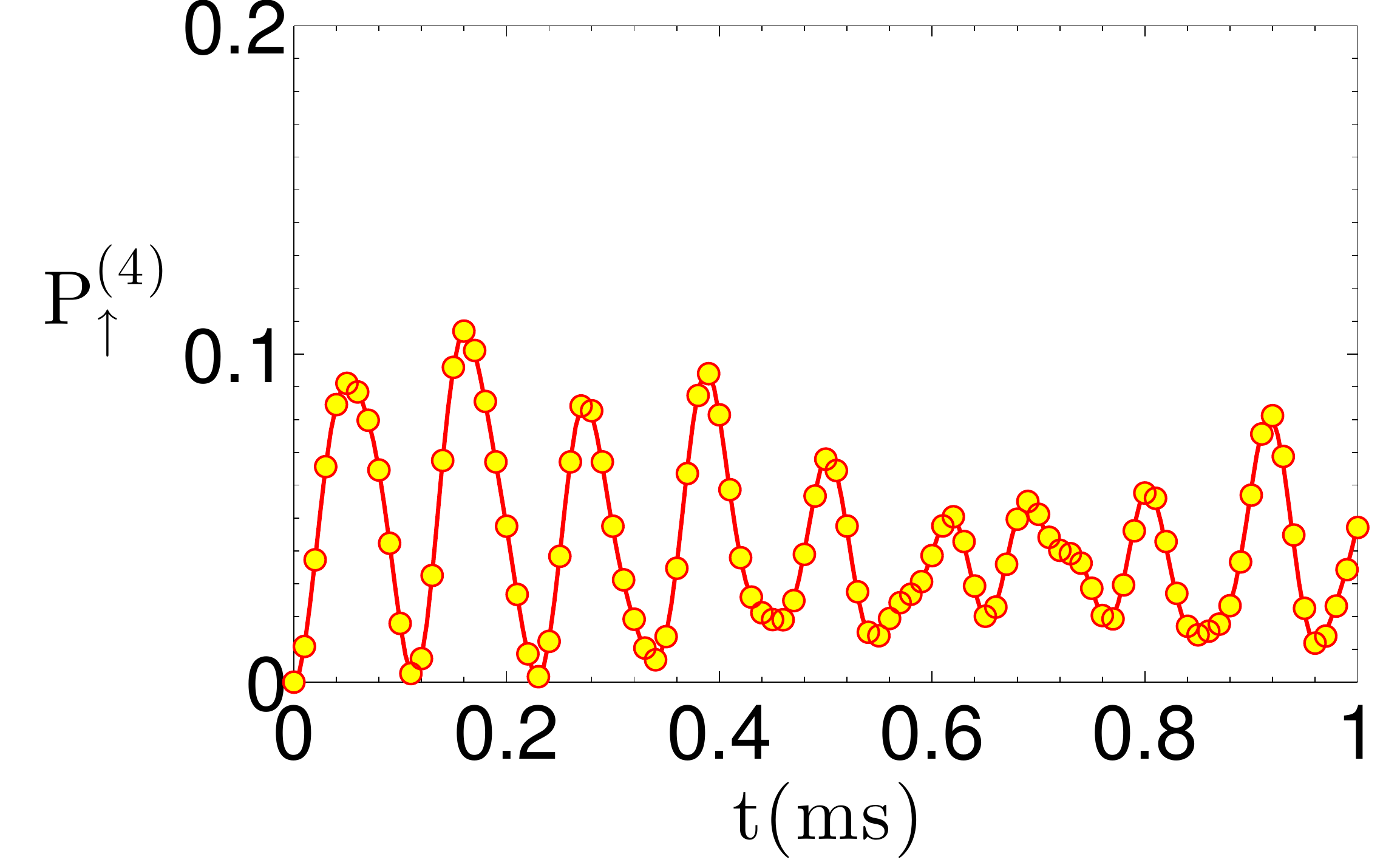}
\hspace{0.2cm}
\includegraphics[width=5cm]{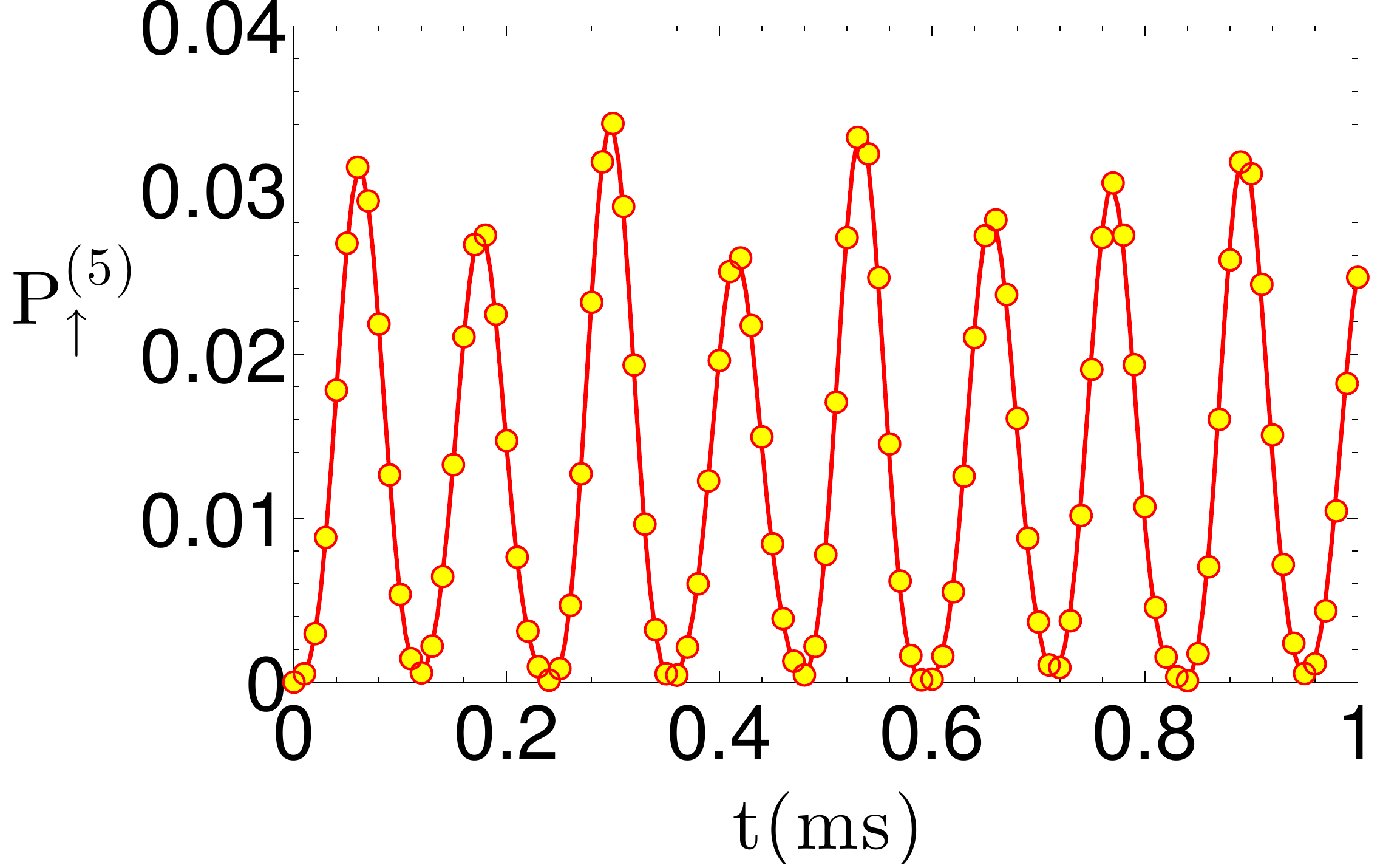}
\hspace{0.2cm}
\includegraphics[width=5cm]{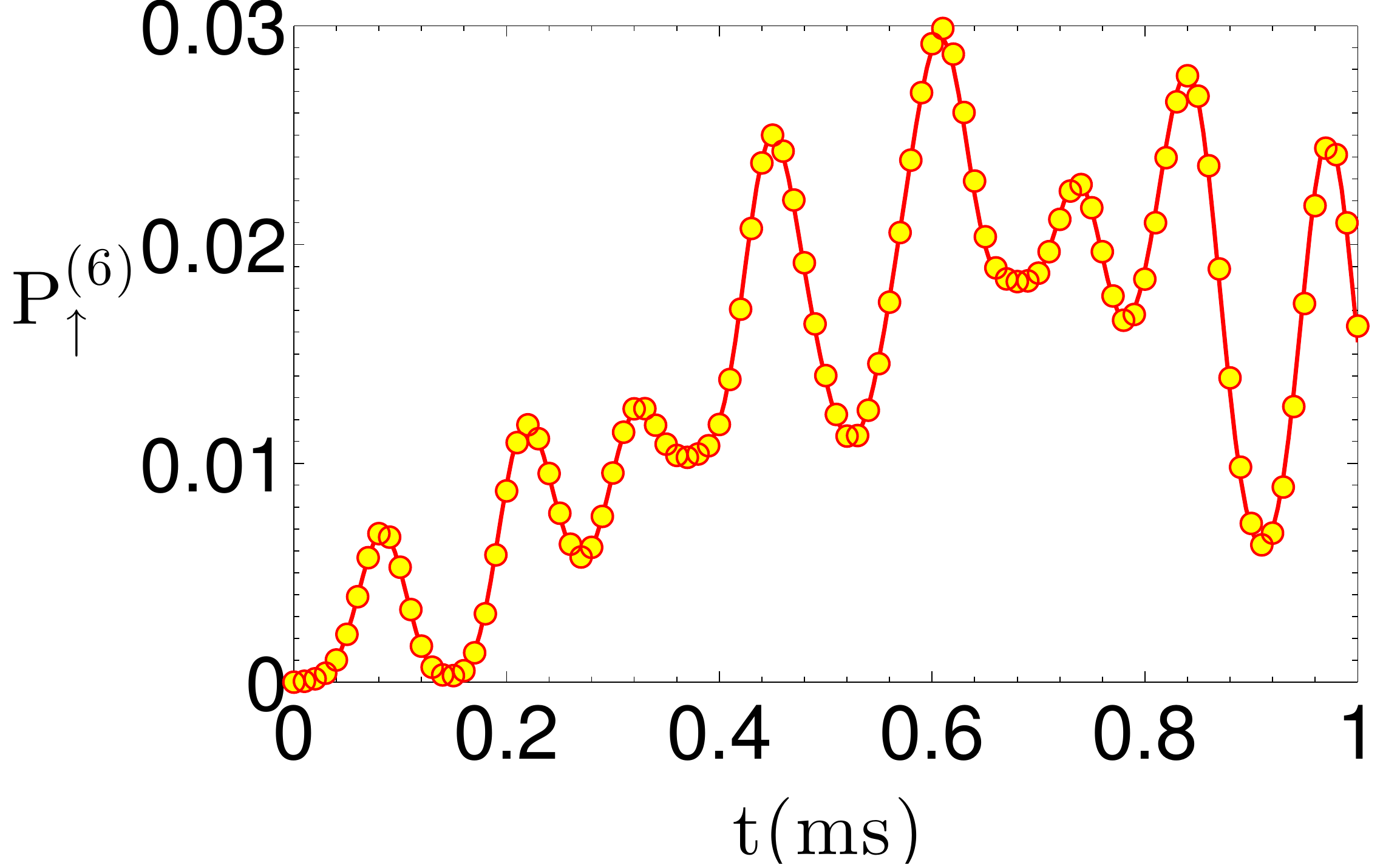}
\end{minipage}
\begin{minipage}{18cm}
\includegraphics[width=5cm]{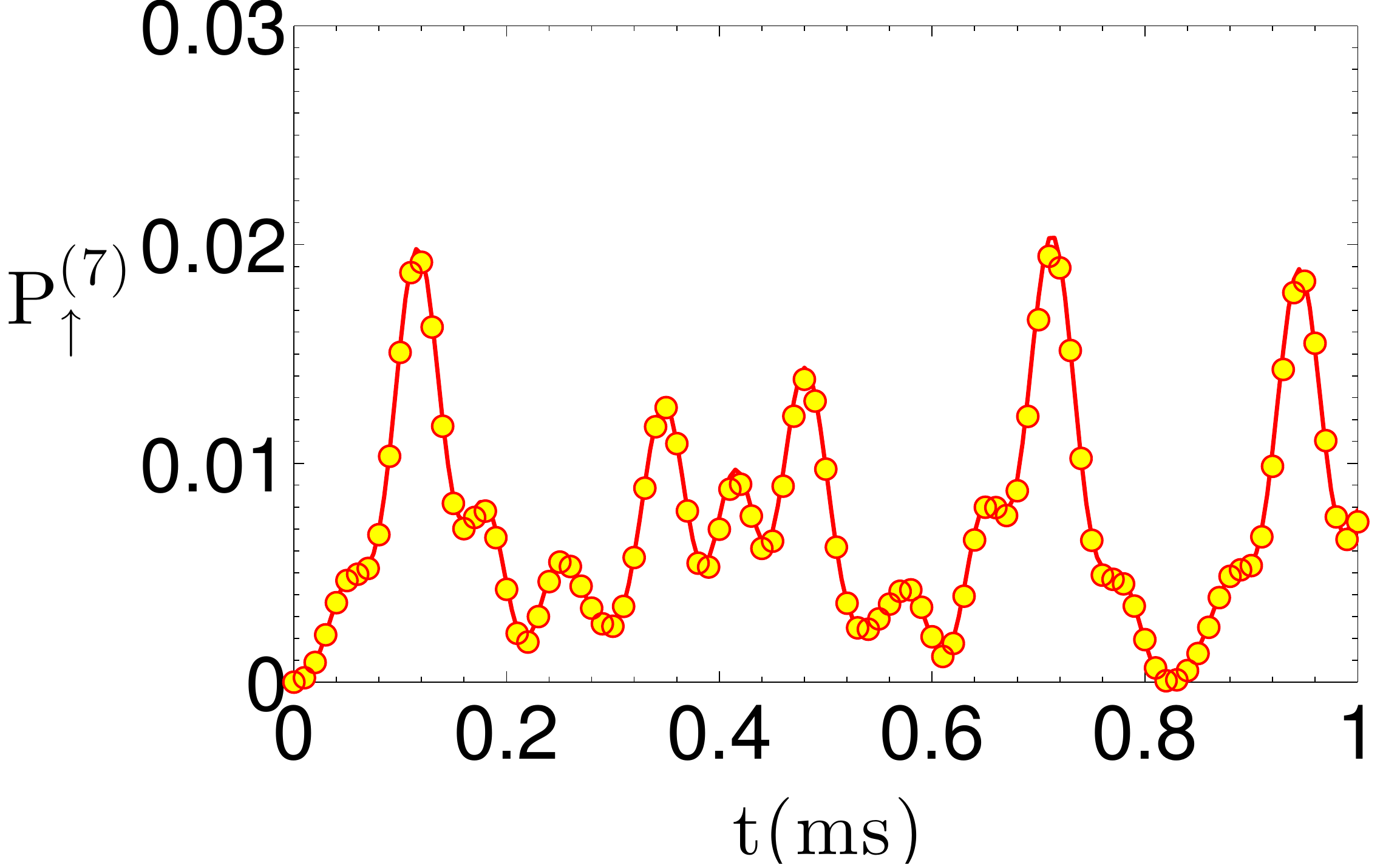}
\hspace{0.2cm}
\includegraphics[width=5cm]{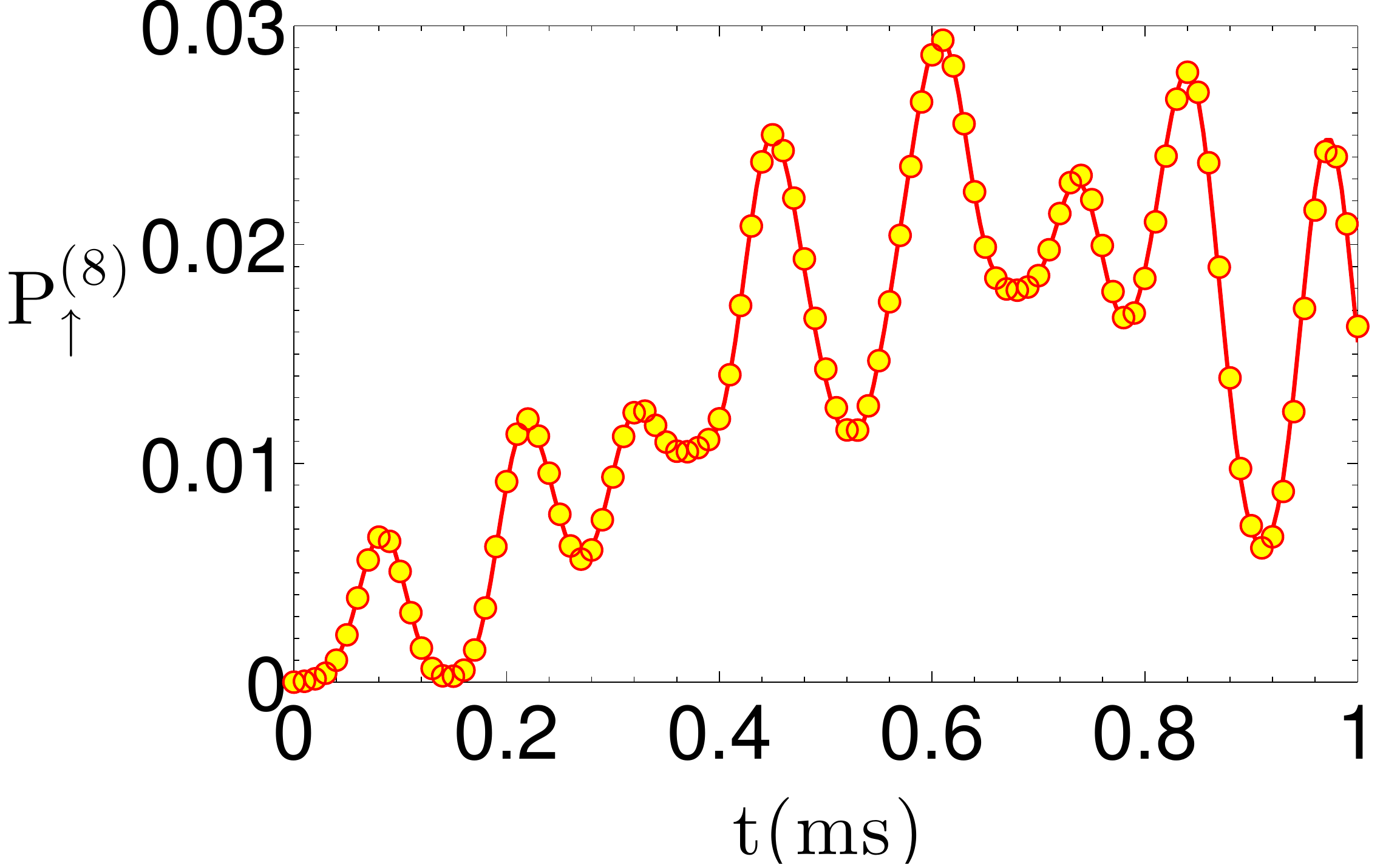}
\hspace{0.2cm}
\includegraphics[width=5cm]{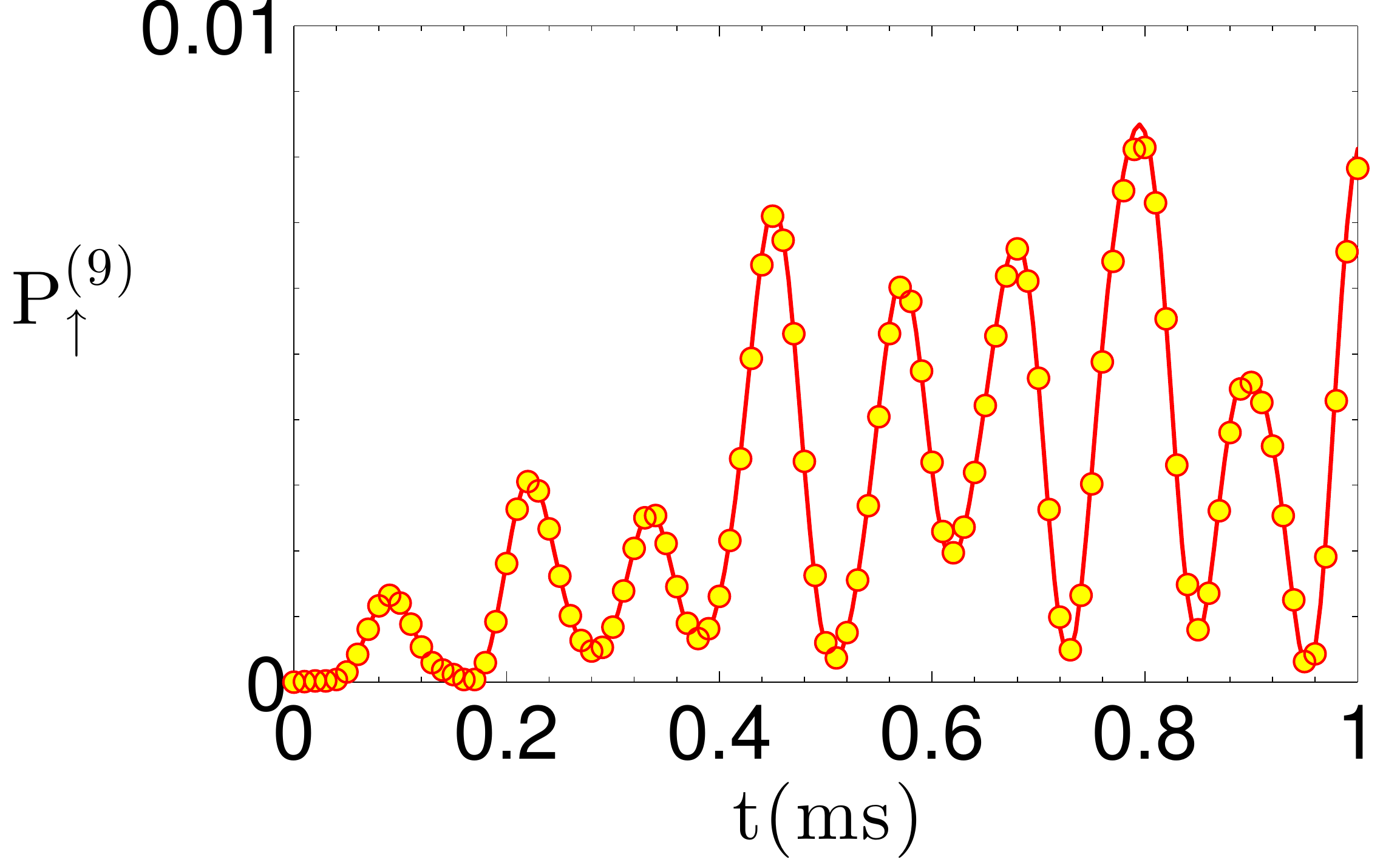}
\end{minipage}
\end{center}
\caption{Dynamics of spin state population $\mbox{P}_{\uparrow}^{(k)}$ for the initial state $\ket{\psi}=\ket{\uparrow}\otimes \ket{\downarrow}\otimes  \cdots \otimes \ket{\downarrow} $: Complete Hamiltonian in Eq.(\ref{HRWA1-s}) (curves) {\it vs.} effective Hamiltonian in Eq.(\ref{HRWA2-s}) (circles). The nuclear spin lattice is $3\times 3$, the magnetic field direction is $\hat{m}=(0,0,1)$ and its strength is $750\mbox{G}$ (corresponding to 3$\mbox{MHz}$ for fluorine nuclear spin). Note that the scales in different panels have different order of magnitude.}\label{fig:RWA-s}
\end{figure}

{\it Validation of the rotating wave approximation.---} We have adopted the rotating wave approximation (RWA) to obtain the XXZ spin Hamiltonian. More specifically, the Hamiltonian describing the magnetic dipole-dipole interaction
\be
H_{\mbox{F}}=\sum_i \gamma_N \mbox{B} \mathbf{s}_i^{z}+\frac{\mu_0}{4\pi} \sum_{i,j} \frac{\gamma_N ^2}{r_{ij}^3} \blb{{\bf s} _i \cdot {\bf s}_j-3\bla{{\bf s}\cdot \hat{\mathbf{r}}_{ij}}\bla{{\bf s}_j\cdot \hat{\mathbf{r}}_{ij}}}
\label{HRWA1-s}
\ee
is approximated by the effective Hamiltonian as
\be
H_S=\sum_i \gamma_N \mbox{B}  \mathbf{s}_i^{z} +\frac{\mu_0}{4\pi} \sum_{i,j} \frac{\gamma_N ^2}{r_{ij}^3}  \bla{1-3( \hat{\mathbf{r}}_{ij}^z)^2} \blb{  {\bf s}_i^z\cdot {\bf s}_j^z - \Delta ({\bf s}_i^x\cdot {\bf s}_j^x+{\bf s}_i^y\cdot {\bf s}_j^y) }
\label{HRWA2-s}
\ee
The rotating wave approximation is valid if the magnetic field is much larger than the spin-spin coupling, i.e. $\gamma_N \mbox{B} \gg \bla{\mu_0 \gamma_N ^2}/\bla{4\pi r_{ij}^3}$. We have verified the rotating wave approximation by comparing the dynamics from the Hamiltonian of Eq.(\ref{HRWA1-s}) and Eq.(\ref{HRWA2-s}). With the initial state $\ket{\psi}=\ket{\uparrow}\otimes \ket{\downarrow}\otimes  \cdots \otimes \ket{\downarrow} $, we plot in Fig.\ref{fig:RWA-s} the state population for each nuclear spin. One can see that the dynamics of the complete Hamiltonian agrees well with the one from the effective Hamiltonian. \\

{\it Isolating nuclear spins for cooling and measurement.---}  Due to the very small energy 
splitting of nuclear spins subjected to a magnetic field, it is very challenging to cool nuclear 
spins to a sufficiently low entropy state, such that they exhibit quantum properties. 
To overcome this problem, we propose to use the electron spin of NV centers to transfer
polarization to the nuclear spins. Hence, we can take advantage of the fact that the NV center electron 
spin can be easily polarized even at room temperatures. Due to the large zero field splitting of 
the NV center electron spin, direct polarization transfer to nuclear spins is not possible due to 
the large energy mismatch. One possible way to bypass this problem is to continuously drive the NV 
electron spin to induce microwave dressed states $\{\ket{{\uparrow_x}},\ket{{\downarrow_x}}\}$ (i.e. the 
eigenstates of the effective Hamiltonian in Eq.(\ref{eq:HNV-x}) realizing an effective spin$-\frac{1}{2}$), the polarization of which can now efficiently be transferred to the nuclear spins when the Hartmann-Hahn condition is fulfilled (i.e. when the driving Rabi frequency 
matches the Larmor frequency of the nuclei) \cite{DS1112-s,Hahn62-s}. 
It is important to note that the coupling strength between nearest-neighbour fluorine 
nuclear spins is $6.8 \mbox{kHz}$, while the interaction between the NV center and the fluorine nucleus is much 
smaller (e.g. $\sim 0.5 \mbox{kHz}$ for a distance of $5 \mbox{nm}$). Therefore, it is rather complicated to achieve the Hartman-Hahn 
condition with each energy gap as required for the polarization exchange. The most promising solution consists of 
decoupling the nuclear spins from each other. This is also important 
for the read-out of quantum simulation result in order to avoid significant changes of nuclear spin states 
during the measurement. To achieve this goal, we apply a RF-field with the amplitude denoted as 
$\Omega_{\mbox{p}}$ whose frequency is off-resonant with the Larmor frequency of the nuclear spins 
by the detuning $\Delta_{\mbox{p}}$. The total Hamiltonian is written as follows
\bea
H_{\mbox{p}}&&= \frac{\Omega_{\mbox{nv}}}{2}  \sigma_{x}  \qquad \qquad \qquad \qquad \qquad \qquad \qquad \qquad \qquad \qquad \qquad \qquad \qquad \qquad \qquad \qquad \qquad   (H_{\mbox{nv}}^{\mbox{gs}})\\
&&+ \sum_i \gamma_N \mbox{B} \mathbf{s}_i^{z}
+\frac{\mu_0}{4\pi} \sum_{i,j} \frac{\gamma_N ^2}{r_{ij}^3} \blb{{\bf s} _i \cdot {\bf s}_j-3\bla{{\bf s}_i\cdot \hat{\mathbf{r}}_{ij}}\bla{{\bf s}_j\cdot \hat{\mathbf{r}}_{ij}}} +\Omega_{\mbox{p}} \cos{\blb{(\gamma_N  \mbox{B} - \Delta_{\mbox{p}})t}}\sum_i \mathbf{s}_i^{x} \qquad  (H_{\mbox{F}}) \\
&&+ \frac{1}{2}(\mathbf{I}+\sigma_{z}) \otimes \sum\limits_i \bla{a_x \mathbf{s}_i^{x}+a_y \mathbf{s}_i^{y}+a_z \mathbf{s}_i^{z}}  \qquad  \qquad \qquad \qquad \qquad \qquad \qquad  \qquad  \qquad \qquad \quad  (H_{\mbox{nv}-\mbox{F}}).
\label{HNVF_18-s}
\eea
We choose the magnetic field such that the energy splitting of the nuclear spin states is much larger than the nuclear coupling strength. This allows us to simplify the above Hamiltonian and obtain
\bea
H_{\mbox{p}}^{(1)}&&= \frac{\omega_{\mbox{nv}}}{2}  \sigma_{x}  \qquad  \qquad \qquad  \qquad \qquad\qquad  \qquad \qquad\qquad \qquad \qquad \qquad \qquad \qquad \qquad   (H_{\mbox{nv}}^{\mbox{gs}})\label{eq:pxynz-s1}\\
&&+\frac{\mu_0}{4\pi} \sum_{i,j} \frac{\gamma_N ^2}{r_{ij}^3}  \bla{1-3( \hat{\mathbf{r}}_{ij}^z)^2} \blb{  {\bf s}_i^z {\bf s}_j^z - \frac{1}{2} ({\bf s}_i^x {\bf s}_j^x+{\bf s}_i^y {\bf s}_j^y) }  +\frac{\Omega_{\mbox{p}}}{2} \sum_i \mathbf{s}_i^{x}+\Delta_{\mbox{p}} \sum_i \mathbf{s}_i^{z} \qquad  (H_{\mbox{F}}) \label{eq:pxynz-s}\\
&&+ \frac{1}{2}\sum\limits_i a_z \mathbf{s}_i^{z} +\frac{1}{4} \{  \sigma^+_{(x)} \otimes \sum\limits_i \blb{ (a_i^x)^2+(a_i^y)^2}^{1/2} \mathbf{s}_i^{-}+h.c.\} \qquad  \qquad \qquad \qquad \quad    (H_{\mbox{nv}-\mbox{F}}) \label{eq:pxynz-s2}
\eea
where $\omega_{\mbox{nv}}=\Omega_{\mbox{nv}}-\bla{\gamma_N\mbox{B}-\Delta_{\mbox{p}}}$, $ \sigma^+_{(x)} =\ketbra{{\uparrow_x}}{{\downarrow_x}}$ and $ \sigma^-_{(x)} =\ketbra{{\downarrow_x}}{{\uparrow_x}}$ are the raising and lowering operator for the dressed spin-$\frac{1}{2}$ of the NV center, and the term $\frac{1}{2}\sum\limits_i a_z \mathbf{s}_i^{z} $ gives the additional field on nuclear spins as created by the NV center. We note that the local part of nuclear spin Hamiltonian $H_{\mbox{F}}=\frac{\Omega_{\mbox{p}}}{2} \sum_i \mathbf{s}_i^{x}+\Delta_{\mbox{p}} \sum_i \mathbf{s}_i^{z} $ introduces a new nuclear spin basis which depends on the ratio $\Delta_{\mbox{p}}/\Omega_{\mbox{p}}$. In particular, if
\be
\Omega_{\mbox{p}}=2\sqrt{2}\Delta_{\mbox{p}}
\ee
\\
the new nuclear spin basis is
\bea
\ket{\tilde{\uparrow}}&=&\cos(\phi)\ket{\uparrow}+\sin(\phi)\ket{\downarrow}\\
\ket{\tilde{\downarrow}}&=&\sin(\phi)\ket{\uparrow}-\cos(\phi)\ket{\downarrow}
\eea
with
\be
\cos(2\phi)=\sqrt{\frac{1}{3}} \qquad \mbox{and} \qquad \sin(2\phi)=\sqrt{\frac{2}{3}}.
\ee
Written in such a basis, the excitation number conserving terms from ${\bf s}_i^z {\bf s}_j^z $ 
and ${\bf s}_i^x {\bf s}_j^x+{\bf s}_i^y {\bf s}_j^y$ cancel each other due to their opposite 
signs, see Eq.(\ref{eq:pxynz-s}), additionally the effects of the other non-energy conserving terms 
are suppressed by the large energy mismatch as long as $\omega_{\mbox{f}}\equiv\sqrt{\Delta_{\mbox{p}}^2+(\Omega_{\mbox{p}}/2)^2}$ is much larger the 
nuclear spin interaction. For the fluorine nuclear spins, RF-fields with an amplitude as strong as 
$200\mbox{kHz}$ are available, which are much larger than the nearest-neighbor nuclear spin 
interaction (i.e. $6.8\mbox{kHz}$). The effective Hamiltonian therefore becomes
\be
H_{\mbox{p}}^{(2)}=\frac{\omega_{\mbox{nv}}}{2}  \sigma_{x} +\omega_{\mbox{f}} \sum_{i} \tilde{\mathbf{s}}_i^z+
\sum_i g_i^{\parallel} \tilde{\mathbf{s}}_i^z  +\sum_i g_i^{\perp} \bla{\sigma_{(x)}^+  \tilde{\mathbf{s}}^-_i+h.c}
\label{eq:HP2-s}
\ee
where $\tilde{\mathbf{s}}_i^z=\ketbra{\tilde{\uparrow}}{\tilde{\uparrow}}-\ketbra{\tilde{\downarrow}}{\tilde{\downarrow}}$ and $ \tilde{\mathbf{s}}^+=  \ketbra{\tilde{\uparrow}}{\tilde{\downarrow}}$, $ \tilde{\mathbf{s}}^-=  \ketbra{\tilde{\downarrow}}{\tilde{\uparrow}}$ are the Pauli operator and raising (lowering) operator 
for nuclear spins in the effective spin basis. The coefficients $ g_i^{\parallel}=\frac{1}{2}\cos(2\varphi)
a_i^z  $ represent the additional field induced by the NV center on the nuclear spins. $g_i^{\perp}=\frac{1}{8}\blb{1+\cos(2\varphi)} \blb{(a_i^x)^2+(a_i^y)^2}^{1/2}$ gives 
the rate of polarization exchange between the NV center and the nuclear spins, where $a_i^\alpha$ 
are the dipole-dipole interaction strength as in Eq.(\ref{HNVF_18-s}). Therefore, nuclear 
spins now evolve independently of each other and couple with the NV center individually, see 
Eq.(\ref{eq:HP2-s}).\\

{\it Polarization dynamics of the nuclear spins.---} The whole polarization process consists of 
repetitive cycles. In each cycle, we initialize NV center in the $\ket{m_s=0}＝\equiv \ket{{\downarrow}}$ 
state with a green laser ($537$ $ \mbox{nm}$), and prepare it in the state $\ket{{\downarrow_x}}=\sqrt{\frac{1}{2}} (\ket{{\uparrow}}+\ket{{\downarrow}})$ with 
a $\frac{\pi}{2}$ microwave pulse. After one polarization cycle, the nuclear spin state evolve as
\be
\rho_{\mbox{f}} (n\tau+\tau)\equiv \mbox{tr}_{\mbox{nv}} [\exp{(-itH_{\mbox{p}})} \rho_{\mbox{f}} (n\tau) \exp{(itH_{\mbox{p}})}]
\label{eq:rhof-s}
\ee
where $\tau$ is the time of each cycle. It can be seen from the effective Hamiltonian in 
Eq.(\ref{eq:HP2-s}) that, when the effective spin-$\frac{1}{2}$ is on resonance with nuclear 
spins, polarization can be efficiently transferred from the NV center to the nuclear spins. The 
appropriate resonant condition, called Hartman-Hahn condition \cite{Hahn62-s}, turns out to be
\be
\omega_{\mbox{nv}}=\Omega_{\mbox{nv}}-\bla{\gamma_N\mbox{B}-\Delta_{\mbox{p}}} = \omega_f\equiv\sqrt{\Delta_{\mbox{p}}^2+(\Omega_{\mbox{p}}/2)^2}\label{eq:HHC-s}.
\ee
The dynamical nuclear polarization process will finally polarize nuclear spins towards the product state $\ket{\tilde{\downarrow}}\otimes \cdots \otimes \ket{\tilde{\downarrow}}$. We have performed exact 
numerical simulations for a 3$\times$3 lattice with a Chebyshev expansion to calculate $U(\tau)=\exp{(-itH_{\mbox{p}})}$ where $H_{\mbox{p}}$ is the Hamiltonian in Eq.(\ref{eq:pxynz-s1}-\ref{eq:pxynz-s2}) with the rotating wave approximation, by assuming that the magnetic field is much stronger than the nuclear spin coupling, which is easily satisfied in experiments. The result is 
shown in Fig.3(a-b) of the main text, which shows that nuclear spins are indeed isolated and our 
polarization scheme works very well for the present system. Due to the limited computational power, to 
estimate the polarization efficiencies in large systems, we describe the polarization dynamics with the following master 
equation\\
\be
\frac{d}{d t} \rho_{\mbox{f}}=-[A_z,\rho_{\mbox{f}}] -\frac{\tau} {2}[A_z,[A_z,\rho_{\mbox{f}}] ]-\frac{\tau} {2}(\rho_{\mbox{f}} A_+A_-+A_+A_-\rho_{\mbox{f}}-2 A_-\rho_{\mbox{f}} A_+)
\ee
where $A_z=\sum_i g_i^{\parallel} \tilde{\mathbf{s}}_i^z $ and $A_{\pm} = \sum_i g_i^{\perp}  \tilde{\mathbf{s}}^{\pm}_i$.
The above master equation can be derived from Eq.(\ref{eq:rhof-s}) by assuming that the polarization 
cycle time $\tau$ is sufficiently short (namely satisfies $\tau  \sqrt{\sum_i (g_i^{\perp})^2} =\epsilon \ll 1$) \cite{Christ07-s}. To numerically solve the above master equation, one needs to make some approximations 
on the coherence between nuclear spins. If we neglect coherences between nuclear spins (known as the
spin temperature approximation), then the polarization rate is $\p_i=(g_i^{\perp})^2 \tau =\epsilon (g_i^{\perp})^2/ \sqrt{\sum_k (g_k^{\perp})^2}\approx \frac{1}{N}(\epsilon g_i^{\perp})$. In this case, the polarization time scales linearly with the total number of nuclear spins. To take into account the coherence between nuclear spins, one can approximate the higher correlation terms by a 
Wick-type factorization \cite{Christ07-s} as follows
\be
\frac{1}{2} \langle \mathbf{s}_i^+ \mathbf{s}_j^z\mathbf{s}_k^-\rangle =\frac{1}{2}\langle \mathbf{s}_i^+ \mathbf{s}_k^-\rangle -\langle \mathbf{s}_i^+  \mathbf{s}_j^-\rangle \langle \mathbf{s}_j^+  \mathbf{s}_k^-\rangle+\langle \mathbf{s}_i^+  \mathbf{s}_k^-\rangle \langle \mathbf{s}_j^+  \mathbf{s}_j^-\rangle.
\ee
\begin{figure}[t]
\begin{center}
\includegraphics[width=7cm]{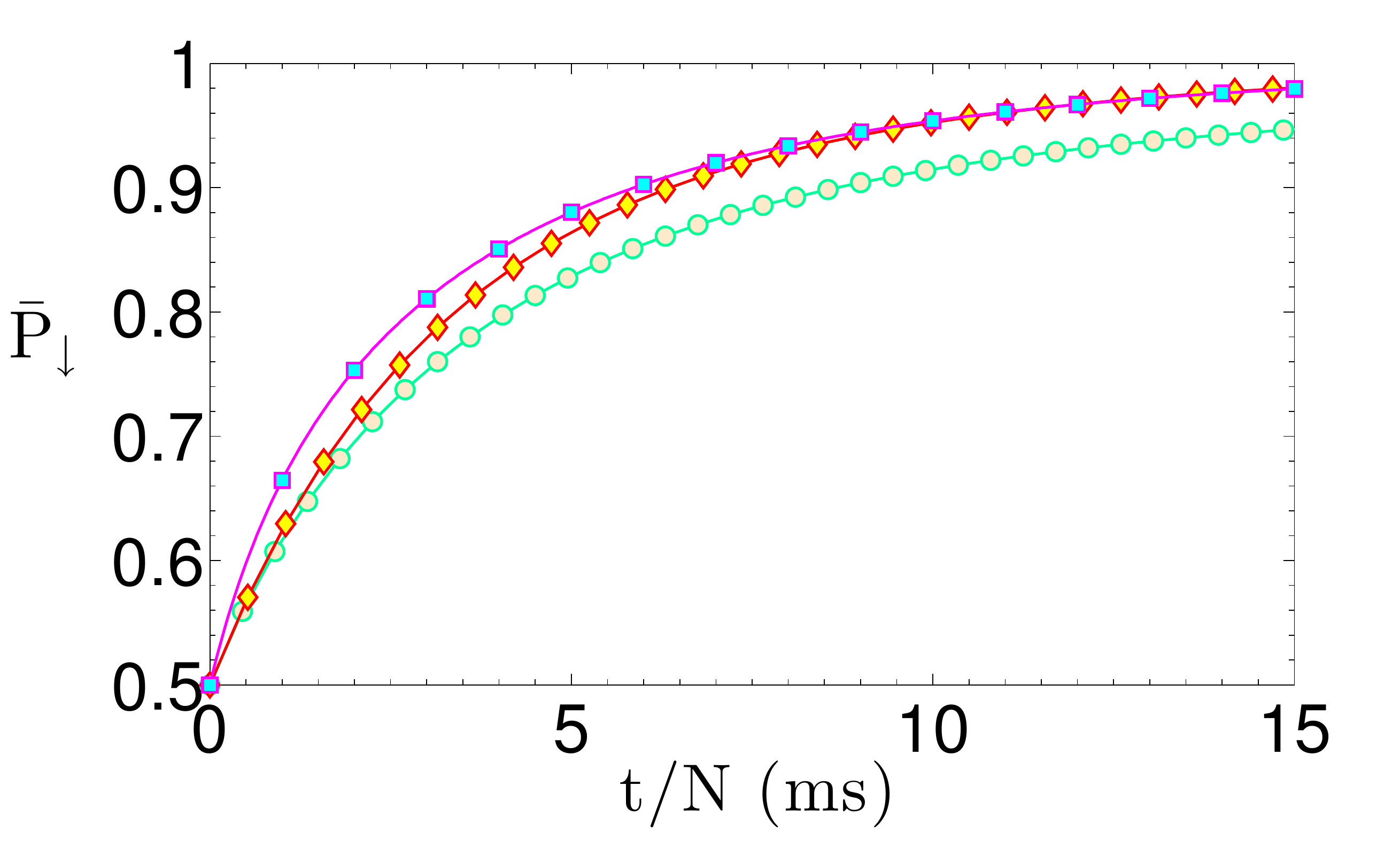}
\end{center}
\caption{The average probability of nuclear spin down state $P_{\downarrow}$ as a function of 
the polarization time $\tau$. Comparison of the spin temperature (red, diamond), Wick-type (green, 
circle) for a $10\times 10$ nuclear spin lattice with the polarization cycle time $\tau=3 \mu s$. 
The spin-temperature limited polarization efficiency can be approached by introducing magnetic noise 
to remove nuclear spin coherence every $20 \mbox{ms}$, the result is calculated with the Wick-type 
approximation (purple, square) with $\tau=5 \mu s$. The magnetic field direction is $\hat{m}=(\sqrt{\frac{1}{2}},0,\sqrt{\frac{1}{2}})$.}\label{fig:POL10x10-s}
\end{figure}
Comparing these approximations with each other and with the exact simulations on small systems can help us to obtain reliable estimations for the polarization efficiencies in large systems. In Fig3.(b) of the main text, we compare the exact numerical calculation with the approximation of spin temperature for a $3\times 3$ nuclear spin lattice. It can been seen that for the chosen $\tau$, 
they show good agreement. We also compare the polarization dynamics for a 10$\times$10 nuclear spin 
lattice under the spin temperature and the Wick-type approximation as shown in Fig.\ref{fig:POL10x10-s}. 
Thus it is possible to estimate the polarization rate based on the spin temperature approximation, 
i.e. the polarization rate is $\p_i\approx \frac{1}{N}(\epsilon g_i^{\perp})$. From these results
we estimate that for the example of a fluorine nuclear spin array with a distance of $5 \mbox{nm}$ 
from the NV center, the polarization time scale for a total number of $N$ nuclear spins is 
$(p_i)^{-1}=N \cdot 10 ( \mbox{ms})$ with $\epsilon=0.1$. Further improvements of the polarization 
efficiency can be expected by optimizing the cycle time.\\

{\it NV center as a quantum probe for measurement of the nuclear spin state.---} Following initialization
and evolution of our nuclear spin quantum simulator, we must be able to determine its properties via 
appropriate measurements. The measurement of individual nuclear spins is challenging due to their
small magnetic moment. In the main text, we propose that an NV center can serve as a measurement interface 
for nuclear spin states. We adopt the same strategy for isolating nuclear spins from each other
during measurements as in the polarization process described in the main text. Thus, we need to apply 
an RF-pulse to map the nuclear spin states from the original spin basis 
$\{\ket{{\uparrow}},\ket{{\uparrow}}\}$ to the basis for measurement $\{\ket{\tilde{\uparrow}},\ket{\tilde{\downarrow}}\}$. The dynamics is again described by the Hamiltonian in Eq.(\ref{eq:HP2-s})
\be
H_{\mbox{p}}^{(2)}=\frac{\omega_{\mbox{nv}}}{2}  \sigma_{x} +\omega_{\mbox{f}} \sum_{i} \tilde{\mathbf{s}}_i^z+
\sum_i g_i^{\parallel} \tilde{\mathbf{s}}_i^z  +\sum_i g_i^{\perp} \bla{\sigma_{(x)}^+  \tilde{\mathbf{s}}^-_i+h.c}.
\ee
Before measurement, we prepare the NV center either in the state $\ket {{\uparrow}_x}$ or 
$\ket {{\downarrow}_x}$. The microwave driving field on the NV center is then tuned to match 
the Hartmann-Hahn condition, see Eq.(\ref{eq:HHC-s}). After time $\tau$, we perform measurement 
on the NV center which is denoted as an observable $A$, which can be written as follows
\bea
A(\tau) =A(0) +i\frac{\tau}{\hbar} [H_{\mbox{p}},A]-\frac{\tau^2}{2\hbar^2} [H_{\mbox{p}},[H_{\mbox{p}},A]]+...
\eea
In particular, we can measure the population of states $\ket{+}\equiv \ket{{\uparrow_x}}$ and $\ket{-}\equiv \ket{{\downarrow_x}}$ of the NV center, and the corresponding observables are $A=\ketbra{+}{+}$ and $A=\ketbra{-}{-}$ respectively. One can obtain
\bea
P_-^+=&&\tau^2 \sum_i\sum_j  ( g_i^{\perp}   g_j^{\perp} )\langle \tilde{\mathbf{s}}^+_i  \tilde{\mathbf{s}}^-_j\rangle=\tau^2 \sum_i (g_i^{\perp} )^2 P_i^{\tilde{\uparrow}}+\tau^2 \sum_{(i,j),i\neq j}  ( g_i^{\perp}   g_j^{\perp} )  \langle \tilde{\mathbf{s}}^+_i  \tilde{\mathbf{s}}^-_j + \tilde{\mathbf{s}}^-_i  \tilde{\mathbf{s}}^+_j\rangle,\\
P_+^-=&&\tau^2 \sum_i\sum_j  ( g_i^{\perp}   g_j^{\perp} ) \langle  \tilde{\mathbf{s}}^-_i  \tilde{\mathbf{s}}^+_j\rangle=\tau^2 \sum_i (g_i^{\perp} )^2 P_i^{\tilde{\downarrow}}+\tau^2 \sum_{(i,j),i\neq j}  ( g_i^{\perp}  g_j^{\perp} ) \langle \tilde{\mathbf{s}}^+_i  \tilde{\mathbf{s}}^-_j + \tilde{\mathbf{s}}^-_i  \tilde{\mathbf{s}}^+_j\rangle
\eea
where $P_i^{\tilde{\uparrow}}$  and $P_i^{\tilde{\downarrow}}$ are the spin up and down population of individual nuclear spins, $ \sum_{(i,j),i\neq j} $ is the sum over all pairs of nuclear spins. Therefore, we obtain
\be
\Delta_P= P_-^+ - P_+^- =\tau ^2 \sum_i (g_i^{\perp} )^2 (P_i^{\tilde{\uparrow}}-P_i^{\tilde{\downarrow}})=2 \tau^2 \sum_i (g_i^{\perp} )^2 \langle \tilde{\mathbf{s}}_z^i \rangle.
\ee
The observable $\Delta_P$ thus provides information about the average magnetization of nuclear spins. We also obtain
\be
\Delta_{XY}= P_-^ + + P_+^-  =\tau^2 \sum_i\sum_j( g_i^{\perp}   g_j^{\perp} ) \langle \tilde{\mathbf{s}}^+_i  \tilde{\mathbf{s}}^-_j + \tilde{\mathbf{s}}^-_i  \tilde{\mathbf{s}}^+_j\rangle
=2\tau^2 \sum_i\sum_j( g_i^{\perp}   g_j^{\perp} ) \langle \tilde{\mathbf{s}}^x_i  \tilde{\mathbf{s}}^x_j + \tilde{\mathbf{s}}^y_i  \tilde{\mathbf{s}}^y_j\rangle
\label{eq:DXY-s}
\ee
which provides information about  ($xx$ and $yy$) correlation functions of nuclear spins. To estimate the other correlation functions, we introduce  the Hadamard operation $O_H$, the phase transformation $O_I$ and $O_P$ as follows
\be
O_H=\frac{1}{\sqrt{2}}\left( \begin{array}{cc}
1& 1  \\
1 &  -1
\end{array} \right), \quad O_I=\left( \begin{array}{cc}
1& 0  \\
0 &  i
\end{array} \right),
\quad O_P=\left( \begin{array}{cc}
1& i  \\
1 &  -i
\end{array} \right).
\ee
The spin operators under the transformations $O_H$ and $O_I$ are listed as follows:
\bea
&&O_H^{\dagger } \tilde{\mathbf{s}}^x O_H=\tilde{\mathbf{s}}^z,\quad O_H^{\dagger } \tilde{\mathbf{s}}^y O_H=-\tilde{\mathbf{s}}^y, \quad O_H^{\dagger } \tilde{\mathbf{s}}^z O_H=\tilde{\mathbf{s}}^x,\\
&&O_I^{\dagger } \tilde{\mathbf{s}}^xO_I=-\tilde{\mathbf{s}}^y, \quad   O_I^{\dagger } \tilde{\mathbf{s}}^y O_I=-\tilde{\mathbf{s}}^x, \quad O_I^{\dagger } \tilde{\mathbf{s}}^z O_I=\tilde{\mathbf{s}}^z,\\
&& O_P^{\dagger } \tilde{\mathbf{s}}^x O_P=\tilde{\mathbf{s}}^z, \quad O_P^{\dagger } \tilde{\mathbf{s}}^y O_P=-\tilde{\mathbf{s}}^x, \quad O_P^{\dagger } \tilde{\mathbf{s}}^z O_P=-\tilde{\mathbf{s}}^y.
\eea
Therefore, we apply an RF-pulse corresponding to the transformation $O=O_H, O_P$ on the nuclear spin state $\ket{\Psi}\rightarrow O \ket{\Psi}$ such that the measurement of the observable, as described in Eq.(\ref{eq:DXY-s}), provides information about the correlation functions $(yy+zz)$ or $(xx+zz)$ of the fluorine nuclear spins.
\bea
\Delta_{YZ}\vert_{\ket{\Psi}} =\Delta_{XY}\vert_{O_H \ket{\Psi}} &=& 2\tau^2 \sum_i\sum_j\bla{ g_i^{\perp}   g_j^{\perp} }\left \langle
\bla{O_H^{\dagger }\otimes O_H^{\dagger }}\bla{\tilde{\mathbf{s}}^y_i  \tilde{\mathbf{s}}^y_j + \tilde{\mathbf{s}}^z_i  \tilde{\mathbf{s}}^z_j} \bla{O_H\otimes O_H }\right\rangle\\
&=&2\tau^2 \sum_i\sum_j\bla{ g_i^{\perp}   g_j^{\perp} } \left\langle \tilde{\mathbf{s}}^x_i  \tilde{\mathbf{s}}^x_j + \tilde{\mathbf{s}}^y_i  \tilde{\mathbf{s}}^y_j\right\rangle,\\
\Delta_{XZ}\vert_{\ket{\Psi}} =\Delta_{XY}\vert_{O_P \ket{\Psi}} &=& 2\tau^2 \sum_i\sum_j\bla{ g_i^{\perp}   g_j^{\perp} } \left\langle
\bla{O_P^{\dagger }\otimes O_P^{\dagger }}\bla{\tilde{\mathbf{s}}^y_i  \tilde{\mathbf{s}}^y_j + \tilde{\mathbf{s}}^z_i  \tilde{\mathbf{s}}^z_j} \bla{O_P\otimes O_P} \right\rangle\\
&=&2\tau^2 \sum_i\sum_j\bla{ g_i^{\perp}   g_j^{\perp}} \left\langle \tilde{\mathbf{s}}^x_i  \tilde{\mathbf{s}}^x_j + \tilde{\mathbf{s}}^z_i  \tilde{\mathbf{s}}^z_j\right\rangle.
\eea
Together with $\Delta_{XY}$ in Eq.(\ref{eq:DXY-s}), it is thus possible to estimate all the correlations of $XX$, $YY$ and $ZZ$ of nuclear spin states. \\

To measure the observable such as nuclear structure factors of nuclear spin state, we can apply a gradient field on the nuclear spins. Accordingly, the nuclear spin at the position $\vec{\mathbf{r}}_i$ experiences a field $b_i=\vec{\mathbf{r}}_i \cdot (b_x,b_y)$, and gains a position-dependent phase $\phi_i=\vec{\mathbf{r}}_i \cdot \vec{\mathbf{q}}$ where $\vec{q}\equiv \bla{q_x,q_y}=\bla{\gamma_N t_p} (b_x,b_y)$. We then perform the same measurement as in the above discussions and obtain
\bea
\Delta_{XY}^{\vec{\mathbf{q}}}+\Delta_{XY}^{-\vec{\mathbf{q}}}&=&2\tau^2 \sum_i\sum_j \bla{g_i^{\perp}   g_j^{\perp} } \cos{\blb{\bla{\vec{\mathbf{r}}_i-\vec{\mathbf{r}}_j}\cdot \vec{\mathbf{q}}}} \left \langle \tilde{\mathbf{s}}^x_i  \tilde{\mathbf{s}}^x_j+\tilde{\mathbf{s}}^y_i  \tilde{\mathbf{s}}^y_j\right\rangle, \label{DXY:1}\\
\Delta_{YZ}^{\vec{\mathbf{q}}}+\Delta_{YZ}^{-\vec{\mathbf{q}}}&=&2\tau^2 \sum_i\sum_j \bla{g_i^{\perp}   g_j^{\perp} } \cos{\blb{\bla{\vec{\mathbf{r}}_i-\vec{\mathbf{r}}_j}\cdot \vec{\mathbf{q}}}} \left \langle \tilde{\mathbf{s}}^y_i  \tilde{\mathbf{s}}^y_j+\tilde{\mathbf{s}}^z_i  \tilde{\mathbf{s}}^z_j\right\rangle, \label{DYZ:1}\\
\Delta_{XZ}^{\vec{\mathbf{q}}}+\Delta_{XZ}^{-\vec{\mathbf{q}}}&=&2\tau^2 \sum_i\sum_j \bla{g_i^{\perp}   g_j^{\perp} } \cos{\blb{\bla{\vec{\mathbf{r}}_i-\vec{\mathbf{r}}_j}\cdot \vec{\mathbf{q}}}} \left\langle \tilde{\mathbf{s}}^x_i  \tilde{\mathbf{s}}^x_j+\tilde{\mathbf{s}}^z_i  \tilde{\mathbf{s}}^z_j\right\rangle. \label{DXZ:1}
\eea
This makes it possible to extract the information on the structure factor defined as follows
\be
S_{\vec{\mathbf{q}}}\bla{q_x,q_y}\equiv \frac{1}{N^2}\left\langle\vert \sum_k e^{i \vec{\mathbf{q}} \cdot \vec{\mathbf{r}}} \mathbf{s}_k^z \vert^2 \right \rangle \propto \bla{\Delta_{XZ}^{\vec{\mathbf{q}}}+\Delta_{XZ}^{-\vec{\mathbf{q}}}}+\bla{\Delta_{YZ}^{\vec{\mathbf{q}}}+
\Delta_{YZ}^{-\vec{\mathbf{q}}}}-\bla{\Delta_{XY}^{\vec{\mathbf{q}}}+\Delta_{XY}^{-\vec{\mathbf{q}}}}.
\ee
The accuracy of the above estimation depends on the difference of the couplings between the NV center and individual nuclear spins. For the relevant nuclei (e.g. with non-zero correlation functions), their mutual distances are usually much smaller than their distances from the NV center. Therefore, their individual couplings with NV center (i.e. $g_i^{\perp}$ and $g_j^{\perp}$ in Eqs.(\ref{DXY:1}-\ref{DXZ:1})) will be similar. For translational invariant systems, the observable estimated from NV measurement is expected to give us information on the average properties of the system state. As the observable is estimated by the measured quantity $P_\mu^{\nu}/\tau^2$ with $\mu,\nu =\pm$, the measurement accuracy will depend on the choice of measurement time $\tau$. In practice, one can perform measurements for different values of $\tau$ and estimate the observables with improved accuracy. In the situation of quantum phase transitions, which are usually accompanied by sudden or non-analytic changes of observables, the measurement via NV centers is expected to be able to identify accurately different phases, e.g. see Fig.5 in the main text.\\

{\it Tuning parameters for quantum magnetism.---} The nuclear spin coupling strength can be controlled by the magnetic field direction as
\be
g(\mathbf{r}_{kl})=\frac{\hbar\mu_0\gamma_N^2}{4\pi r_{ij}^3} \blb{1-3\bla{\hat{\mathbf{r}_{kl}}\cdot \hat{z}}^2}=\frac{\hbar\mu_0\gamma_N^2}{4\pi r_{kl}^3} \bla{1-3\cos^2\theta_{kl}}
\ee
where $\mathbf{r}_{kl}= r_{kl} \hat{\mathbf{r}_{kl}}$ is the vector that connects the nuclear spin $k$ and $l$, and $\theta_{kl}$ is the relative angle between $\hat{\mathbf{r}_{kl}}$ and the magnetic field direction $\hat{z}$.
In the main text, we consider two examples of magnetic fields to tune the relative ratio between the nearest neighbor nuclear spin interactions along the three directions, i.e. $J_{\hat{a}_1}$, $J_{\hat{a}_2}$, $J_{\hat{a}_3}$ with $\hat{a}_1=(1,0,0)$ $\hat{a}_2=(\frac{1}{2},\frac{\sqrt{3}}{2},0)$, $\hat{a}_3=(-\frac{1}{2},\frac{\sqrt{3}}{2},0)$. \\

\begin{figure}[h]
\begin{center}
\hspace{-0.8cm}
\includegraphics[width=4.5cm]{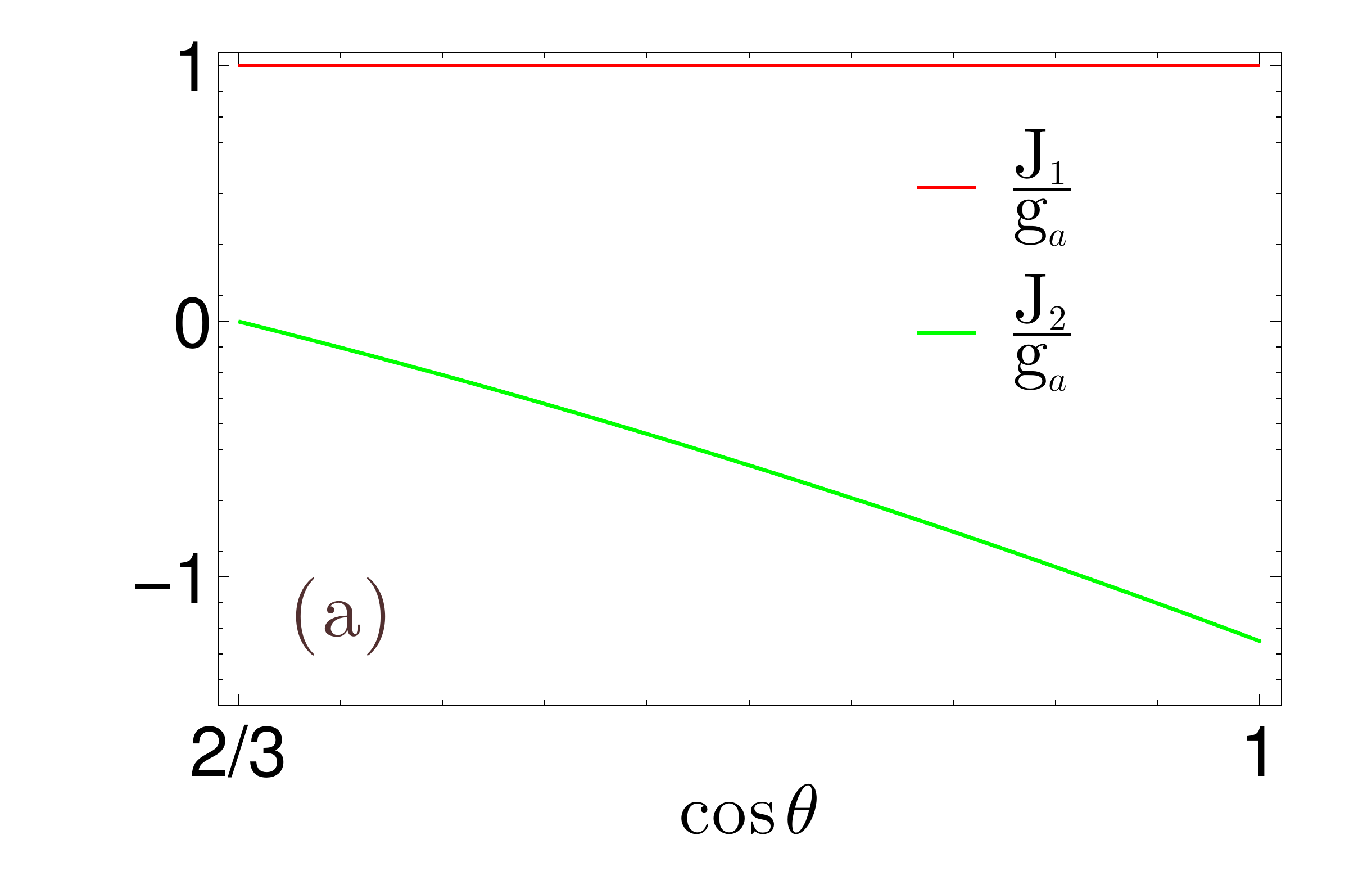}
\hspace{-0.5cm}
\includegraphics[width=4.5cm]{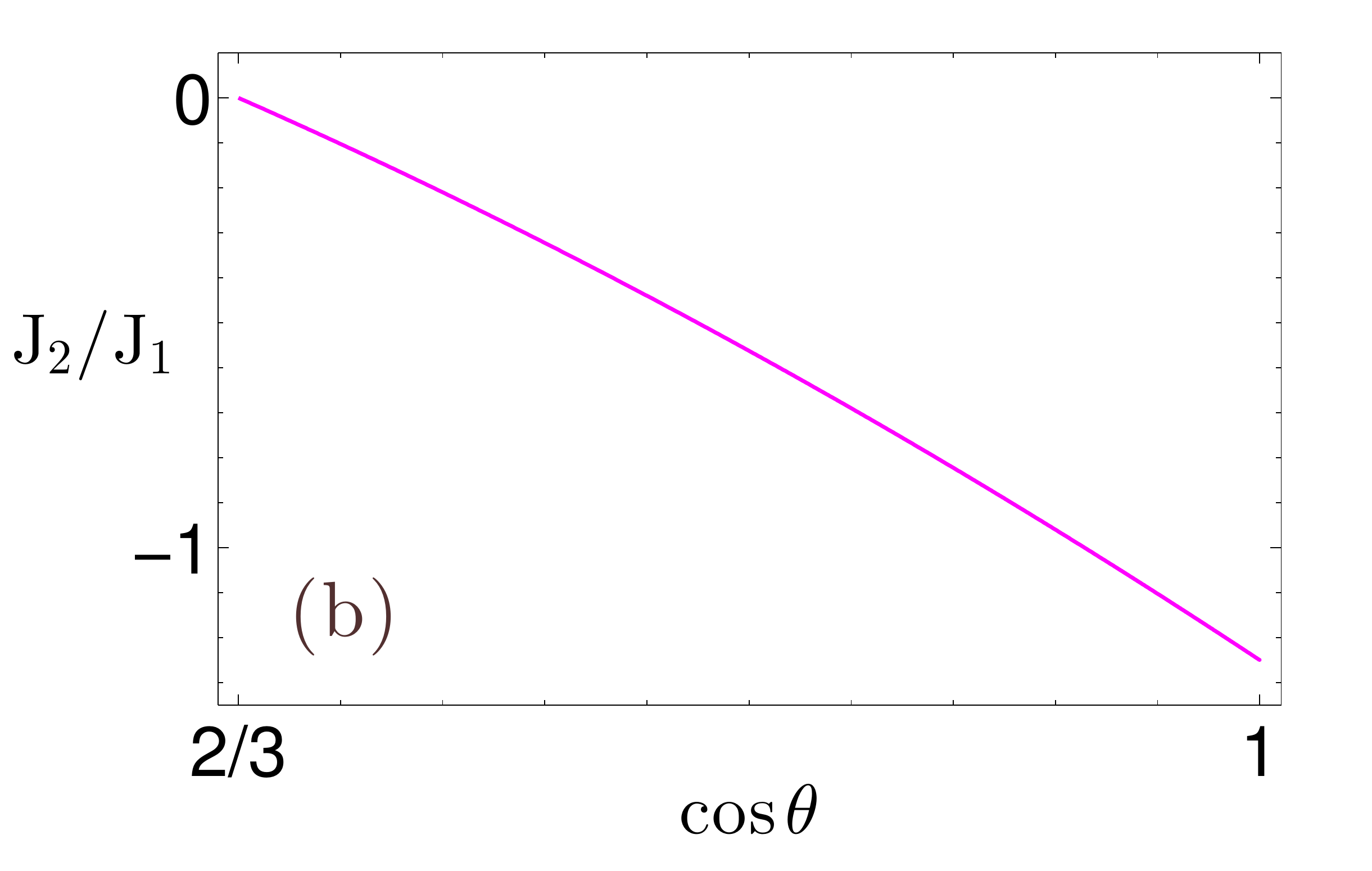}
\hspace{-0.2cm}
\includegraphics[width=4.5cm]{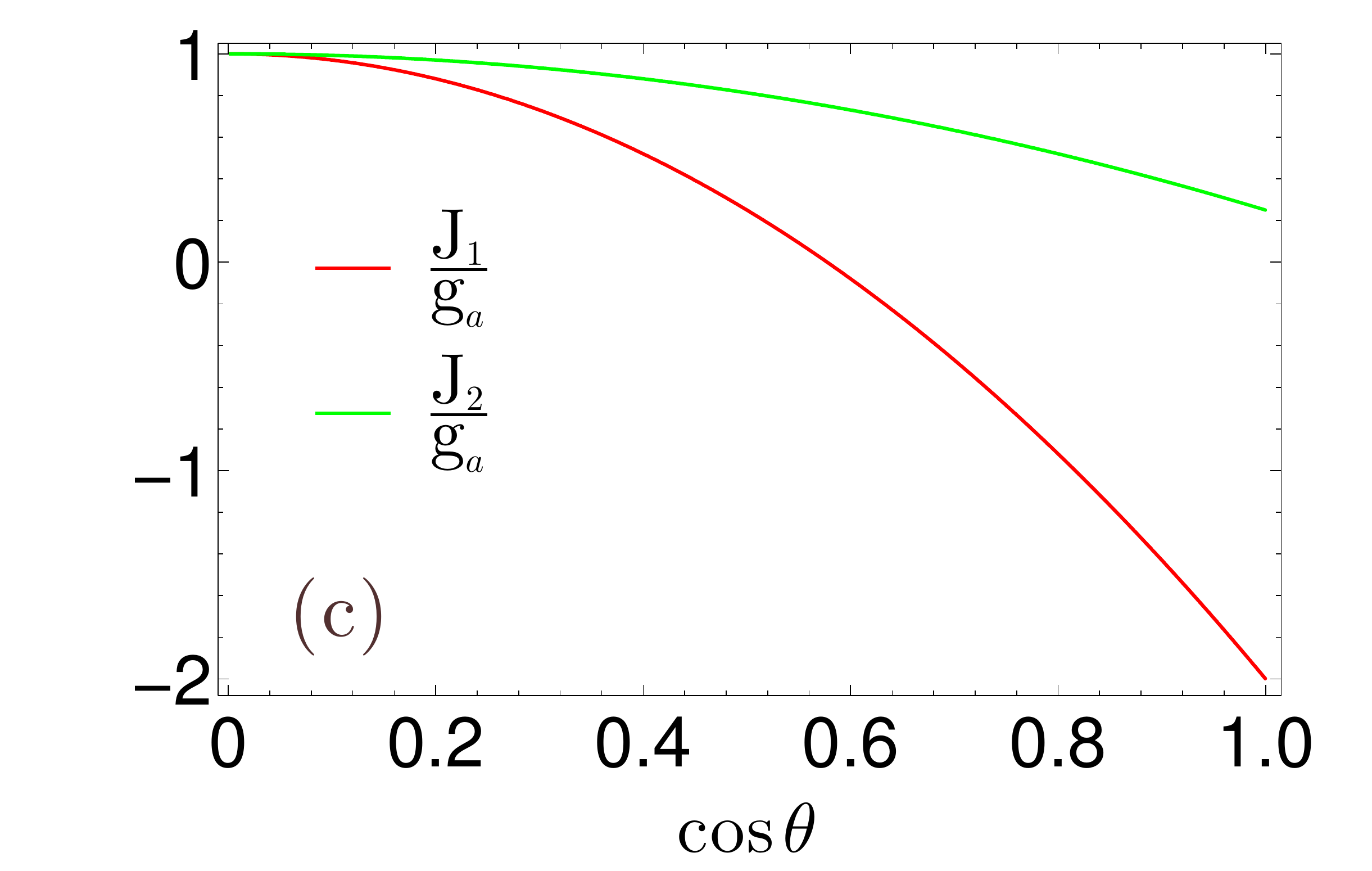}
\hspace{-0.5cm}
\includegraphics[width=4.5cm]{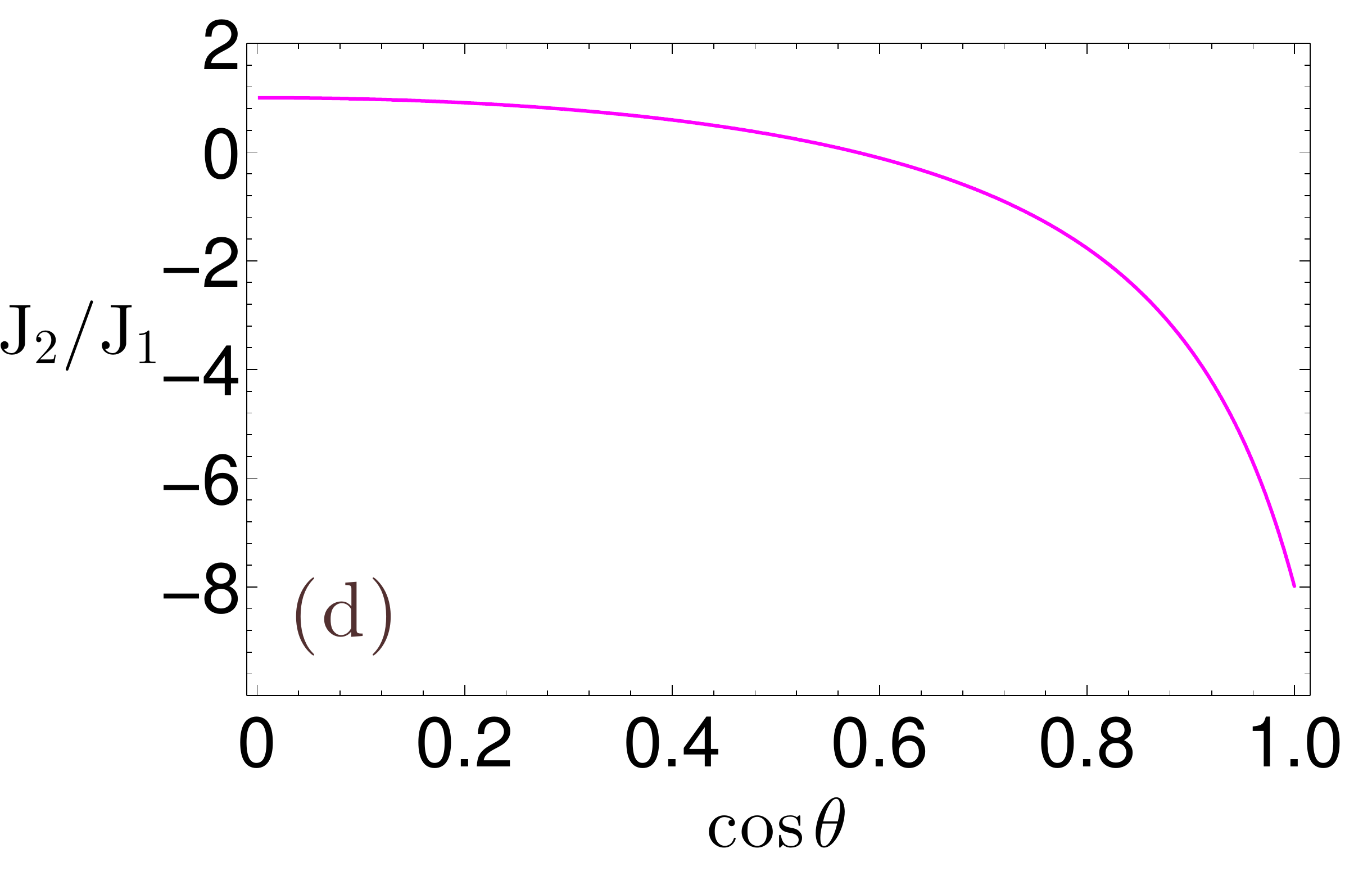}
\end{center}
\caption{The nearest neighbor interaction strength $J_{\hat{a}_k}$ in the unit of $g_a=6.8\mbox{kHz}$ as a function of the direction of magnetic field $\hat{m}=(0,\cos\theta,\sin\theta)$ (a) and $\hat{m}=(\cos\theta,0, \sin\theta)$ (b): $J_{\hat{a}_1}\equiv J_1$ (red), $J_{\hat{a}_2}=J_{\hat{a}_3}\equiv J_2$ (green). The ratio $\frac{J_{\hat{a}_2}}{J_{\hat{a}_1}}= \frac{J_{\hat{a}_3}}{J_{\hat{a}_1}}\equiv \frac{J_2}{J_1}$ as a function of $\cos\theta$ for the magnetic field direction $\hat{m}=(0,\cos\theta,\sin\theta)$ (c) and $\hat{m}=(\cos\theta,0, \sin\theta)$ (d).}\label{fig:cT_S_S1-s}
\end{figure}

In the first example, where the magnetic field direction is $\hat{m}=(0,\cos\theta,\sin\theta)$, the nearest-neighbor couplings are $J_{\hat{a}_1}\equiv J_1=g_a$ (AF), $J_{\hat{a}_2}=J_{\hat{a}_3}\equiv J_2=g_a(1-\frac{9}{4} \cos^2 \theta)$, together with the ratio are plot in Fig.\ref{fig:cT_S_S1-s}(a-b). For $\cos\theta>\frac{2}{3}$, we obtain $J_2<0$ and thereby are ferromagnetic interaction. For the particular value $\cos\theta=\frac{2}{3}$, $J_2=0$, and the system consists of 1D (AF) chains interacting through weaker non-nearest-neighbor couplings. In the other example, where the magnetic field direction is $\hat{m}=(\cos\theta,0,\sin\theta)$, the nearest-neighbour interaction strengths are $J_{\hat{a}_1}\equiv J_1=g(a)(1-3 \cos^2 \theta)$, $J_{\hat{a}_2}=J_{\hat{a}_3}\equiv J_2=g(a)(1-\frac{3}{4} \cos^2 \theta)$, as represented in Fig.\ref{fig:cT_S_S1-s}(c-d). As the magnetic field direction changes, $J_2$ is always positive and thus is antiferromagnetic (AF) , while $J_1$ is negative (ferromagnetic) for $\cos\theta>1/3$.\\

\begin{figure}[h]
\begin{center}
\includegraphics[width=8cm]{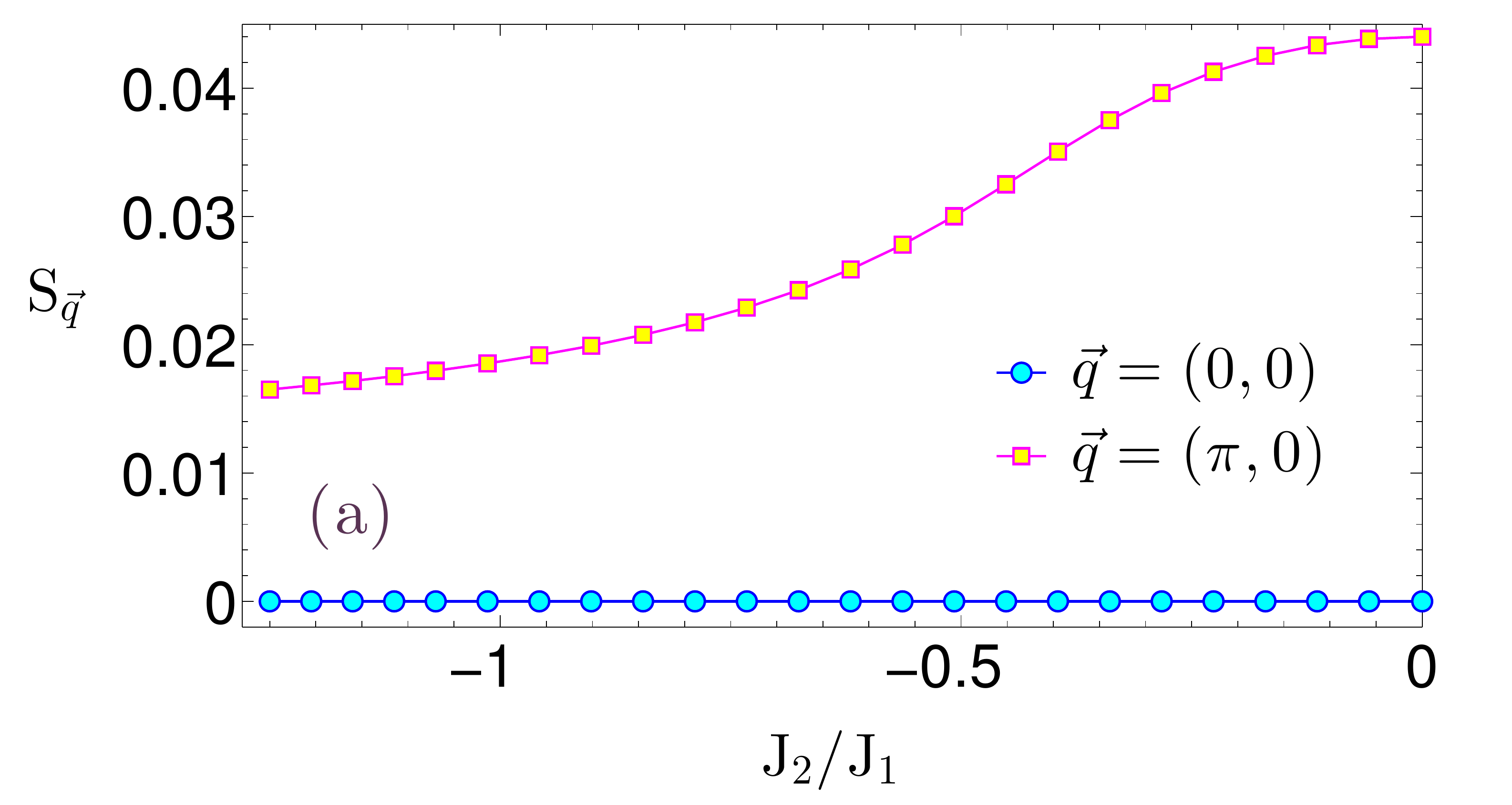}
\includegraphics[width=8cm]{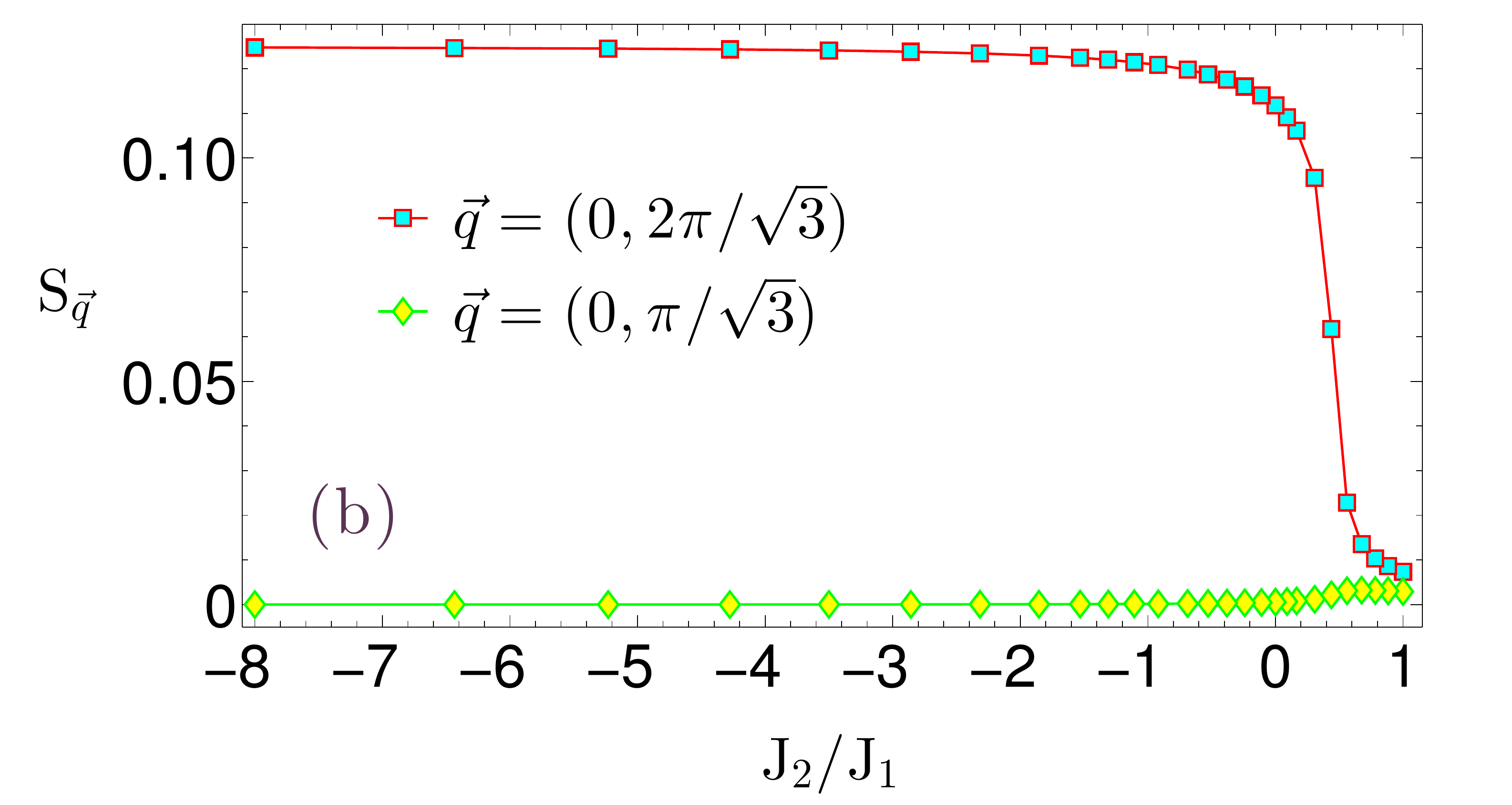}
\end{center}
\caption{Quantum magnetic phase transitions of fluorine quantum simulation on a triangular lattice for short-range models, i.e. only nearest-neighbor interactions are taken into account. (a) The spin structure factors changes with the magnetic field in the direction $\hat{m}=(0,\cos\theta,\sin\theta)$ as compared with Fig.5(a) in the main text. (b) The spin structure factors changes the magnetic field in the direction $\hat{m}=(\cos\theta,0,\sin\theta)$ as compared with Fig.5(b) in the main text. The spin structure factors are calculated using the Lanczos algorithm  for a 6$\times 4$ lattice under periodic boundary condition. }\label{fig:QMPST-s}
\end{figure}

{\it Comparison of quantum magnetic phases with the short-range model.---} In the main text, see Fig.5, we 
present two examples of quantum magnetic phase transitions that are controlled by the orientation of 
the external magnetic field. Here, we have taken into account long-range interactions up to 
the cut-off distance of $\sqrt{13}\mbox{a} \simeq 3.6 \mbox{a}$, where $\mbox{a}$ is the lattice constant .
In the first example with the magnetic field direction $\hat{m}=(0,\cos\theta,\sin\theta)$, we find that the 
competition between nearest-neighbor interactions $J_1$ and $J_2$, together with the strongest 
next-nearest neighbor interaction, leads to a transition towards the ferromagnetic phase as the value of 
$\cos\theta$ increases, see Fig.5(a) in the main text. We remark that the non-nearest neighbor 
interaction in fact plays an essential role in such a phase transition. For comparison, we calculate 
the spin structure factors for the short-range model with only nearest-neighbor interactions. It 
can be seen from Fig.\ref{fig:QMPST-s}(a) that the ferromagnetic phase does not appear in the 
corresponding parameter regime. \\

Let us move to the second example, where the magnetic field direction is $\hat{m}=(\cos\theta,0,\sin\theta)$. By only considering nearest-neighbor interactions (which are all negative values), the system does not display frustration, and we obtain the ferromagnetic order in the $\hat{a}_1$ direction, and the anti-ferromagnetic order in the $\hat{a}_2$ direction (F-AF phase), see the corresponding spin structure factors for this short-range model in Fig.\ref{fig:QMPST-s}(b). Moreover, as the value of $\cos\theta$ becomes smaller, $J_2$ decreases while the non-nearest-neighbor interactions now becomes comparable with $J_2$, see Fig.5(b) in the main text. Then, the triangle consists of $J_2$ and $J_{31}$, $J_{32}$, $J_{33}$ (see Fig.5(a) in the main text) becomes frustrated, and the competition between the nearest-neighbor interaction and long-range interactions leads to a new magnetically order phase. The spins are ferromagnetic in the $\hat{a}_1$ direction, while anti-ferromagnetic in the sublattice every second line in the  $\hat{a}_2$ direction, see Fig.5(b) in the main text. \\

\begin{figure}[h]
\begin{center}
\includegraphics[width=8cm]{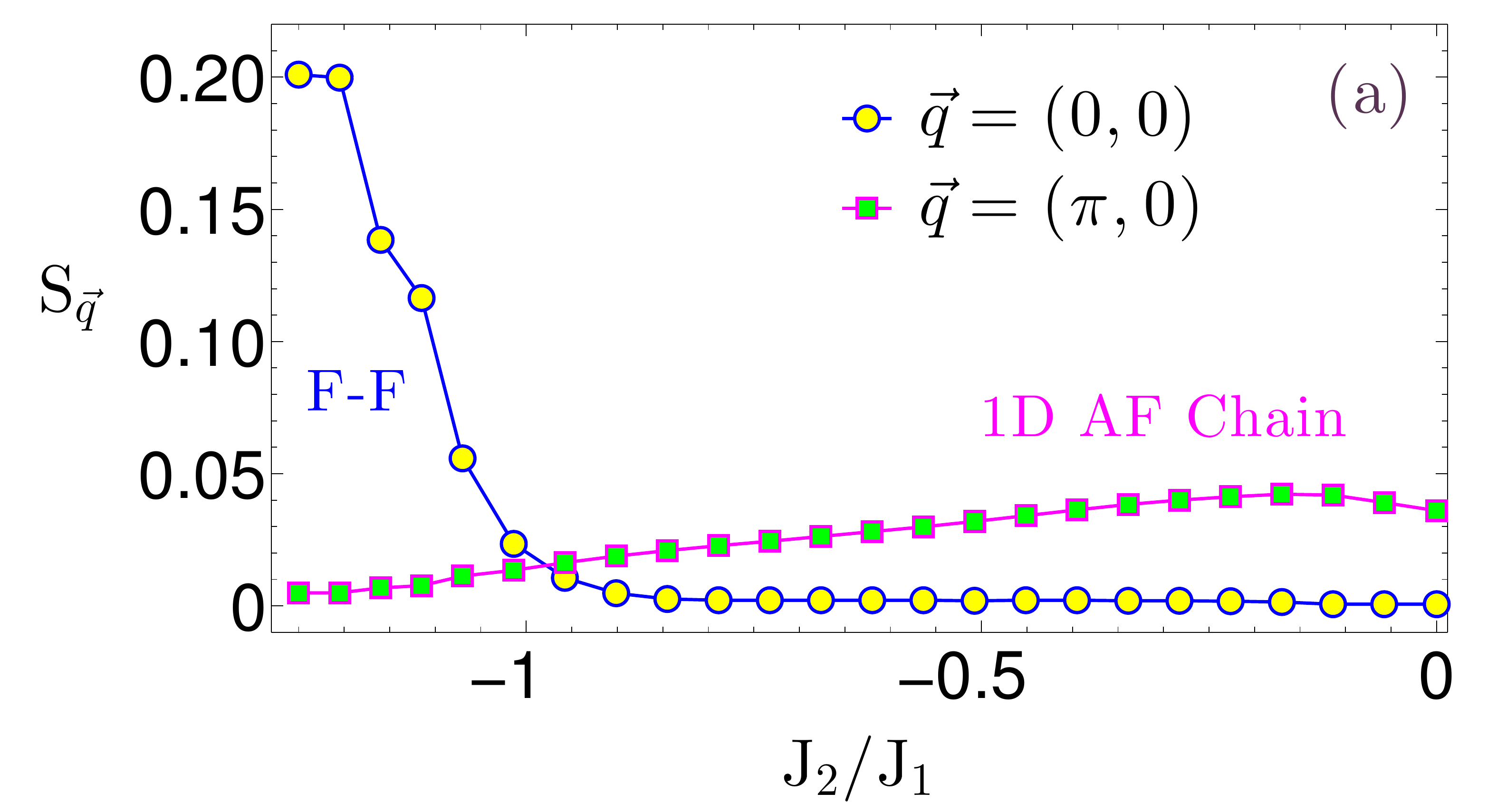}
\hspace{1cm}
\includegraphics[width=8cm]{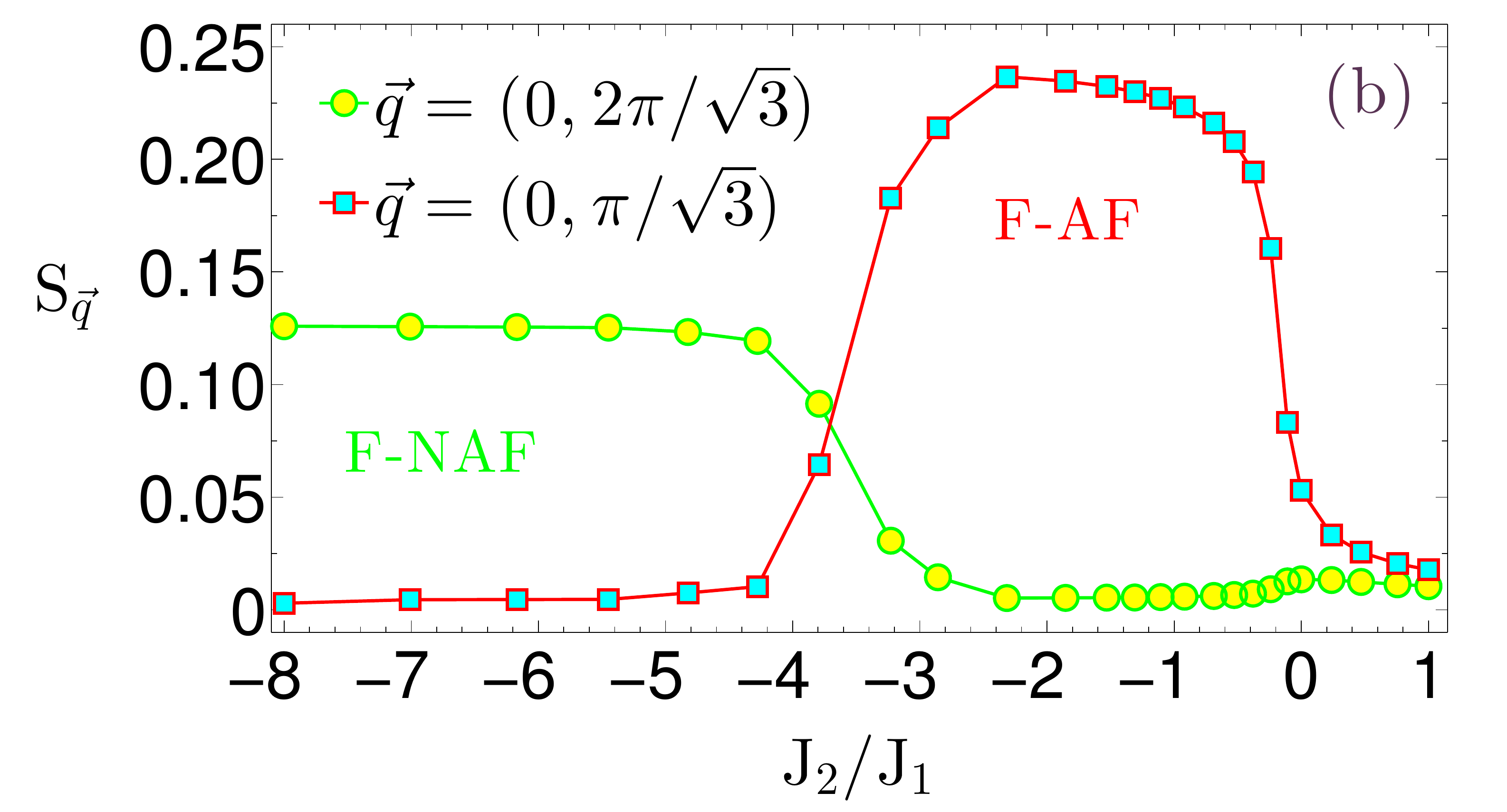}
\end{center}
\caption{Quantum magnetic phase transitions of quantum simulation on a triangular lattice with defects. (a) Tune the phase from 1D anti-ferromagnetic chains to ferromagnetic with the magnetic field in the direction $\hat{m}=(0,\cos\theta,\sin\theta)$ on a 6$\times 4$ lattice with $25 \% $ site depletion. (b) The phase transition from ferromagnetic-antiferromagnetic (F-AF) order to ferromagnetic-(alternative) antiferromagnetic (F-NAF) controlled by the magnetic field in the direction $\hat{m}=(\cos\theta,0,\sin\theta)$ on a 6$\times 4$ lattice with $20 \% $ site depletion. The spin structure factors are calculated using the Lanczos algorithm  under periodic boundary condition.}\label{fig:QMPT-s}
\end{figure}

{\it Quantum magnetic phase transitions with fluorine defects.---} We have demonstrated different 
magnetic phases in Fig.5 of the main text. Here, we demonstrate that key signatures for these phase 
transitions persist even when the lattice has defects, e.g. the dangling bond of carbon on diamond 
surface is terminated by spin-less nuclei rather than fluorine. We calculate the spin structure 
factors under the same magnetic field condition as in the main text, on a 6$\times 4$ lattice with 
more than $20\%$ sites at random positions being absent. The results, averaging over 30 random 
realization of the triangular lattice with defects, are shown in Fig.\ref{fig:QMPT-s}, which provide 
evidence that the transition between different quantum magnetism phases are still observable in the 
presence of significant disorder. We remark that the exact diagonalization is only performed on a 
relatively small lattice due to the limited computational overhead. Further numerical simulations 
will be necessary to determine the exact quantum phase diagram for various levels of defects rates.\\

\begin{figure}[t]
\begin{center}
\includegraphics[width=8cm]{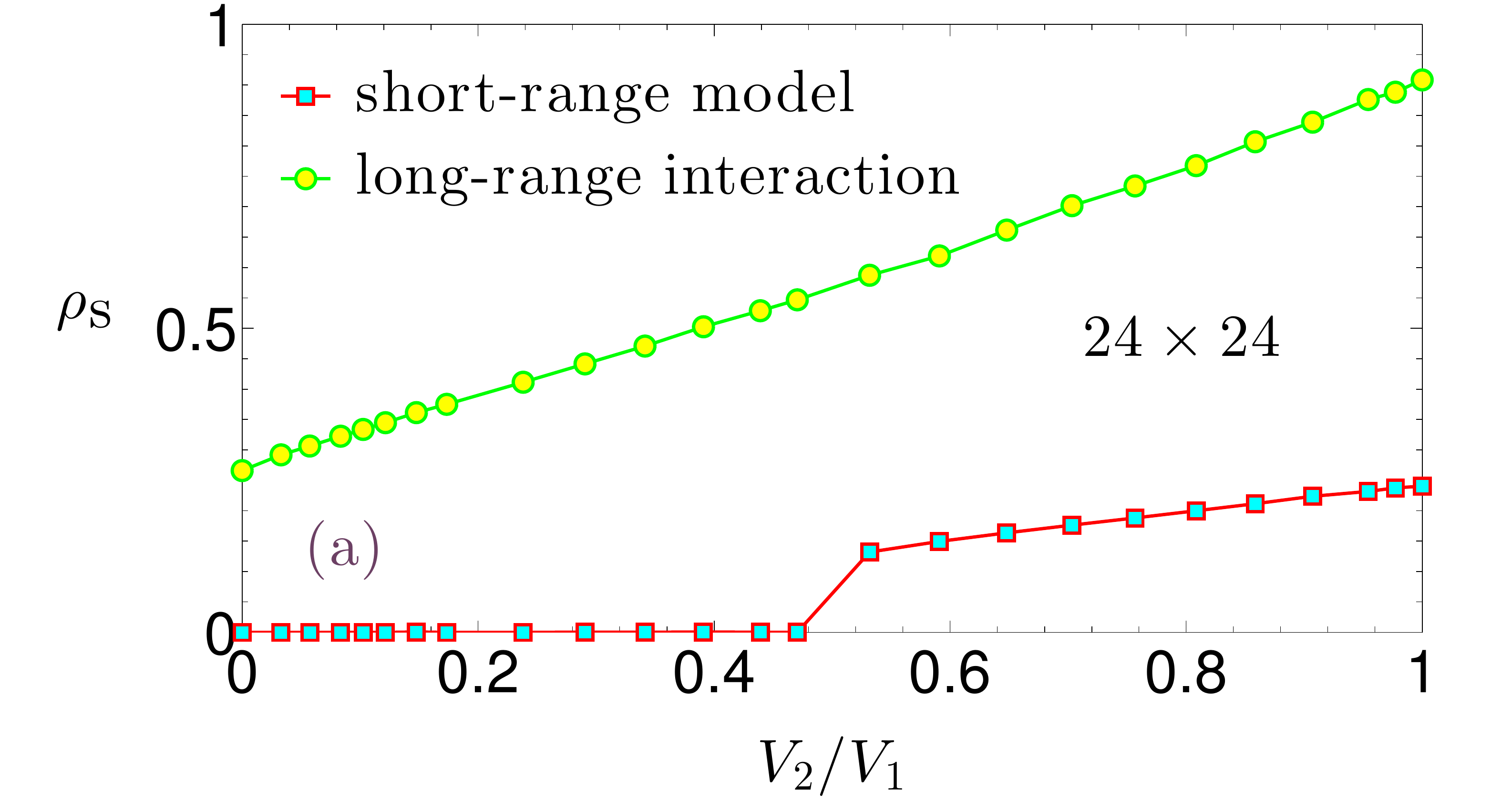}
\hspace{0cm}
\includegraphics[width=8cm]{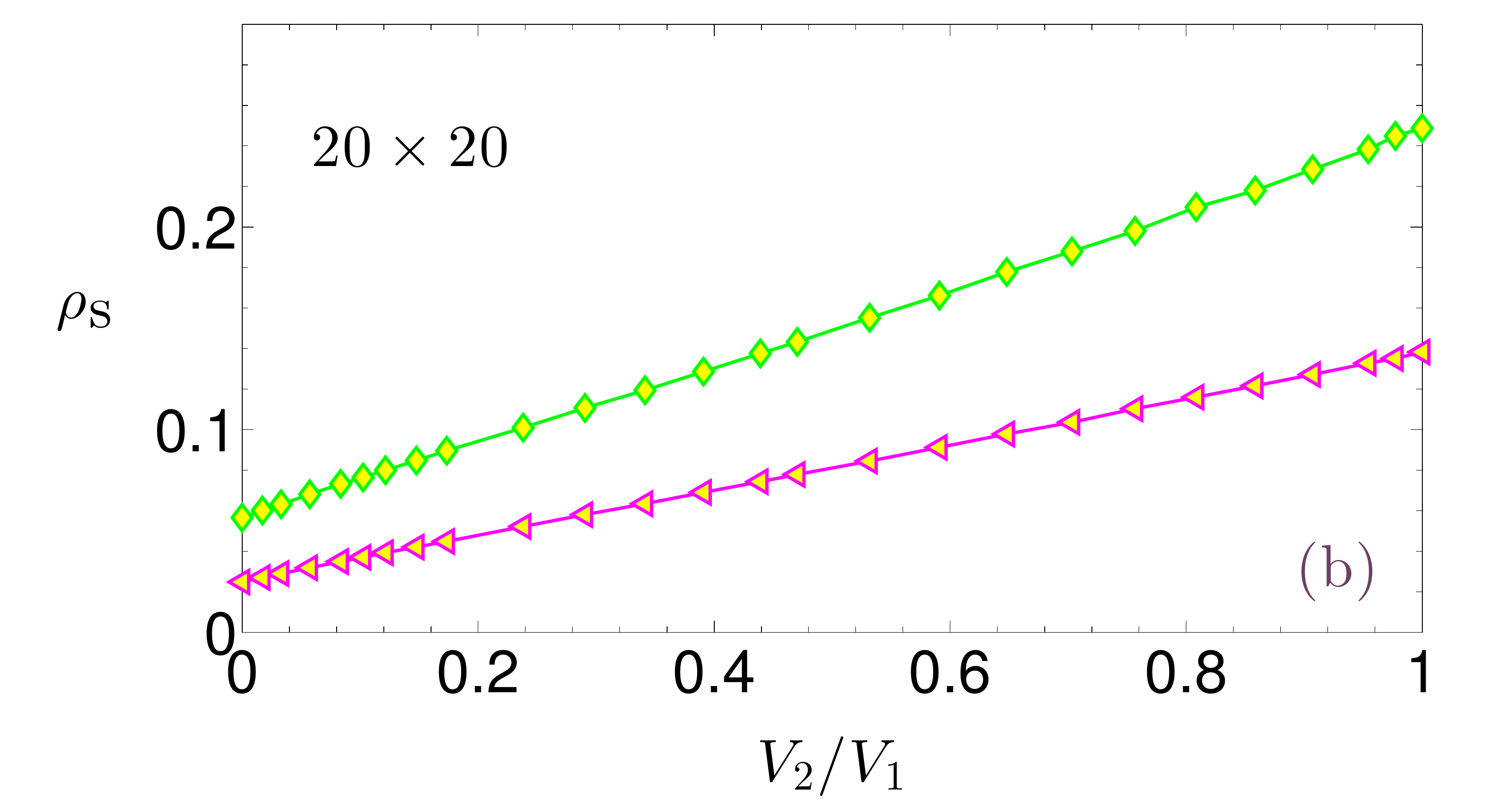}
\end{center}\caption{Enhancement of superfluidity by long-range interaction. (a) Comparison of superfluidity between our model with long-range interaction and the short-range model with the system size $24\times 24$. (b) Robustness of superfluidity with long-range interaction under the depletion of lattice with the system size $20\times 20$: the site depletion probability is $25\%$ (circle) and $50\%$ (square). The geometry frustration (as characterized by the ratio $V_2/V_1$) is controlled by the magnetic field direction as $\hat{m}=\cos\theta \hat{a}_3 +\sin\theta \hat{Z}$ with $\hat{Z}=(0,0,1)$. The temperature used in QMC simulation is $T/g_a=0.1$. }\label{fig:SF-s}
\end{figure}

{\it Hard-core boson superfluid and supersolid.---} The nuclear spin Hamiltonian of our system can be 
mapped to the hard-core boson model as
\be
    H_b=\sum_{\langle i,j \rangle }\left( V_{ij} n_i n_j- t_{ij} \l( a_i^\dagger a_j+ 
    a _i a_j^\dagger \r)\right) +\mu\sum_i  n_i
\ee
by the Holstein-Primakoff transformation \cite{HPT-s} $s_i^z=n_i-\frac{1}{2}$ ($n_i=0,1$ ), 
$s_i^{+}\equiv s_i^x+i s^i_y=a_i ^\dagger \sqrt{1-n_i} $ and $s_i^{-}\equiv s_i^x-i s^i_y = 
\sqrt{1-n_i} a_i $. Here, the chemical potential is $\mu=(\gamma_N B) - \sum_j  g(\mathbf{r}_{ij}) $, 
the repulsive interaction is $V_{ij}=  g(\mathbf{r}_{ij})$, and the hopping rate is $t_{ij} = \frac{\Delta}{2} g(\mathbf{r}_{ij})$. Depending on the ratio between the tunneling, the repulsive interaction $\frac{t}{V}=\frac{\Delta}{2}$ and the lattice geometric frustration, the system may demonstrate interesting phases such as the solid (S), superfluid (SF) or supersolid (SS) phases. The superfluid 
is a phase which possesses long-range off-diagonal order; while the supersolid exhibits both long-range 
off-diagonal and diagonal order, showing thus the features of a superfluid and a solid simultaneously. Our model has a large flexibility in tuning the geometric frustration and the ratio between the repulsive interaction and the hopping, both of which are long-range. It is thus possible to investigate rich phases of hard-core boson models. On the other hand, the model with frustration and long-range interactions poses great challenges to the classical numerical simulations. Quantum Monte Carlo (QMC) simulation which might be efficient for 2D systems are likely to become inefficient due to the sign problem. Here, we present simple examples which nevertheless can provide interesting insights into the properties of superfluid and supersolid phases. \\

\begin{figure}[h]
\begin{center}
\includegraphics[width=8.2cm]{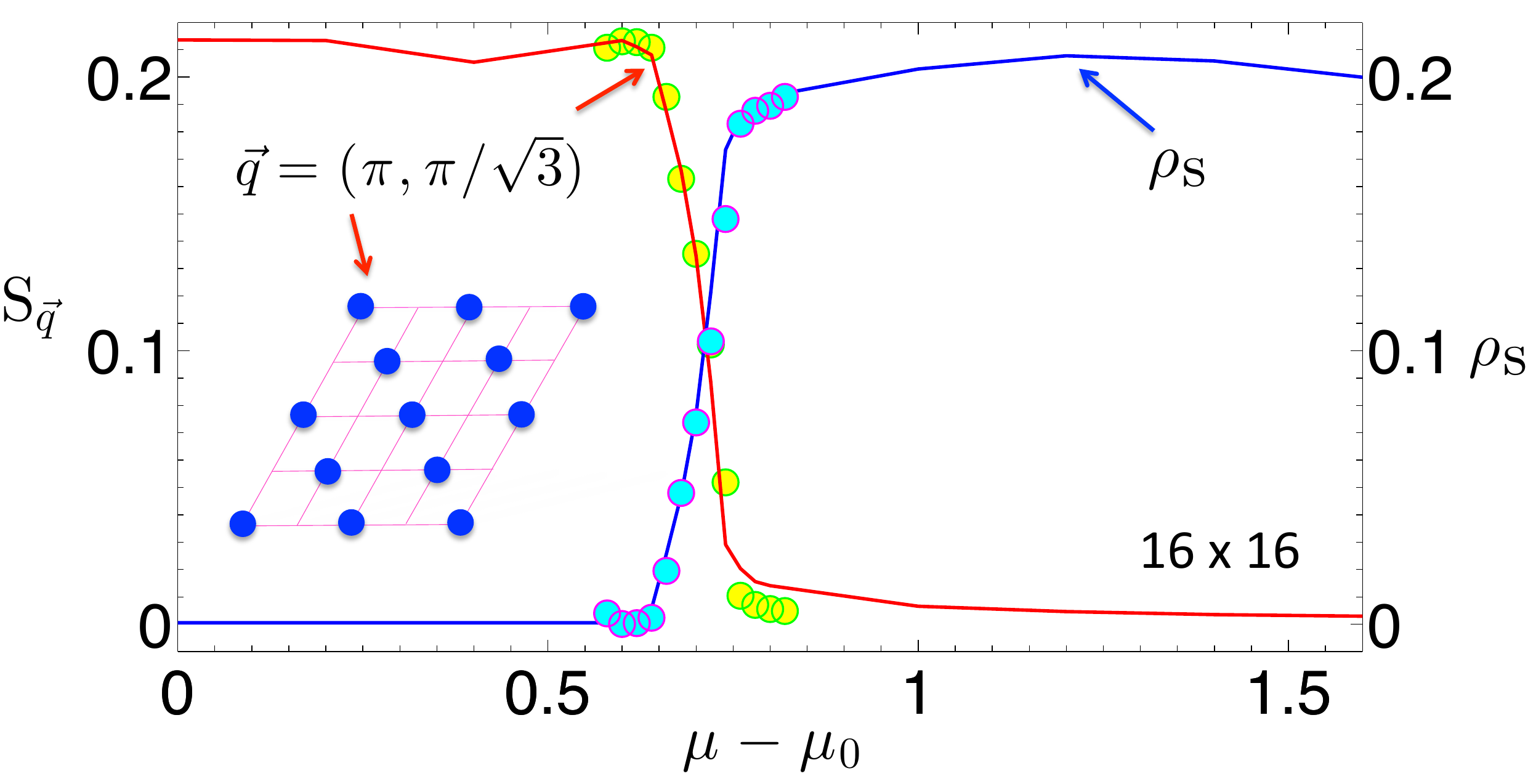}
\includegraphics[width=8.3cm]{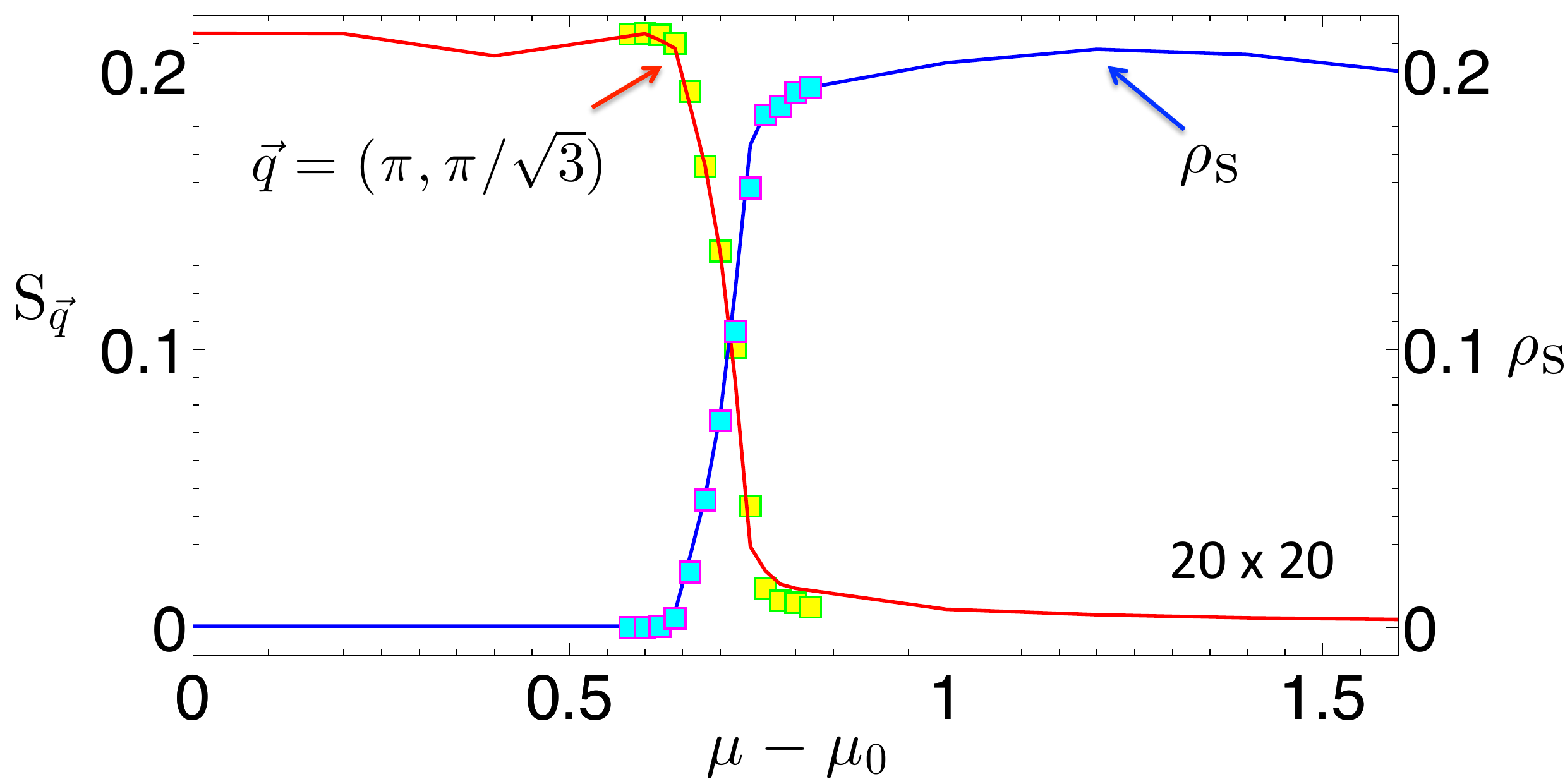}
\end{center}
\caption{The superfluid density $\rho_{\mbox{s}}$ and the normalized structure factor $\mbox{S}_{\vec{q}} (\pi,\pi/\sqrt{3})$ (see the illustration in the inset of a) as a function of chemical potential $\mu-\mu_0$, where $\mu_0$ is the chemical poterntial corresponding to half-filling, for a lattice of size $16\times 16$ (a) and $20\times 20$ (a) with the parameter $t/V=0.2$. We also plot the curves for the results of a lattice of size $12 \times 12$ for visibility. The magnetic field direction is $\hat{m}=\cos\theta \hat{a}_3 +\sin\theta \hat{Z}$ with $\cos\theta=\sqrt{\frac{1}{3}}$ and $\hat{Z}=(0,0,1)$. The temperature used in QMC simulation is $T/g_a=0.1$.}\label{fig:SFSS-s}
\end{figure}

With the magnetic field along the direction $\hat{m}=\cos\theta \hat{a}_3 +\sin\theta \hat{Z}$, where $\hat{Z}=(0,0,1)$, the nearest-neighbor interactions become $V_{\hat{a}_1}=V_{\hat{a}_2}\equiv V_1 =g_a(1-\frac{3}{4}\cos^2\theta)$, and $V_{\hat{a}_3}\equiv V_2=g_a(1-3\cos^2\theta)$. By changing the value of the magnetic direction angle $\theta$, we can gradually tune the geometric frustration, as quantified by the ratio $V_2/V_1$. The geometric frustration is important for the emergence of the superfluid, as can be seen from the result for the short-range model shown in Fig.\ref{fig:SF-s}(a). It can also be seen that taking long-range interactions into account will significantly enhance the superfluidity, which is robust under a relatively high level of lattice depletion (defects), see Fig.\ref{fig:SF-s}(b).\\

\begin{figure}[t]
\begin{center}
\includegraphics[width=7.5cm]{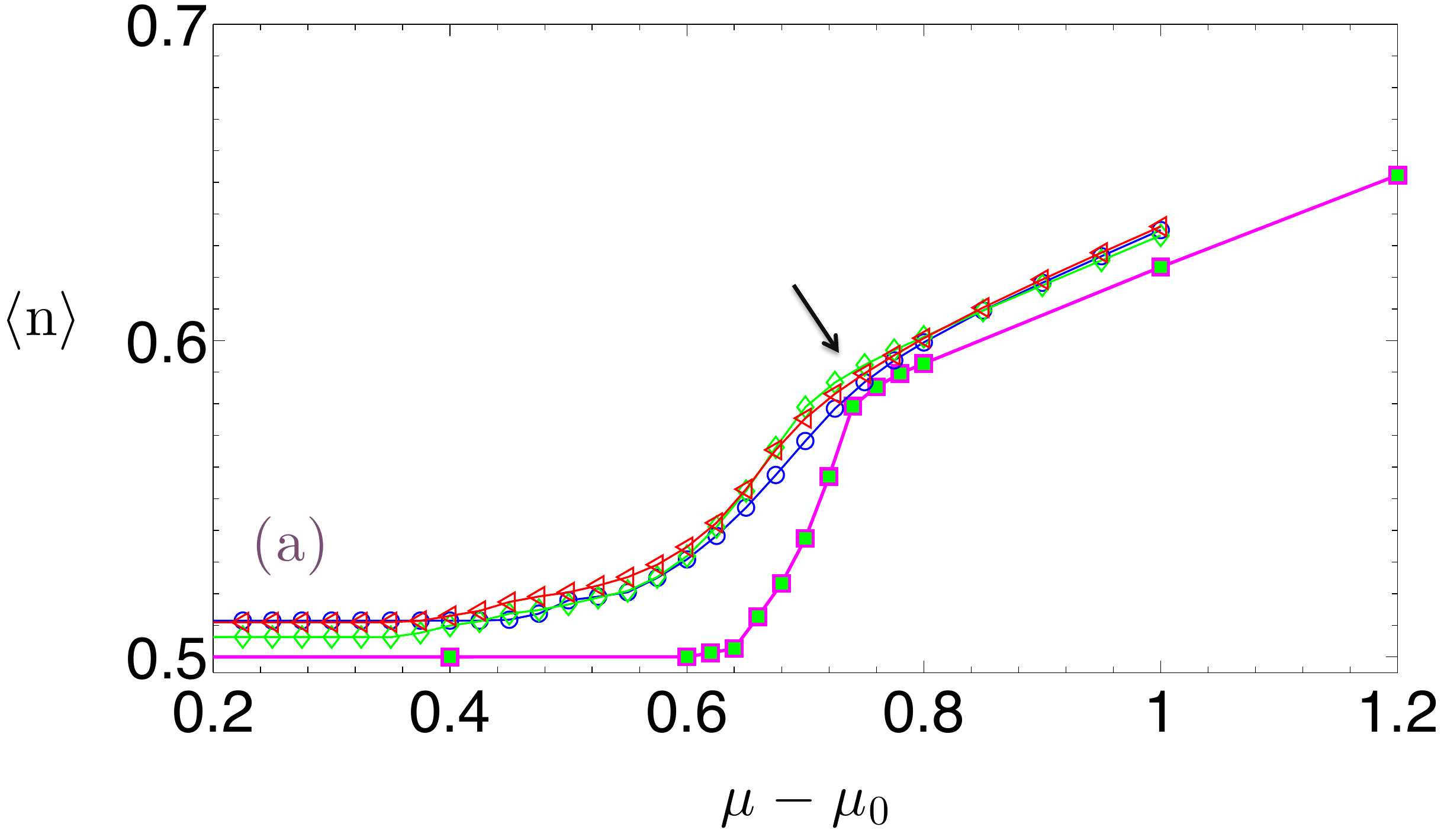}
\hspace{0.2cm}
\includegraphics[width=8.2cm]{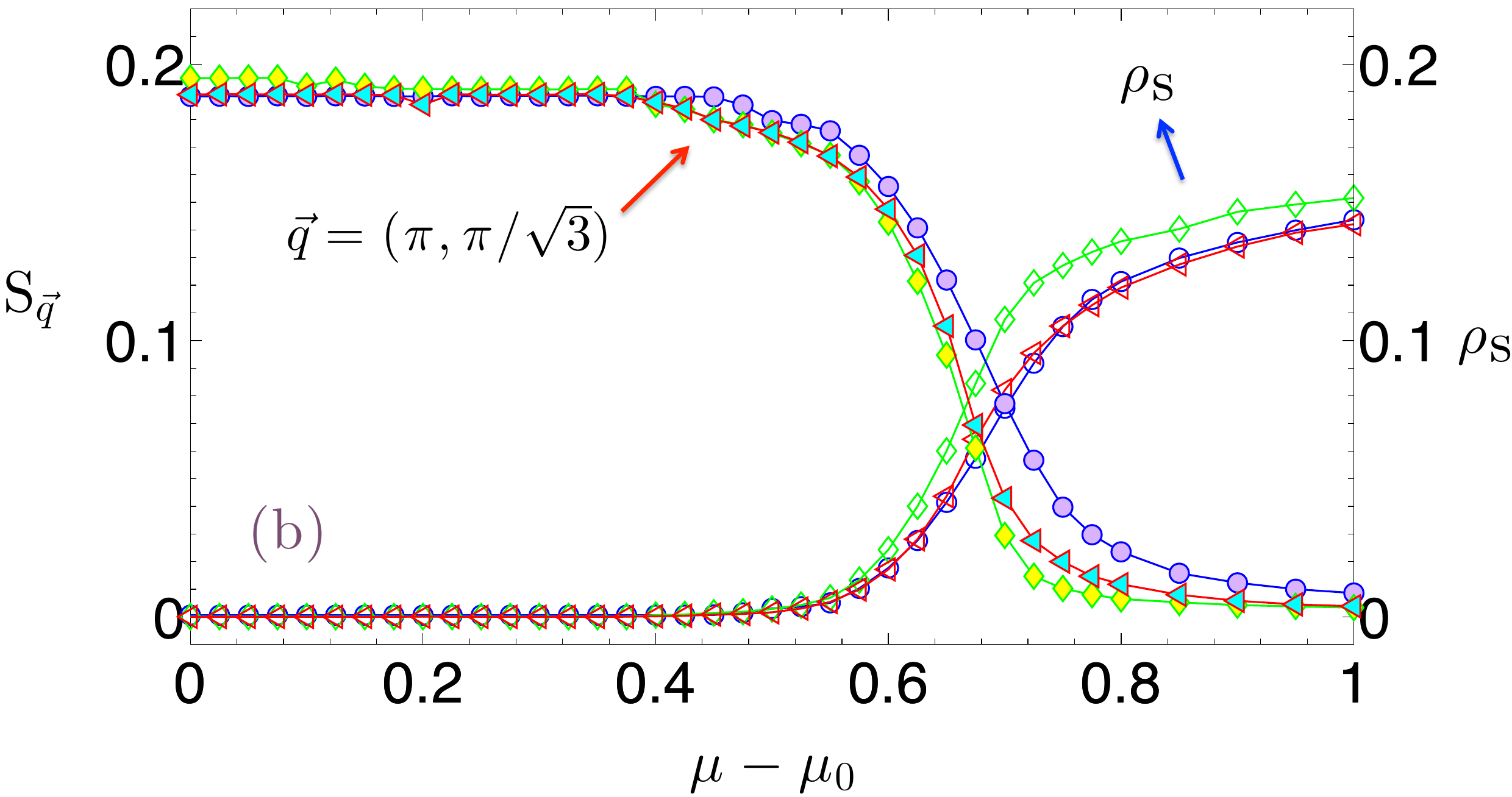}
\end{center}
\caption{Quantum phase transition of superfluid (SF) and supersolid (SS) with $10\%$ lattice depletion. (a) The average filling $\langle \mbox{n}\rangle$ as a function of chemical potential $\mu-\mu_0$, where $\mu_0$ is the chemical potential corresponding to half-filling. We also plot the curve for the same model without lattice depletion for visibility (square). The kink is indicated by the arrow. (b) The superfluid density $\rho_{\mbox{s}}$ and the normalized structure factor $\mbox{S}_{\vec{q}} (\pi,\pi/\sqrt{3})$. The system size is $12\times 12$ (blue, circle), $16 \times 16$ (green, diamond) and $20 \times 20$ (red, triangle). The magnetic field direction is $\hat{m}=\cos\theta \hat{a}_3 +\sin\theta (0,0,1)$ with $\cos\theta=\sqrt{\frac{1}{3}}$. The parameters are the same as Fig.6 in the main text: $t/V=0.2$, and the temperature used in our simulation is $T/g_a=0.1$.}\label{fig:SFSSLD-s}
\end{figure}

If we choose $\cos\theta=\sqrt{\frac{1}{3}}$ for the magnetic field direction $\hat{m}=\cos\theta \hat{a}_3 +\sin\theta \hat{Z}$, the lattice is equivalent to a 2D rectangular lattice. The model incorporates long-range repulsive and hopping interactions. For the ratio $t/V=0.2$, we have calculated the superfluid density $\rho_{\mbox{s}}$ and the normalized structure factor $\mbox{S}_{\vec{q}} (\pi,\pi/\sqrt{3})$ for a lattice of size $12 \times 12$, which evidences the solid (S), superfluid (SF) and supersolid (SS) phases, see Fig.6 in the main text. In Fig.\ref{fig:SFSS-s}, we show the results for a lattice of size $16\times 16$ and $20\times 20$, and find that both $\rho_{\mbox{s}}$ and $\mbox{S}_{\vec{q}} (\pi,\pi/\sqrt{3})$ are finite and size independent in the regime $\mu -\mu_0 \in [0.6,0.8]$. This provides evidence supporting the existence of the supersolid phase. We have also performed simulations including the lattice site depletion. The results are shown in Fig.\ref{fig:SFSSLD-s}. As compared with Fig.6 in the main text, the lattice depletion would downgrade the superfluidity. Nevertheless, the supersolid phase may still exist with a certain level of lattice depletion ($10\%$ in our calculation). The full study of quantum phases with lattice depletion for the present quantum simulation will be very interesting, which is beyond the scope of the present work and will be addressed in detail in a future publication.  \\

\begin{figure}[b]
\begin{center}
\begin{minipage}{16cm}
\hspace{-0.3cm}
\includegraphics[width=7cm]{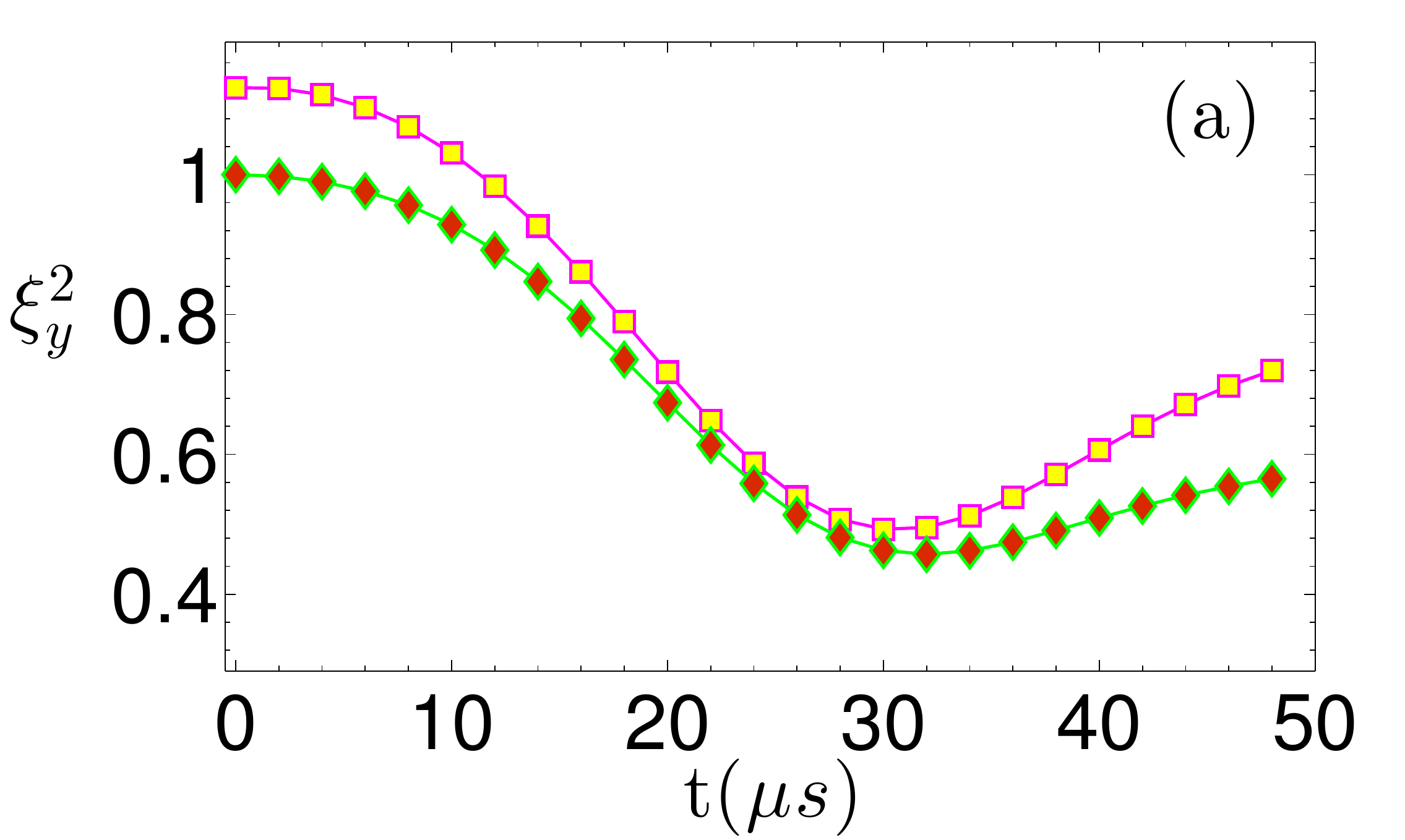}
\hspace{1cm}
\includegraphics[width=7cm]{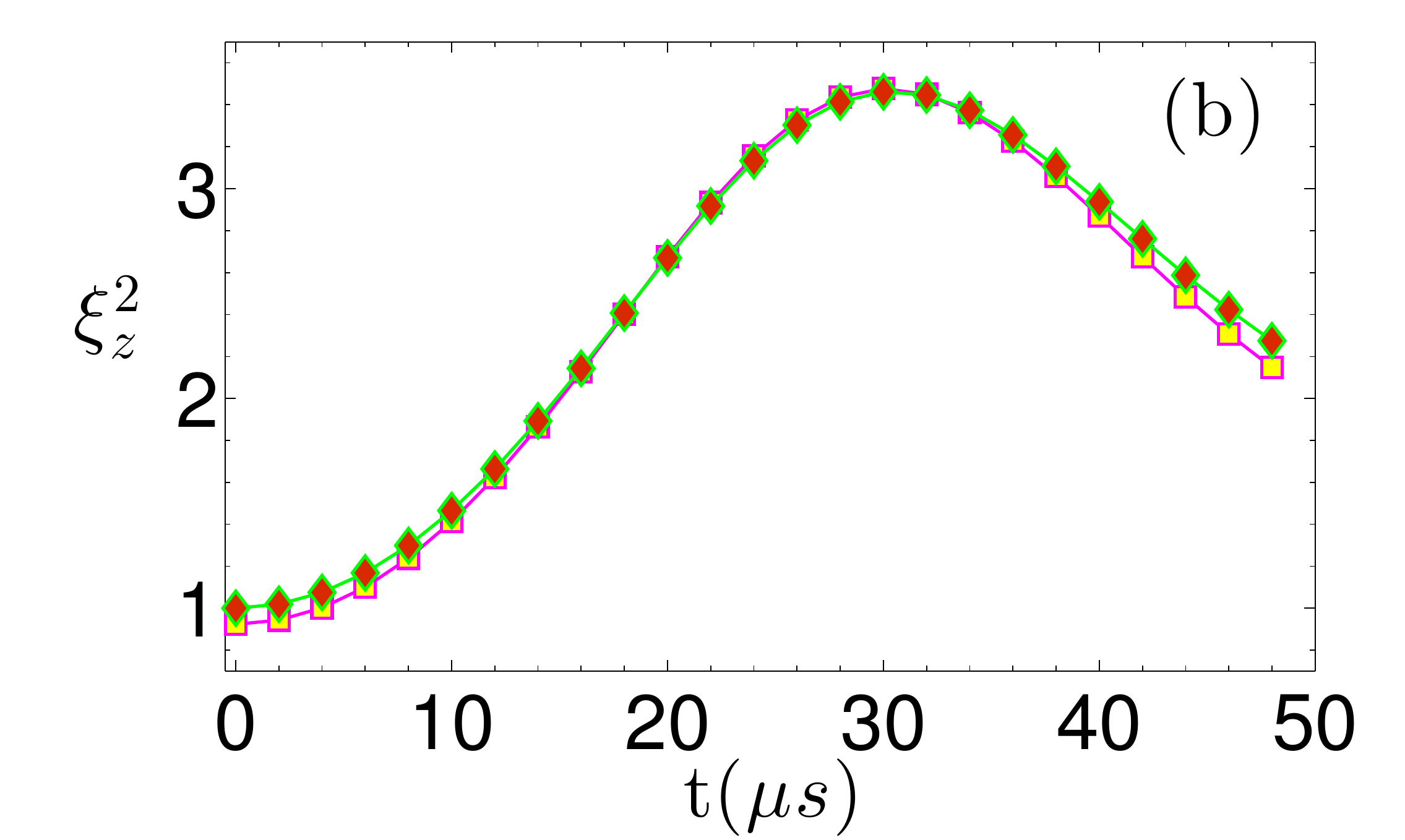}
\end{minipage}
\end{center}
\caption{Squeezing parameter $\xi^2_{y} $ (a) and $\xi^2_{z} $ (b) as a function of the evolution time $\mbox{t}$ for the fluorine simulation on a 4$\times$4 triangular lattice. We compare the estimation with NV center measurement (purple square) with the exact values (green diamond).  }\label{fig:SPE}
\end{figure}

{\it Quantum quench and spin squeezing.---} To illustrate our ideas, we consider the dynamics of a fluorine quantum simulator on a triangular lattice which undergoes a quantum quench. After the polarization, the system is prepared in the product state $\ket{\psi(0)} = \ket{{\uparrow_x}} \otimes \cdots \otimes \ket{{\uparrow_x}}$, which is the eigenstate of the Hamiltonian with a large resonant RF-field $H_{sq}=\sum_{(i,j)}  g(\mathbf{r}_{ij})\left[ \mathbf{s}_i^z \mathbf{s}_j^z-\Delta \l(\mathbf{s}_i^x \mathbf{s}_j^x+\mathbf{s}_i^y\mathbf{s}_j^y\r)\right] +\Omega_{sq}  \sum_i \mathbf{s}_i^x$. Then we suddenly decrease $\Omega_{sq}$ to a value that is comparable with $g(\mathbf{r}_{ij})$ (in our numerical calculation, we use the value of $10 \mbox{kHz}$). After such a quantum quench, the state $\ket{\psi(0)}$ is not anymore the eigenstate of the new Hamiltonian and the nuclear spins start to entangle with each other. We remark that during the dynamics, the NV center is polarized to the state $m_s=0$ to avoid its effect on the nuclear spins. We characterise the system dynamics by the quantity of spin squeezing, which shows a reduced variance of the collective angular momentum even when the mean angular momentum in an orthogonal direction is large. The squeezing parameter is $\xi^2_{\alpha}\equiv 2 \l( \Delta J_{\alpha}\r )^2 /|\langle J_x \rangle |$, where the collective angular momentum is $J_{\alpha}=\sum_k I_k^{\alpha}$ and $\l( \Delta J_\alpha \r )^2=\langle J_\alpha^2\rangle -\langle J_\alpha \rangle ^2$ with $\alpha=x,y,z$. Spin states with $\xi^2_{\alpha} < 1$ are referred to as spin squeezed states and necessarily entangled \cite{Toth09-s}. We perform exact numerical simulation for a 4$\times$4 triangular lattice and calculate the evolution of squeezing parameters, see Fig.\ref{fig:SPE}. This shows that the angular momentum variance in the $\hat{y}$ direction is reduced. To measure the squeezing parameters with the NV center, we apply the Hadamard operation $O_H$ and then measure the nuclear spin magnetization which gives us the estimation of $\langle J_{x} \rangle$. The value of $\langle J_\alpha^2\rangle $ can be approximated by $(\Delta_{\alpha\beta}+\Delta_{\alpha\gamma}-\Delta_{\beta\gamma})/4$. For the simplicity of numerical calculation, we choose the measurement time $\tau=\min\{P_\mu^{\nu}\vert_\tau =10\%, 42 \mu s\}$, and the observable is estimated by the measured quantity $P_\mu^{\nu}/\tau^2$ with $\mu,\nu =\pm$.

\end{document}